\documentclass[iop]{emulateapj}
\usepackage{natbib}
\usepackage{amsmath}
\usepackage{xfrac}
\usepackage[utf8]{inputenc}

\providecommand{\e}[1]{\ensuremath{\times10^{#1}}}

\def \logTd6 {\hbox{log$( T/6 \kev)$} }

\def \mag	{\,\mathrm{mag}}
\def \Ry	{\,\mathrm{Ry}}
\def \pMpc	{\,\mathrm{pMpc}}
\def \cMpc	{\,\mathrm{cMpc}}
\def \Mly	{\,\mathrm{Mly}}	
\def \Myr	{\,\mathrm{Myr}}
\def \um	{\,\mathrm{\mu{}m}}
\def \Gqso	{\Gamma^\mathrm{HeII}_\mathrm{QSO}}
\def \Gsl	{\Gamma^\mathrm{HeII}_\mathrm{SL}}
\def \tHeEQ	{t^\mathrm{HeII}_\mathrm{eq}}
\def \tDC	{t_\mathrm{dc}}
\def \tQ	{t_\mathrm{Q}}
\def \tTO	{t_\mathrm{age}}
\def \xiTPE	{\xi_\mathrm{\,TPE}}

\def \arcmin     { ^{\prime} }
\def \arcsec    {^{\prime\prime}}

\def \yr        {{\rm\ yr}}

\def \kev       {{\rm\ keV}}

\newcommand{\lya}{{\rm Ly}\alpha}

\DeclareMathAlphabet{\mathsc}{OT1}{cmr}{m}{sc}
\DeclareRobustCommand{\ion}[2]{%
  \relax
  \ifmmode
    \ifx\testbx\f@series
      {\mathbf{#1\,\mathsc{#2}}}
    \else
      {\mathrm{#1\,\mathsc{#2}}}
    \fi
  \else
    \textup{#1\,{\mdseries\textsc{#2}}}%
  \fi
 }

\begin{document}

\lefthead{Statistical Detection of the \ion{He}{ii} Transverse Proximity Effect}\righthead{T. M. Schmidt et al.}
\title{
Statistical Detection of the \ion{He}{ii} Transverse Proximity Effect: Evidence for Sustained Quasar Activity for $>$25 Million Years
}
\author{Tobias M. Schmidt\altaffilmark{1}\altaffilmark{*}, Gabor Worseck\altaffilmark{1}, Joseph F. Hennawi\altaffilmark{1,2}, J. Xavier Prochaska\altaffilmark{3}, Neil H. M. Crighton\altaffilmark{4}}
 
\altaffiltext{*}{e-mail: tschmidt@mpia.de}
\altaffiltext{1}{Max-Planck-Institut f\"ur Astronomie, K\"onigstuhl 17, D-69117 Heidelberg, Germany}
\altaffiltext{2}{Department of Physics, University of California, Santa Barbara, CA 93106, USA}
\altaffiltext{3}{Department of Astronomy and Astrophysics, UCO/Lick Observatory, University of California, 1156 High Street, Santa Cruz, CA 95064, USA}
\altaffiltext{4}{Centre for Astrophysics and Supercomputing, Swinburne University of Technology, PO Box 218, VIC 3122, Australia}

\begin{abstract}

The \ion{He}{ii} transverse proximity effect -- enhanced \ion{He}{ii} $\lya$~transmission in a background sightline caused by the ionizing radiation of a foreground quasar -- offers a unique opportunity to probe the morphology of quasar-driven \ion{He}{ii} reionization. 
We conduct a comprehensive spectroscopic survey to find $z\sim3$ quasars in the foreground of 22 background quasar sightlines with \textit{HST}/COS \ion{He}{ii} $\lya$~transmission spectra. 
With our two-tiered survey strategy, consisting of a deep pencil-beam survey and a shallow wide-field survey, we discover 131 new quasars, which we complement with known SDSS/BOSS quasars in our fields.
Using a restricted sample of 66 foreground quasars with inferred \ion{He}{ii} photoionization rates greater than the expected UV
background at these redshifts ($\mathrm{\Gqso > 5 \times 10^{-16}\,{s}^{-1}}$) we perform the first statistical analysis of the \ion{He}{ii} transverse proximity effect. Our results show qualitative evidence for a large object-to-object variance: among the four foreground quasars with the highest $\Gqso$ only one (previously known) quasar is associated with a significant \ion{He}{ii} transmission spike. We perform a stacking analysis to average down these fluctuations, and detect an excess in the average \ion{He}{ii} transmission near the  foreground quasars at $3\sigma$ significance.
This statistical evidence for the transverse proximity effect is corroborated by a clear dependence of the signal strength on $\Gqso$.
Our detection places a purely geometrical lower limit on the quasar lifetime of $\tQ > 25\,\mathrm{Myr}$.
Improved modeling would additionally constrain quasar obscuration and the mean free path of \ion{He}{ii}-ionizing photons.

\end{abstract}


\section{Introduction}

The double ionization of helium at $z\sim 3$, known as \ion{He}{ii} reionization, is the final phase transition of the intergalactic medium (IGM)
and substantially influences the thermal and ionization state of the baryons in the Universe.
According to the currently accepted picture \citep{HaardtMadau2012, Planck2016}
hydrogen was reionized at redshifts $z \sim 8$ by the UV photons emitted from stars.
Their spectra, however, were not hard enough to supply sufficient numbers of photons with energies $>4\:\mathrm{Ry}$ required to doubly ionize helium.
Therefore, \ion{He}{ii} reionization was delayed until the quasar era, culminating in the completion of helium reionization at $z\approx2.7$ \citep{Madau1994, Reimers1997, MiraldaEscude2000, McQuinn2009, Faucher-Giguere2009, Worseck2011b, HaardtMadau2012,Compostella2013, Compostella2014, Worseck2016}.
Since quasars are bright but rare sources, it is expected that \ion{He}{ii} reionization is a very patchy process. The general picture is that quasars create ionized bubbles around them which expand with time, and eventually overlap to form 
the relatively homogeneous UV background that keeps the IGM in photoionization equilibrium up to the present day \citep{Bolton2006, Furlanetto2008, McQuinn2009, Furlanetto2010, Furlanetto2011, HaardtMadau2012,  Meiksin2012, Compostella2013, Compostella2014}.

The exact structure of \ion{He}{ii} reionization crucially depends on the duration of the bright quasar phase.
To match a given luminosity function, quasars either have to be very common objects which individually shine only for a short time or rare objects if the lifetime is long \citep[e.g.][]{Martini2004}.
The quasar lifetime also places important constraints on the physics and accretion processes of active galactic nuclei (AGN).
All massive galaxies are expected to host a supermassive black hole (SMBH) in their center but most of them are inactive and do not accrete matter most of the time. 
It is however assumed that all of them went through an active AGN phase at least once in their history and SMBHs at least partially grew to their current mass during bright quasar phases \citep{Soltan1982, Shankar2009, Kelly2010}. 
Also, quasars might have a substantial influence on their host galaxies, and simulations of galaxy formation usually invoke quasar feedback to match the galaxy luminosity function at the massive end. 
Constraining the duration of the quasar phase might shed light on the largely unknown triggers for AGN activity and thereby test quasar models (e.g. \citealt{Springel2005, Hopkins2005, Schawinski2015}).

There have been several attempts for measuring the quasar lifetime
\citep[e.g.][and references therein]{Martini2004}. 
An upper limit of  $10^9\yr$ comes from demographic arguments and the evolution of the AGN population. 
Methods based on the quasar luminosity function and clustering statistics are in principle able to determine the average 
time galaxies host luminous quasars, described as the duty cycle ($t_\mathrm{dc}$) of quasars. 
However, studies so far only deliver weak constraints of $t_\mathrm{dc} \sim 10^6 - 10^9\yr$  \citep{Adelberger2005, Croom2005, Shen2009, White2012, Conroy2013, LaPlante2015} due to the uncertainty in how quasars populate dark matter halos.
Studies focusing on the mass assembly of SMBHs are sensitive to the total black hole growth time which might also include non-luminous or obscured phases. Constraints from such studies \citep[e.g.][]{Yu2002, Kelly2010} are between $30$ and $150\Myr$.

These methods determine the quasar duty cycle or growth time for a full population, reflecting the total time quasars are active. This might be distinct from the duration of single bursts if quasars go through several activity periods over a Hubble time. The duration of these can be described by the episodic lifetime $t_\mathrm{Q}$.
Methods which aim to constrain the episodic lifetime usually focus on individual quasars and use the presence of a tracer that is sensitive to the quasar luminosity at some time in the past. This measures for how long an individual quasar has been active prior to the observation, basically its age. We denote this as $\tTO$ and stress that it is distinct from the quasar episodic lifetime $\tQ$ but clearly sets a lower limit on it ($\tTO \leq \tQ$).

Constraints for sustained activity over at least $10^6\yr$ comes e.g. from the presence of enormous $\lya$~nebulae around luminous quasars 
\citep[][]{Cantalupo2014, Hennawi2015}.
On slightly larger scales geometric constraints of $t_\mathrm{Q}\lesssim8\Myr$%
\footnote{%
Given the luminosity of hyperluminous quasars ($m_\mathrm{r} \approx 16\mag$ at $z=3$) and the typical sensitivity of narrowband surveys ($m_\mathrm{NB} \lesssim 26.5\mag$, $\mathrm{5\times10^{-18}\:erg\:s^{-1}\:cm^{-2}}$ for a $\mathrm{\Delta \lambda=50\:\AA}$ filter) achieved  in long integrations ($\mathrm{\sim5\:h}$) on 10\:m class telescopes (e.g. \citealt{Trainor2013}), the fluorescent emission from $1.0\arcsec$ sized galaxies can only be detected for separations from the illuminating quasar smaller than $\mathrm{\approx1.2\:Mpc}$. This allows one
to constrain quasar lifetimes $\lesssim 7.8\Myr$. The required integration time increases as the fourth power of the probed separation.} might be derived from fluorescent $\lya$ emission of galaxies caused by the UV radiation
of a nearby quasar as claimed by \citet[][]{Cantalupo2012}, \citet{Trainor2013} or \citet{Borisova2015}. 
\citet{Goncalves2008} investigate metal absorption systems in a $z=2.5$ quasar triplet and claim overionization due to the foreground
quasars' radiation and hence lifetime constraints of $t_\mathrm{Q} > 25\Myr$ and $15 < t_\mathrm{Q} < 33\Myr$, but do not place empirical constraints on the gas density, consider the effects of collisional ionization, or explore the effects of
non-solar relative abundances. Therefore, current constraints on the quasar episodic lifetime suggest values between $10^6$ and $10^8\yr$ (but see, \citealt{Schawinski2015}). 
All studies giving more specific constraints are either based on single or very few objects, or are indirect and therefore rely on assumptions.

The direct influence of quasars on the IGM can be clearly detected in quasar spectra as statistically lower IGM $\lya$ absorption in the vicinity of a quasar. This enhanced ionization is called the line-of-sight proximity effect \citep{Carswell1982, Bajtlik1988, Scott2000, DallAglio2008b, Calverley2011}. The presence of a line-of-sight proximity effect in quasar spectra sets a lower limit on the episodic lifetime of quasars. To cause an observable effect in the IGM they have to be active for at least the equilibration timescale. This is for \ion{H}{i} of the order of $t^\mathrm{HI}_\mathrm{eq} \sim 10^4\yr$ ($\mathrm{\Gamma^{HI}_{UVB}} \sim 10^{-12}\,\mathrm{s}^{-1}$, \citealt{Becker2013}).
A very similar effect is expected when a background sightline intersects the proximity zone of a foreground quasar, called the transverse proximity effect. 
It would enable robust estimates of the quasar lifetime purely based on geometric arguments and the light travel time from the foreground quasar to the background sightline \citep[e.g.][]{Dobrzycki1991, Adelberger2004}.
However, up to now there is no unambiguous detection of the transverse proximity effect the \ion{H}{i} $\lya$ forest,
which might be attributed to the cosmic overdensities in which quasars are hosted or the obscuration of the quasar radiation \citep{Liske2001, Schirber2004, Croft2004, Hennawi2006, Hennawi2007, Kirkman2008}.

In addition to the hydrogen $\lya$ forest, there is also a helium $\lya$ forest at frequencies $4\times$ higher, caused by singly ionized helium (\ion{He}{ii}).
Studying proximity effects in helium is for several reasons advantageous over probing the same effect in hydrogen.
At $z\lesssim 5.5$, the Universe is transparent to $1\Ry$ photons, resulting in a high and quasi-homogeneous UV background
\citep[e.g.][]{MiraldaEscude2000, Meiksin2004, Bolton2007c, Worseck2014}. A single quasar therefore causes a significant increase over the background only within a relatively small zone of influence.
At $z > 2.7$, before \ion{He}{ii} reionization is complete, helium offers a much larger contrast since the background is low, the mean free path short and single quasars can produce a stronger enhancement over the background, resulting in a much larger region where the total ionization rate is higher than
the UV background, far beyond the region in which the host halo causes a substantial enhancement of the cosmic density field \citep{Khrykin2016}.
Highly saturated \ion{He}{ii} spectra reaching effective optical depths $\tau_\mathrm{eff} \approx 5$ on scales $\Delta z = 0.04$ \citep{Worseck2016} constrain the \ion{He}{ii} fraction to $x_\mathrm{HeII}\approx 0.02$ at $z\approx 3$ \citep{Khrykin2016, Khrykin2017}. Under these conditions, a relatively small decrease in the \ion{He}{ii} fraction can cause a large increase in the transmission.

Limited by the Galactic and extragalactic \ion{H}{i} Lyman continuum \citep[e.g.][]{Worseck2011b}, the $z>2$ \ion{He}{ii} $\lya$ absorption is observable from space with far UV (FUV) spectrographs.
Over the last two decades the Hubble Space Telescope (\textit{HST}) has obtained \ion{He}{ii} Ly$\alpha$
spectra for a limited number of \ion{He}{ii}-transparent quasars \citep{Jakobsen1994, Reimers1997, Heap2000, Shull2010, Worseck2011b, Syphers2013, Syphers2014, Zheng2015, Worseck2016}. Many of these exhibit prominent line-of-sight proximity effects which itself sets a constraint on the quasar lifetime, similar to the hydrogen case \citep{Khrykin2016}. For \ion{He}{ii}, this is however approximately three orders of magnitude longer ($t_\mathrm{Q}\gtrsim10\Myr$) than for \ion{H}{i} due to the lower \ion{He}{II} photoionizing background ($\mathrm{\Gamma^{HeII}_{UVB}} \sim 10^{-15}\,\mathrm{s}^{-1}$, \citealt{Faucher-Giguere2009, HaardtMadau2012}) and therefore longer equilibration timescale \citep{Khrykin2016}.

The first evidence for a \ion{He}{ii} transverse proximity effect was found in the \ion{He}{ii} sightline to Q\,0302$-$003 that exhibits 
a transmission spike at $z=3.05$ \citep{Heap2000} and a foreground quasar at practically the same redshift \citep{Jakobsen2003}.  
This prominent example represents the prototype case for the \ion{He}{ii} transverse proximity effect. From the required transverse light crossing time in this 
association, one can infer a geometrical limit 
of the quasar lifetime of $t_\mathrm{Q} \approx 10\:\mathrm{Myr}$.
Although this estimate is only based on a single object, it demonstrates the feasibility of deriving lifetime constraints 
using the  \ion{He}{ii} transverse proximity effect.
In the same sightline \citet{Worseck2006} compared \ion{He}{ii} and \ion{H}{i} spectra and computed the hardness of the radiation field, based on the relative absorption strength of the two ions, sensitive to ionization at $1\Ry$ and $4\Ry$. They detect the proximity effect for at least one other foreground quasar which sets a lower lifetime limit of $17\Myr$.
\citet{Syphers2014} claim a transverse proximity effect for another quasar $34\Mly$ away from the Q\,0302$-$003 sightline, but this case is degenerate with the proximity effect of Q\,0302$-$003 itself.
In the limited sample of \ion{He}{ii} spectra and foreground quasars \citet{Furlanetto2011} see indications against very long ($t_\mathrm{Q} > 10^8\yr$) and very short ($t_\mathrm{Q}<3\times10^6\yr$) quasar lifetimes, by simply counting \ion{He}{ii} transmission spikes under the assumption that they are associated with  foreground quasars.

The installation of the Cosmic Origins Spectrograph (COS) on \textit{HST} with unprecedented FUV sensitivity offers for the first time the opportunity for an extended survey of \ion{He}{ii} transparent quasars \citep{Worseck2011b, Syphers2012, Zheng2015, Worseck2016}. 
We use a dataset of 22 \ion{He}{ii} sightlines and complement it with a dedicated optical spectroscopic survey to discover foreground quasars around these sightlines. These allow us for the first time to statistically quantify the \ion{He}{ii} transverse proximity effect and to derive a constraint on the quasar episodic lifetime. The survey is described in \S\,\ref{Sec:Survey} and a few individual objects are discussed in \S\,\ref{Sec:Special_Objects}. In \S\,\ref{Sec:StatisticalAnalysis} we describe our statistical search for the presence of a \ion{He}{ii} transverse proximity effect and constrain the quasar lifetime. In \S~\ref{Sec:Single_QSO_Stat} we investigate the object-to-object variation of the transverse proximity effect within our sample. We discuss the implications of our findings in \S\,\ref{Sec:Discussion} and summarize in \S\,\ref{Sec:Summary}.

Throughout the paper we use a flat $\Lambda$CDM cosmology with $H\mathrm{_0 =70\,km \: s^{-1} \: Mpc^{-1}}$, $\mathrm{\Omega_m = 0.3}$ and $\mathrm{\Omega_\Lambda = 0.7}$ which is broadly consistent with the \citet{Planck2015} results. We use depending on the situation proper distances or comoving distances and denote the corresponding units as $\pMpc$ and $\cMpc$, respectively. We denote the different quasar lifetimes with $\tDC$ for the duty cycle (total activity over Hubble time), $\tQ$ for the quasar episodic lifetime, and $\tTO$ for the age of a quasar. Magnitudes are given in the AB system.

\section{Description of the Survey}
\label{Sec:Survey}

As part of a comprehensive effort to study \ion{He}{ii} reionization, we conducted
an extensive imaging and spectroscopic survey to identify foreground quasars around 22 \ion{He}{ii}-transparent quasar sightlines for which science-grade \textit{HST}/COS spectra are available.
The survey consisted of a deep narrow survey covering the immediate vicinity of the \ion{He}{ii} sightline ($\Delta\theta \le 10\arcmin$) down to a magnitude of $r\le24.0$~mag based on deep imaging and multi-object spectroscopic follow-up on 8\,m-class telescopes, as well as a wider survey targeting individual quasars on 4\,m telescopes. 
Finding quasars that have a separation from the \ion{He}{ii} sightline of more than $\approx10\arcmin$ is in particular important
to constrain the quasar lifetime. At redshift $z\approx3$, this angular separation corresponds to a physical distance of only $4.7\pMpc$ or a light crossing time of $15\Myr$. Therefore, we also conducted the shallower and wider survey on 4\,m~class telescopes extending out to $\Delta\theta\approx90\arcmin$.

The sample from our own surveys was complemented by quasars from the Sloan Digital Sky Survey \citep[SDSS,][]{York2000} and the Baryon Oscillation Spectroscopic Survey \citep[BOSS,][]{Eisenstein2011, Dawson2013}, specifically from the twelfth data release \citep[DR12,][]{Alam2015} spectroscopic quasar catalog \citep{Paris2016}.
In the context of our study, SDSS and BOSS cover a similar parameter space (magnitude $r \lesssim 21\,\mathrm{mag}$) as our wide survey. However, the SDSS quasar selection is substantially incomplete at $z\approx3$  \citep{Fan1999, Richards2002a, Richards2006b, Worseck2011a}. This makes it necessary to conduct our own wide survey and find the quasars not identified by SDSS.

\subsection{\ion{He}{ii} Sightlines}
\label{sec:HeII_Sightlines}

\begin{deluxetable*}{llrrcl}
\tablecolumns{6}
\tablewidth{0.95\linewidth}
\tablecaption{
Overview of the FUV spectra used for this work.
}
\tablehead{
\colhead{Quasar} & \colhead{Instrument}	& \colhead{$R$}	& \colhead{Program}	& \colhead{PI}	& \colhead{References}
}
\startdata
PC\,0058$+$0215		& COS G140L	& 2000	& 11742		& Worseck	& \citealt{Worseck2016}						\\
HE2QS\,J0233$-$0149	& COS G140L	& 2000	& 13013		& Worseck	& Worseck et al. in prep., this paper				\\
Q\,0302$-$003		& COS G130M	& 18000	& 12033		& Green		& \citealt{Syphers2014}, this paper				\\
SDSS\,J0818$+$4908	& COS G140L	& 2000	& 11742		& Worseck	& \citealt{Worseck2016}						\\
HS\,0911$+$4809		& COS G140L	& 2000	& 12178		& Anderson	& \citealt{Syphers2011, Syphers2012, Worseck2016}		\\
HE2QS\,J0916$+$2408	& COS G140L	& 2000	& 13013		& Worseck	& Worseck et al. in prep, this paper				\\
SDSS\,J0924$+$4852	& COS G140L	& 2000	& 11742		& Worseck	& \citealt{Worseck2011b, Worseck2016}				\\
SDSS\,J0936$+$2927	& COS G140L	& 2000	& 11742		& Worseck	& \citealt{Worseck2016}						\\
HS\,1024$+$1849		& COS G140L	& 2000	& 12178		& Anderson	& \citealt{Syphers2012, Worseck2016}				\\
SDSS\,J1101$+$1053	& COS G140L	& 2000	& 11742		& Worseck	& \citealt{Worseck2011b, Worseck2016}				\\
HS\,1157$+$3143		& STIS G140L	& 1000	& 9350		& Reimers	& \citealt{Reimers2005, Worseck2016}				\\
SDSS\,J1237$+$0126	& COS G140L	& 2000	& 11742		& Worseck	& \citealt{Worseck2016}						\\
SDSS\,J1253$+$6817	& COS G140L	& 2000	& 12249		& Zheng		& \citealt{Syphers2011, Zheng2015, Worseck2016}			\\
SDSS\,J1319$+$5202	& COS G140L	& 2000	& 12249		& Zheng		& \citealt{Zheng2015, Worseck2016}				\\
Q\,1602$+$576		& COS G140L	& 2000	& 12178		& Anderson	& \citealt{Syphers2012, Worseck2016}				\\
HE2QS\,J1630$+$0435	& COS G140L	& 2000	& 13013		& Worseck	& Worseck et al. in prep., this paper				\\
HS\,1700$+$6416		& COS G140L	& 2000	& 11528		& Green		& \citealt{Syphers2013, Worseck2016}				\\
SDSS\,J1711$+$6052	& COS G140L	& 2000	& 12249		& Zheng		& \citealt{Zheng2015, Worseck2016}				\\
HE2QS\,J2149$-$0859	& COS G140L	& 2000	& 13013		& Worseck	& Worseck et al. in prep., this paper				\\
HE2QS\,J2157$+$2330	& COS G140L	& 2000	& 13013		& Worseck	& Worseck et al. in prep., this paper				\\
SDSS\,J2346$-$0016	& COS G140L	& 2000	& 12249		& Zheng		& \citealt{Syphers2011, Zheng2015, Worseck2016}			\\
HE\,2347$-$4342		& COS G140L	& 2000	& 11528		& Green		& \citealt{Shull2010, Worseck2016}	
\enddata
\label{Tab:FUV_Spectra}
\end{deluxetable*}

Our foreground quasar survey targeted fields around 22 \ion{He}{ii}-transparent quasars observed with \textit{HST}/COS or \textit{HST}/STIS (Table~\ref{Tab:FUV_Spectra}). 
\citet{Worseck2016} describe the homogeneous data reduction and analysis of 16 of these, including a much improved \textit{HST}/COS background subtraction (dark current, quasi-diffuse sky emission, scattered light) and suppression of geocoronal contamination compared to the default CALCOS pipeline reductions from the \textit{HST} archive. This improved reduction ensures that weak excess \ion{He}{ii} transmission due to the transverse proximity effects is not affected by zero-level calibration errors.
Five \ion{He}{ii}-transparent quasar sightlines observed in \textit{HST} Cycle~20 were reduced analogously (Worseck et al., in preparation). Almost all spectra (20/22) have been taken with the \textit{HST}/COS G140L grating ($R=\lambda/\Delta\lambda\sim 2000$ at 1150\,\AA, $\simeq 0.24$\,\AA{} per Nyquist-binned pixel). Their signal-to-noise ratio per binned pixel at \ion{He}{ii} Ly$\alpha$ of the background quasar varies between 2 and 15, mostly depending on whether the sightline was known before to be \ion{He}{ii}-transparent \citep{Shull2010, Syphers2013, Syphers2014, Zheng2015}, 
or had been discovered in recent \textit{HST}/COS surveys \citep{Worseck2011b, Syphers2012, Worseck2016}.
The sightline to HS~1157$+$3143 was observed with the \textit{HST}/STIS G140L grating \citep[$R\sim 1000$, 0.6\,\AA\,pixel$^{-1}$;][]{Reimers2005}. For the Q~0302$-$003 sightline we used the higher-quality \textit{HST}/COS G130M data ($R\approx 18,000$, $\simeq 0.03$\,\AA{} per Nyquist-binned pixel; \citealt{Syphers2014}) instead of the \textit{HST}/STIS data presented in \citet{Worseck2016}. We checked our reduction by comparing the measured \ion{He}{ii} effective optical depths in the Q~0302$-$003 sightline to those presented by \citet{Syphers2014}, finding very good agreement.

As detailed in \citet{Worseck2016} we suppressed geocoronal emission lines by considering only data taken during orbital night or at restricted Earth limb angles in the affected spectral ranges. Nevertheless, decontaminated regions had to be excluded from our analysis due to the vastly reduced sensitivity to strong \ion{He}{ii} absorption and sometimes extremely weak geocoronal residuals in the coadded data. Geocoronal Ly$\alpha$ emission was excluded. 
In addition we apply a liberal signal-to-noise (S/N) cut. Since most of our pathlength is within Gunn-Peterson troughs where the flux is consistent with zero we cannot apply a real limit on the S/N but instead require the continuum level to correspond to at least five counts
per spectral bin. This mostly affects the short wavelength end of the spectra where the efficiency of the instrument drops dramatically. 
The background quasar proximity zones were excluded by measuring the dropping \ion{He}{ii} transmission from the proximity zone 
to the strongly saturated \ion{He}{ii} absorption in the IGM (e.g. \citealt{Zheng2015}). By excluding the entire background quasar proximity zone we may have excluded the transverse proximity zones of foreground quasars at similar redshifts \citep{Worseck2006, Syphers2014}, but as these cases are degenerate and therefore require detailed modeling we make sure that our sample is not affected by any background quasar proximity zone.

\subsection{Deep Survey on 8\,m~class Telescopes}

For our deep imaging survey we used the Large Binocular Cameras at the Large Binocular Telescope (LBT/LBC, \citealt{Speziali2008, Giallongo2008}) to obtain optical multiband photometry ($U_\mathrm{Spec}$, $g$, $r$ and $i$) over an area of $23'\times25'$ approximately centered on the targeted \ion{He}{ii} sightline. Imaging for 10 \ion{He}{ii} sightlines was obtained over several runs in 2009, 2011 and 2013 (Table~\ref{Tab:Table1}).
We observed in binocular mode with $U_\mathrm{Spec}$ and $g$ ($r$ and $i$) filters on the blue (red) camera.
Individual exposure times were short (around 120 seconds) to limit saturation of bright stars. Depending on the field and the observing conditions, total exposure times were 70 to 100 minutes in the $U_\mathrm{Spec}$ filter, 10 to 30 minutes in $g$, 18 to 35 minutes in $r$ and 38 to 54 in the $i$ filter.
With this strategy we reached a homogeneous depth in $U_\mathrm{Spec}$, which is the limiting factor for our target selection due to the expected colors of $z\sim 3$ quasars ($U_\mathrm{Spec}-g\sim 2$). Due to the observation in binocular mode, the red filters naturally reached a sufficient depth.

Due to declination and scheduling constraints, the fields of HE\,2347$-$4342 and
SDSS\,J1237$+$0126 were not observed with LBT/LBC. 
For the field of HE\,2347$-$4342 we obtained multiband imaging ($U\,g\,r\,i$) with the $36\arcmin\times 36\arcmin$ Mosaic~II camera at the 4\,m Blanco Telescope at the Cerro Tololo Inter-American Observatory \citep{Muller1998}. The field of SDSS\,J1237$+$0126 was imaged in $g\,r\,i$ with Mosaic~1.1 at the 4\,m Mayall Telescope at the Kitt Peak National Observatory, and in $U$ with Magellan/Megacam \citep{McLeod2015}.
Exposure times were increased to achieve an imaging depth similar to our LBC observations.

\begin{figure*}
\centering
\includegraphics[width=.84\linewidth]{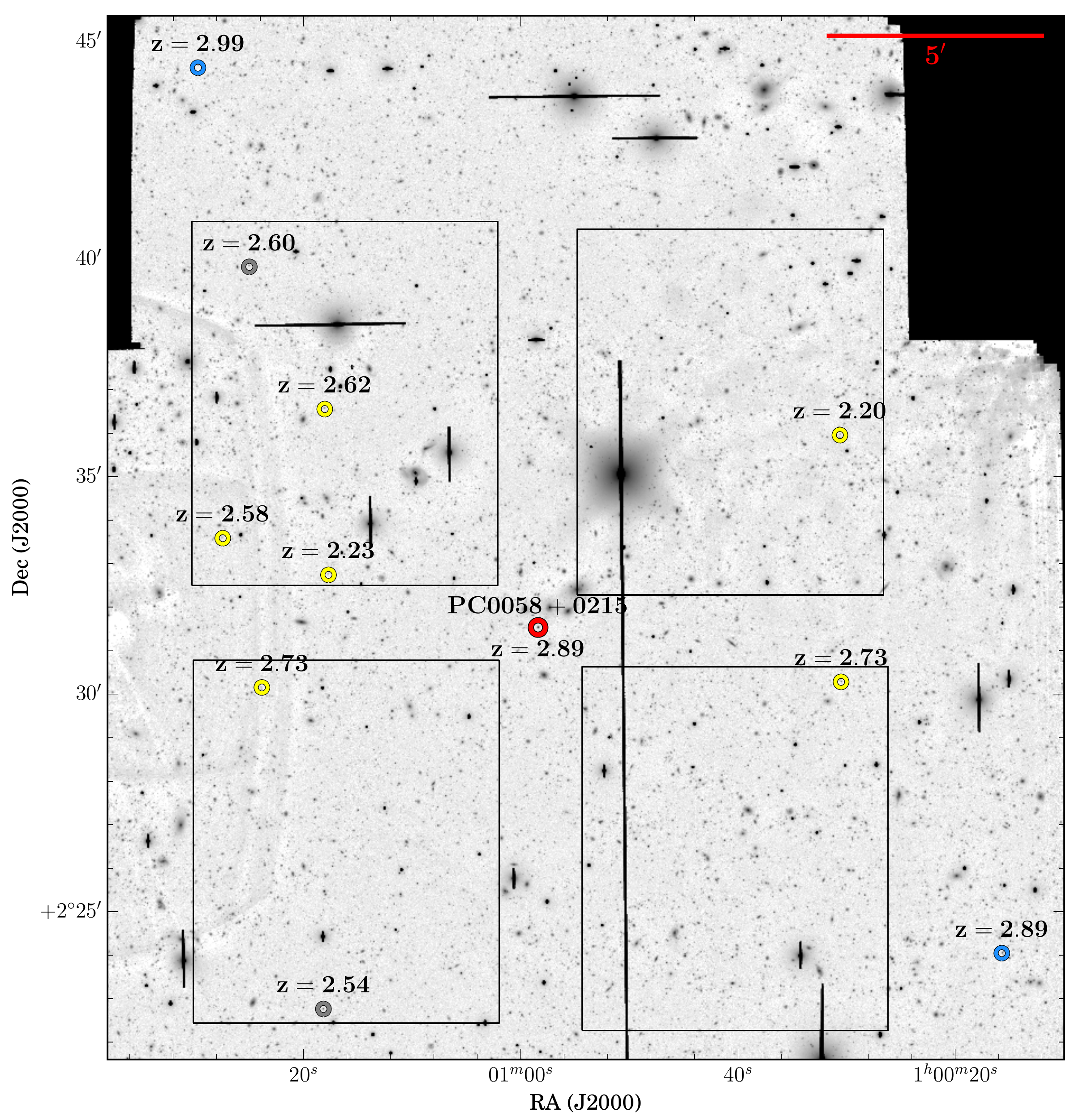}
\caption{LBT/LBC $r$-band image of the field around PC\,0028$+$0215 ($23'\times 25'$, $r\,<\,26\mag$ at $5\,\sigma$). 
The \ion{He}{ii}-transparent background quasar is marked in the center (red). The approximate positions of the four quadrants of the VIMOS field of view are indicated. In this area our deep spectroscopic survey discovered six foreground quasars (yellow). Two additional quasars outside the VIMOS footprint were found by our wide survey with NTT/EFOSC2 (blue) and two quasars are from SDSS (gray).}
\label{LBT_imaging}
\end{figure*}

Data reduction, mosaicing of the individual dithered exposures, stacking, astrometry and photometry was done using our own custom pipeline based on IRAF routines and SCAMP, SWarf and SExtractor\footnote{\url{http://www.astromatic.net/software}} \citep{Bertin2006, Bertin2002, Bertin1996}. An example for a reduced $r$ band image is given in Figure~\ref{LBT_imaging}. For fields covered by SDSS the astrometric solution is tied to the SDSS reference frame using SDSS star positions, while the photometric calibration is tied to SDSS ubercal photometry \citep{Padmanabhan2008} to define the zero point and to correct for non-photometric observations. For HE\,2347$-$4342 that lies outside the SDSS footprint we had to rely on the USNO-B1.0 catalog and photometric standards.
We used the reddest bands ($r$ or $i$) for source detection and used forced photometry to extract fluxes at the identical positions in the other bands. 
Magnitudes were corrected for Galactic extinction assuming the reddening terms from SDSS \citep{Stoughton2002} and $E(B-V)$ for the background quasar from \citet{Schlegel1998}, i.e. not accounting for reddening variations across the field.
Star-galaxy classification was also done in the reddest observed bands since they had the best image quality with a full width at half maximum (FWHM) of $\simeq 0.8\arcsec$.
The $5\sigma$ point source imaging depth of our LBT/LBC images is typically $\simeq 26.5\mag$ in $U_\mathrm{Spec}$ and $g$, and $\simeq 26.0\mag$ in $r$ and $i$, respectively.

\begin{figure}
\includegraphics[width=\linewidth]{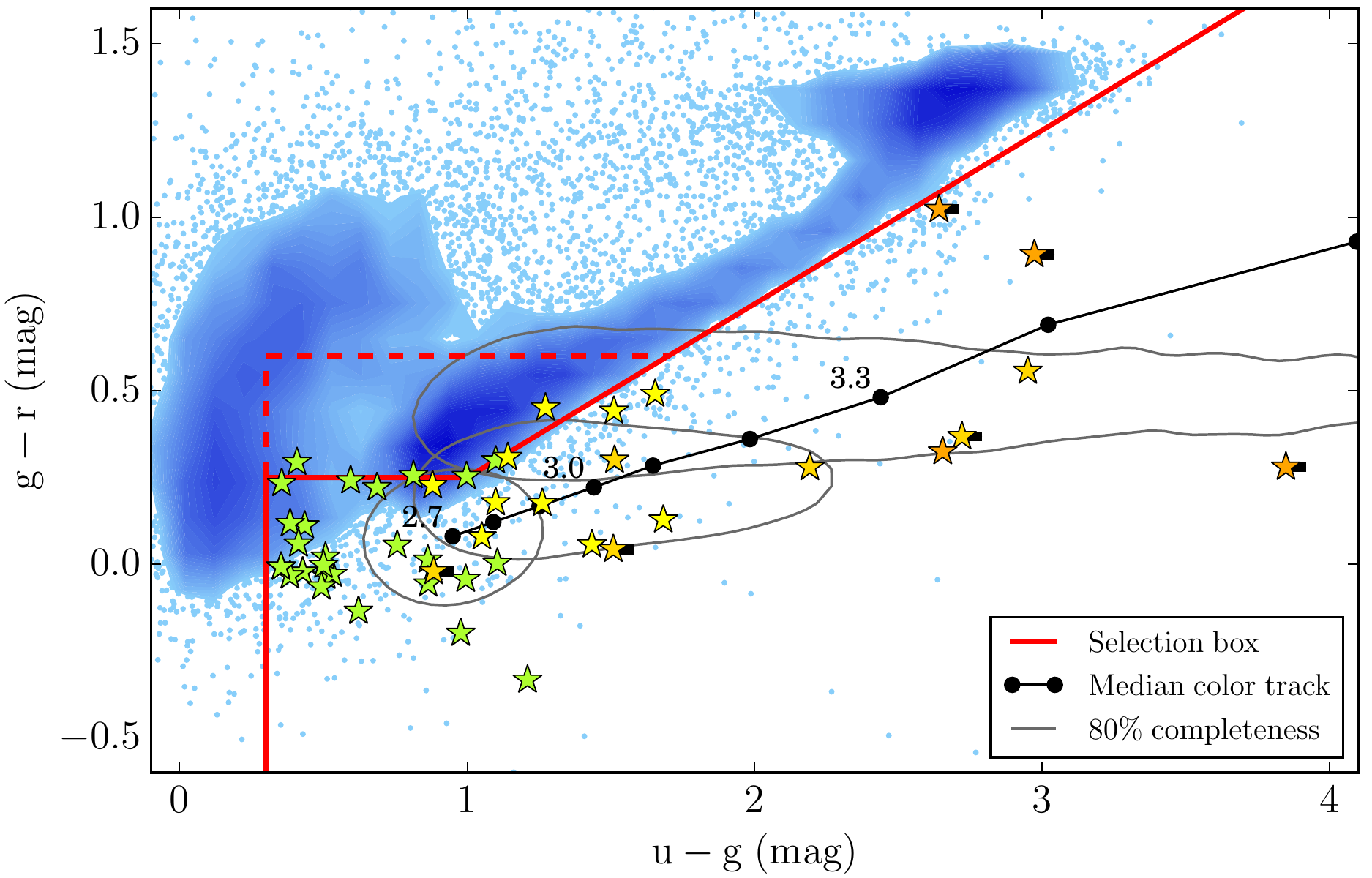}
\caption{Candidate selection for our deep survey. Shown are all point-like objects with photometric detections in all bands (S/N$>5$) as dots or contours (blue). A theoretical color track for $2.7<z<3.5$ quasars and corresponding completeness contours are shown in black \citep{Worseck2011a}. We selected high-priority quasar candidates from a box shown in red and lower-priority candidates from the dashed region. Confirmed quasars are overplotted as star symbols. Colors from green to orange indicate the redshift ($z<2.5$, $2.5<z<3$, $3<z<3.5$, $z>3.5$).
Quasars having only a limit in $U$ (S/N$<5$) have a black tick to the right.
}
\label{color-selection}
\end{figure}

Selection of quasar candidates was done by applying cuts in the $U_\mathrm{Spec}-g$~vs.~$g-r$ color space as shown in Figure~\ref{color-selection}.
A theoretical color track and contours have been computed from SDSS mock photometry of quasars including the spread in color due variations in the spectral energy distribution (SED) and IGM absorption \citep{Worseck2011a}.
The stochastic IGM Lyman continumm absorption leads to a large scatter around the median track, and in particular for $z \approx 3$ the range of expected quasar colors overlaps substantially with the stellar locus, highlighting again the difficulties of quasar color selection at these redshifts \citep{Richards2002a, Worseck2011a}.
We used a selection box of the form
\begin{equation}
\begin{split}
&(U_\mathrm{Spec}-g) > 0.3 \; \land \\
&\left[ \, (g-r) < 0.25 \; \lor \; (g-r) < 0.5 \, (U_\mathrm{Spec}-g) - 0.25 \, \right]
\end{split}
\end{equation}
which is visualized in Figure~\ref{color-selection}. 
It accounts for the expected range in color for $z\sim3$ quasars while limiting the stellar contamination.
Lower-priority candidates were selected from an extended selection box $(g-r)<0.6$ that overlapped with the stellar locus (Figure~\ref{color-selection}, dashed region).

\begin{figure}
\includegraphics[width=\linewidth]{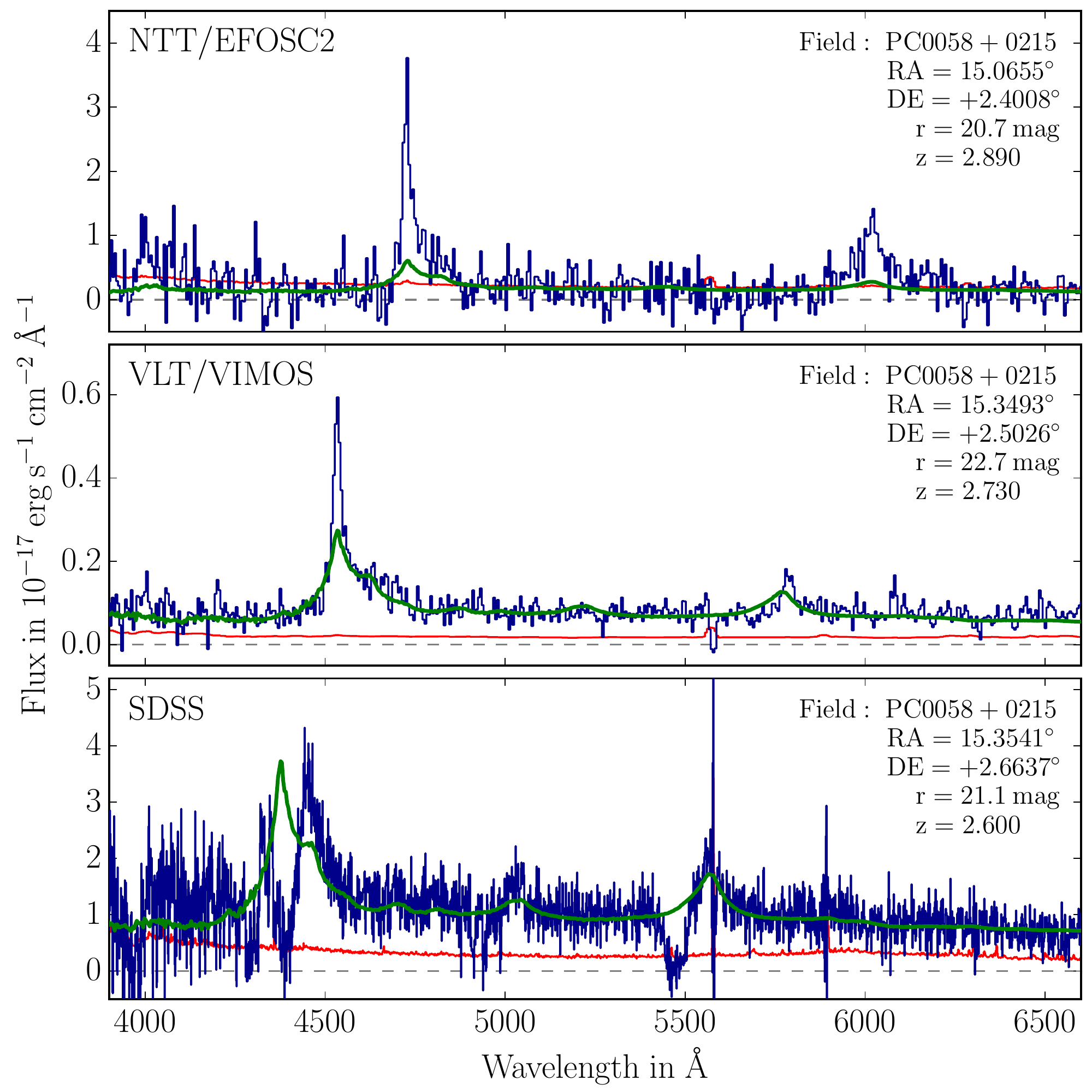}
\caption{Spectra of three representative foreground quasars observed with different telescopes and instruments used for this study. All three quasars are located around the \ion{He}{ii} sightline toward PC\,0058$+$0215 and marked in Figure~\ref{LBT_imaging}.
The observed flux density is shown in blue and the $1\sigma$ error array in red. Overplotted is a quasar template \citep{vandenBerk2001} shifted to the redshift of the observed quasars but not adapted to match the different line strengths of the shown quasars.
}
\label{Fig:Raw_Gallery}
\end{figure}

Spectroscopic verification of the quasar candidates was done with the VIsible MultiObject Spectrograph \citep[VIMOS,][]{LeFevre2003} at the Very Large Telescope (VLT), whose four $7\arcmin\times8\arcmin$ quadrants cover most of the $23\arcmin\times25\arcmin$ LBC field of view (Figure~\ref{LBT_imaging}). 
Custom designed focal-plane slit masks were used to simultaneously take low-resolution spectra (LR Blue grism, $R \approx180$, wavelength range 3700--6700\,\AA) of $\approx32$ candidates per quadrant. Each of our 10 imaged fields was covered by a single VIMOS pointing with a typical exposure time of $2\times 30$ minutes.
The data were reduced with the standard EsoRex VIMOS pipeline\footnote{\url{http://www.eso.org/sci/software/cpl/esorex.html}} to which we added custom masking of zeroth-order contamination which is unavoidably present in the raw frames. An example spectrum of a quasar discovered by our VIMOS survey is shown in the middle panel of Figure~\ref{Fig:Raw_Gallery}.

For the field of Q\,0302$-$003, additional quasar candidates outside our VIMOS pointing were observed with the DEep Imaging Multi-Object Spectrograph (DEIMOS, \citealt{Faber2003}) at Keck Observatory. We however did not find any additional quasar in this attempt.

\subsection{Wide Survey on 4\,m Class Telescopes}

Our deep multi-object spectroscopy was complemented by individual longslit observations of brighter quasar candidates at larger angular separations ($10\arcmin \lesssim \Delta\theta \lesssim 90\arcmin$). Due to the large area on the sky and sparse target distribution, longslit spectroscopy of single targets is preferable to multi-object spectroscopy. It requires, however, a much higher selection efficiency than possible for $z\approx3$ quasars from optical photometry alone.
We selected candidates from the \textit{XDQSOz} catalog \citep{DiPompeo2015} based on the extreme deconvolution technique \citep{Bovy2011, Bovy2012} and the KDE catalog \citep{Richards2015}. Both catalogs are based on SDSS $ugriz$ imaging but also incorporate infrared photometry from the \textit{Wide-field Infrared Survey Explorer} \citep[\textit{WISE,}][]{Wright2010}.
The \textit{WISE} $3.6\um$ and $4.5\um$ bands are sensitive to the emission of hot dust surrounding AGN \citep{Stern2012, Assef2013}.
This characteristic feature allows for very efficient separation of quasars from galaxies and stars. Both catalogs give a photometric redshift estimate or even a probability distribution. We used this information to maximize the probability for candidates to be confirmed with a redshift covered by the \ion{He}{ii} $\lya$ absorption spectra of the background quasars and prioritized targets accordingly.
We also designed our wide survey to maximize the expected \ion{He}{ii} transverse proximity effect.
Based on the position, brightness and photometric redshift, we estimated the photoionization rate $\Gqso$ that every of these putative quasars would cause at the background sightline (see \S~\ref{Sec:Gamma_QSO} for a definition of $\Gqso$). We then selected and prioritized according to cuts in $\Gqso$, primarily targeting objects with expected $\Gqso > 0.5 \times 10^{-15}\,\mathrm{s}^{-1}$.

For spectroscopic confirmation we used the ESO 3.5\,m New Technology Telescope Faint Object Spectrograph and Camera (NTT/EFOSC2, \citealt{Buzzoni1984}) and the Calar Alto Observatory (CAHA) 3.5\,m telescope TWIN spectrograph. We used the EFOSC2 grating g782 ($R=180$--450, wavelength range 3700--9000\,\AA). For TWIN we used only the blue arm with grating T13 ($R=620$--1000, wavelength range 3900--7000\,\AA). A slit width between $1.2$ and $1.5\arcsec$ was used and the slit oriented at the parallactic angle. With these setups the limiting magnitude for both instuments was $r \approx 21\mag$ for the longest used integration times of one hour.
The CAHA/TWIN spectra were taken in 32 nights between November 2014 and August 2015, while NTT/EFOSC2 observations were performed during 5 nights in December 2014.

Data were reduced using the XIDL Low-Redux package\footnote{\url{http://www.ucolick.org/~xavier/LowRedux/}}. An example spectrum of a quasar confimed with NTT/EFOSC2 is shown in Figure~\ref{Fig:Raw_Gallery}. Overall, 36\% of the observed targets were confirmed as quasars and 11\% had a redshift within the covered \ion{He}{ii} $\lya$ forest of the background quasar.

As part of the wide survey we also verified a quasar with an uncertain redshift in the vicinity of HS\,1700$+$6416 \citep{Syphers2013}. We used the Keck Low Resolution Imaging Spectrometer (Keck/LRIS, \citealt{Oke1995, McCarthy1998}) to confirm its redshift. Data were as well reduced using the XIDL Low-Redux package.

\subsection{Selection from SDSS and BOSS}

For \ion{He}{ii} sightlines within the SDSS footprint we also use quasars from the SDSS DR12 catalog \citep{Alam2015, Paris2016}. This has the advantage that we can include large numbers of quasars out to very large separation from the background sightline. For our initial input catalog we selected all quasars within $240\arcmin$ of the background sightlines and with $z>2.5$. For HS\,1700$+$6416 whose COS spectrum covers lower redshifts we adapted the latter criterion accordingly. This ensures that we include all objects that may contribute significantly to the ionizing background at the location of the background sightline. At later stages in our analysis we will impose cuts on the expected photoionization rate at the background sightline. The vast majority of the selected quasars have $r < 21\mag$.

We note that Q\,0302$-$003 and SDSS\,J2346$-$0016 lie within SDSS Stripe 82 that was imaged multiple times, offering photometry approximately two magnitudes deeper than the standard SDSS imaging \citep{Abazajian2009}, and was also targeted by additional spectroscopy, using different selection algorithms \citep[e.g. variability, see][]{Butler2011, Palanque-Delabrouille2011}. We actually find a higher density of foreground quasars near the SDSS\,J2346$-$0016 sightline but not for Q\,0302$-$003.

\subsection{Systemic Quasar Redshifts}

Quasar redshifts determined from the rest-frame ultraviolet emission lines (redshifted into the optical at $z \sim 3$) can differ by up to $\mathrm{1000\,km\,s^{-1}}$ from the systemic frame, because of outflowing/inflowing material in the broad line regions of quasars \citep{Gaskell1982, Tytler1992, vandenBerk2001, Richards2002b, Shen2007, Shen2016, Coatman2017}. We estimate systemic redshifts by combining the line-centering procedure used in \citet{Hennawi2006} with the recipe in \citet{Shen2007} for combining measurements from different emission lines. The resulting typical redshift uncertainties using this technique are in the range $\sigma_{z} \simeq 270–770\mathrm{\,km\,s^{-1}}$ depending on which emission lines  are used.  But given the low S/N ratio of $\approx 5$ of many of our spectra, we conservatively assume our estimates of the systemic quasar redshift to be not better than $\mathrm{1000\,km\,s^{-1}}$.

\subsection{Estimate of the \ion{He}{ii} Photoionization Rate}
\label{Sec:Gamma_QSO}

To estimate the impact a foreground quasar has on the ionization state of the IGM we calculate the \ion{He}{ii}
photoionization rate at the location of the background sightline. We use the \citet{Lusso2015} quasar template which is based on \textit{HST} UV grism spectroscopy, corrected for IGM absorption and covers the restwavelength range down to $600\,\mathrm{\AA}$. We redshift and scale this template to match our $r$~band photometry which always falls redwards of $\lya$ and measures the quasar continuum flux. From the scaled template we infer the flux at $912\,\mathrm{\AA}$ and extrapolate to the \ion{He}{ii} Lyman limit at $228\,\mathrm{\AA}$ assuming the specific luminosity to follow a power law $L_\nu \propto \nu^{\alpha}$. The quasar spectral slope $\alpha$ beyond $912\,\mathrm{\AA}$ is not very well constrained since the frequencies between the extreme UV and soft X-rays are basically unobservable. We adopt a value of $\alpha = -1.7$ as determined by \cite{Lusso2015} (however with a large uncertainty of $\pm0.6$), which is consistent with the independent measurement of \citet{Stevans2014}, 
as well as the slope between UV and X-ray regime \citep{Lusso2015}, but differs from the value of $\alpha = -0.73 \pm 0.26$ reported by \citet{Tilton2016}. The uncertainty in $\alpha$ and the long range of extrapolation from $912\,\mathrm{\AA}$ beyond $228\,\mathrm{\AA}$ causes substantial uncertainty in the inferred photoionization rate, e.g.\ up to a factor of $2.5$ for $\alpha=-1.7 \pm 0.6$. However, this mostly affects the absolute scaling of $\Gqso$. A relative comparison of different foreground quasars will not be severely affected.

We convert quasar luminosity to flux density $F_\nu$ at the background sightline according to
\begin{equation}
F_\nu = L_\nu \; \frac{1}{4\pi\:D^2}\;e^{-\frac{D}{\lambda_\mathrm{mfp}}}\;.
\label{Eq:GeometricDilution}
\end{equation}

This is a function of the transverse distance $D$ between the foreground quasar and the background sightline measured at the redshift of the foreground quasar, and the mean free path to \ion{He}{ii}-ionizing photons in the IGM $\lambda_\mathrm{mfp}$. However, quasars change the ionization state of the surrounding IGM by creating large proximity zones. Therefore, the mean free path relevant for us is not the one at random locations in the IGM but an effective mean free path within the proximity zone \citep[e.g.][]{McQuinn2014, Davies2014, Khrykin2016}. The \ion{He}{ii} transverse proximity effect should in principle be able to constrain the mean free path, but at present it is not well constrained by observation or simulation. Current studies \citep[e.g][]{Davies2014} suggest $\lambda_\mathrm{mfp} \gtrsim 50\cMpc$ at $z\approx3$, longer than the scales probed by our study ($D \lesssim 40\cMpc$). We therefore ignore IGM absorption for now and assume pure geometrical dilution of the radiation by setting $\lambda_\mathrm{mfp} = \infty$.

Implicitly, we have assumed isotropic emission and infinite quasar lifetime with constant luminosity. Although current constraints suggest $\tQ\lesssim10^8\,\mathrm{yr}$, we assume for simplicity no time dependence in our fiducial model and constrain $\tQ$ later. Also, the widely used unified AGN models \citep[see, e.g.][]{Antonucci1993, Elvis2000} assume a large-scale anisotropy of the UV and optical emission and relate the dichotomy between broad line quasars (Type\,I) and AGN displaying only narrow emission lines (Type\,II) to the presence of obscuring material that blocks the direct view on the accretion disc and broad line region if observed from certain directions. Studies focusing on the numbers of Type\,Is vs. Type\,IIs suggest (with large uncertainties) approximately equal numbers \citep[e.g.][]{Brusa2010, Lusso2013, Marchesi2016} which suggests opening angles of $\approx120\,\mathrm{degrees}$ if one assumes a bi-conical emission. In consequence, the Type\,I quasars from our survey might be obscured towards parts of the background sightlines. However, we have no way to either infer the orientation or the exact opening angle of the foreground quasar and therefore no other choice than to assume isotropic emission for our fiducial model. A detailed modeling of obscuration effects is beyond the scope of this paper.

We therefore remain for now with the simplest isotropic model and convert UV flux density to \ion{He}{ii} photoionization rate by
\begin{equation}
 \Gqso = \int_{\nu_o}^\infty \frac{ F_\nu \; \sigma_{\nu,\,\mathrm{HeII}} }{ \mathit{h}_\mathrm{P} \, \nu } \: d\nu\; \approx \frac{ F_{\nu_o} \; \sigma_\mathrm{\,HeII} }{\mathit{h_\mathrm{P}} \; (3 - \alpha)}
 \label{Eq:Q_QSO}
\end{equation}
in which $\mathit{h}_\mathrm{P}$ denotes Planck's constant and $\nu_0$ the frequency of the \ion{He}{ii} ionization edge.
For the second part we have assumed the quasar spectra to follow the power law description
$F_\nu = F_{\nu_0} \times \left( \sfrac{\nu}{\nu_0} \right) ^\alpha$
introduced above and for the \ion{He}{ii} cross-section the approximation $\sigma_{\nu,\,\mathrm{HeII}} \approx \sigma_\mathrm{\,HeII,0} \times \left( \sfrac{\nu}{\nu_0} \right) ^{-3}$ with $\sigma_\mathrm{\,HeII,0} = 1.58 \times 10^{-18}\;\mathrm{cm}^2$ \citep{Verner1996}.

This calculation condenses the observed parameters (angular separation from background sightline, apparent magnitude, redshift) into one convenient number which can be used to estimate the possible influence of a foreground quasar on the background sightline. 
Although there is substantial uncertainty in the spectral slope and our calculation of the photoionization rate assumes simplifications like infinite lifetime, no IGM absorption and completely ignores obscuration effects, $\Gqso$ represents an important physical quantity and is (on average) a good proxy for the strength of the transverse proximity effect (\S~\ref{Sec:Gamma_Dependence}). We will use $\Gqso$ extensively throughout the paper to select subsamples from our quasar catalog. Although we impose no universal threshold, we typically consider foreground quasars with $\mathrm{\Gqso > 2 \times 10^{-15}\,{s}^{-1}}$, i.e.\ similar to or exceeding current estimates of the UV background \citep[$\Gamma^\mathrm{HeII}_\mathrm{UVB}\approx10^{-15}\,\mathrm{s}^{-1}$ at $z \sim 3$, ][]{HaardtMadau2012, Khrykin2016}. We caution that a direct comparison of our $\Gqso$ estimates to global UV background models is affected by substantial uncertainty due to different quasar SED parameterizations (see \citealt{Lusso2015} for a discussion). It gives, however, a rough indication of the value above which we expect a foreground quasar to substantially impact the IGM ionization structure. We also include objects with $\mathrm{0.5 \times 10^{-15}\,{s}^{-1} < \Gqso < 2 \times 10^{-15}\,{s}^{-1}}$ to explore the effect of weaker quasars. See \S~\ref{Sec:Gamma_Dependence} for a detailed analysis and \S~\ref{Sec:Single_QSO_Stat} for the impact of individual objects.
\label{sec:SED}

\subsection{Final Quasar Sample}

\begin{deluxetable*}{lcccccccc}
\tablecolumns{7}
\tablewidth{0pc}
\tablecaption{
Overview of the \ion{He}{ii} sightlines and the number of foreground quasars with \ion{He}{ii} Ly$\alpha$ coverage in the background sightline.}
\tablehead{
\colhead{Name}	& \colhead{R.A. (J2000)}	& \colhead{Decl. (J2000)}	& \colhead{$z_\mathrm{BG}$}	& \colhead{LBT/LBC}	&	\colhead{\small{NTT/}}	& \colhead{\small{CAHA/}}	& \colhead{SDSS\tablenotemark{a}}	& \colhead{Literature\tablenotemark{b}}	\\
\colhead{}   	& \colhead{hh\,:\,mm\,:\,ss}	& \colhead{hh\,:\,mm\,:\,ss}	& \colhead{}			& \colhead{VLT/VIMOS}				& 			\colhead{EFOSC2}				& \colhead{TWIN}				& \colhead{}						& \colhead{}
}
\startdata
PC\,0058$+$0215     & $01:00:58.40$ & $+02:31:32.0$ & $2.890$ & $    2$       & $    1$       & $    0$       & $    1$       & $     $       \\
HE2QS\,J0233$-$0149 & $02:33:06.01$ & $-01:49:50.5$ & $3.314$ & $    1$       & $    0$       & $    0$       & $    4$       & $     $       \\
Q\,0302$-$003       & $03:04:49.85$ & $-00:08:13.4$ & $3.286$ & $    1$       & $    0$       & $     $       & $    2$       & $    2$       \\
SDSS\,J0818$+$4908  & $08:18:50.02$ & $+49:08:17.2$ & $2.957$ & $  -  $       & $  -  $       & $    0$       & $    1$       & $     $       \\
HS\,0911$+$4809     & $09:15:10.00$ & $+47:56:59.0$ & $3.350$ & $  -  $       & $  -  $       & $    2$       & $    2$       & $     $       \\
HE2QS\,J0916$+$2408 & $09:16:20.85$ & $+24:08:04.6$ & $3.440$ & $     $       & $    2$       & $     $       & $    3$       & $     $       \\
SDSS\,J0924$+$4852  & $09:24:47.36$ & $+48:52:42.8$ & $3.027$ & $  -  $       & $  -  $       & $     $       & $    1$       & $     $       \\
SDSS\,J0936$+$2927  & $09:36:43.51$ & $+29:27:13.6$ & $2.930$ & $     $       & $     $       & $     $       & $    1$       & $     $       \\
HS\,1024$+$1849     & $10:27:34.13$ & $+18:34:27.6$ & $2.860$ & $    0$       & $    0$       & $    0$       & $    0$       & $     $       \\
SDSS\,J1101$+$1053  & $11:01:55.74$ & $+10:53:02.3$ & $3.029$ & $    0$       & $    1$       & $    0$       & $    4$       & $     $       \\
HS\,1157$+$3143     & $12:00:06.25$ & $+31:26:30.8$ & $2.989$ & $     $       & $     $       & $    2$       & $    5$       & $     $       \\
SDSS\,J1237$+$0126  & $12:37:48.99$ & $+01:26:07.0$ & $3.154$ & $    1$       & $     $       & $    0$       & $    0$       & $     $       \\
SDSS\,J1253$+$6817  & $12:53:53.70$ & $+68:17:14.4$ & $3.481$ & $  -  $       & $  -  $       & $    5$       & $    7$       & $     $       \\
SDSS\,J1319$+$5202  & $13:19:14.19$ & $+52:02:00.3$ & $3.930$ & $  -  $       & $  -  $       & $    0$       & $    1$       & $     $       \\
SBS\,1602$+$576     & $16:03:55.93$ & $+57:30:54.5$ & $2.862$ & $  -  $       & $  -  $       & $    0$       & $    1$       & $     $       \\
HE2QS\,J1630$+$0435 & $16:30:56.34$ & $+04:35:59.4$ & $3.788$ & $     $       & $     $       & $    0$       & $  -  $       & $     $       \\
HS\,1700$+$6416     & $17:01:00.61$ & $+64:12:09.0$ & $2.751$ & $  -  $       & $  -  $       & $    1$       & $    0$       & $    1$       \\
SDSS\,J1711$+$6052  & $17:11:34.40$ & $+60:52:40.5$ & $3.834$ & $  -  $       & $  -  $       & $    1$       & $    0$       & $     $       \\
HE2QS\,J2149$-$0859 & $21:49:27.77$ & $-08:59:03.6$ & $3.259$ & $    1$       & $     $       & $    2$       & $    1$       & $     $       \\
HE2QS\,J2157$+$2330 & $21:57:43.63$ & $+23:30:37.3$ & $3.143$ & $    1$       & $     $       & $    3$       & $    1$       & $     $       \\
SDSS\,J2346$-$0016  & $23:46:25.66$ & $-00:16:00.4$ & $3.512$ & $    0$       & $    0$       & $    0$       & $    6$       & $     $       \\
HE\,2347$-$4342     & $23:50:34.21$ & $-43:25:59.6$ & $2.887$ & $    0$       & $    0$       & $    0$       & $  -  $       & $    1$
\enddata
\tablecomments{For each telescope/instrument or catalog we only give the number of detected or used quasars that are included in our statistical analysis, therefore only counting quasars for which we have usable coverage in the \ion{He}{ii} Ly$\alpha$ spectrum (e.g. position not contaminated, not in the proximity zone or behind the \ion{He}{ii} quasar, see Figure~\ref{Fig:HeSpectra1}) and which cause an estimated photoionization rate at the \ion{He}{ii} sightline of $\mathrm{\Gqso > 0.5 \times 10^{-15}\,s^{-1}}$ (see \S\,\ref{Sec:Gamma_QSO}).}  
\tablenotetext{a}{SDSS DR12, \citet{Alam2015, Paris2016}}
\tablenotetext{b}{Literature objects from \citet{Jakobsen2003, Steidel2003, Hennawi2006, Worseck2006, Worseck2007, Syphers2013}. }
\label{Tab:Table1}
\end{deluxetable*}

\begin{figure}
\includegraphics[width=\linewidth]{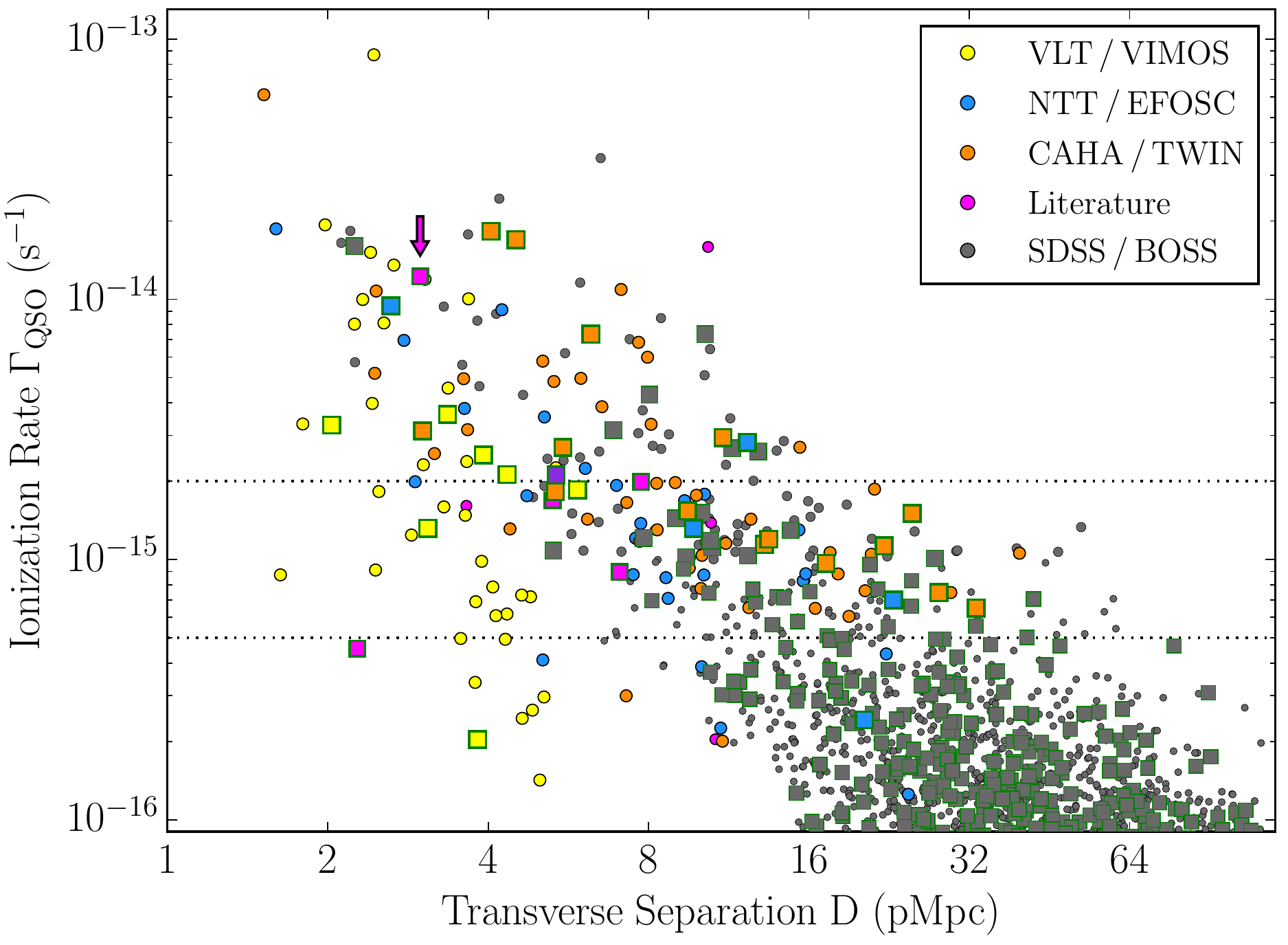}
\caption{
Properties of our foreground quasar sample. 
The panel shows the transverse separation of the foreground quasars from the background sightline and their estimated photoionization rate $\Gqso$ at the location of the background sightline. Colors indicate the origin of the objects. Horizontal dotted lines give the ranges for which we expect the objects to have a substantial impact on the background sightline ($\mathrm{ \Gqso > 2 \times 10^{-15}\,{s}^{-1}}$) or at least include them for the extended selections ($\mathrm{\Gqso > 0.5 \times 10^{-15}\,{s}^{-1}}$). Only objects shown as green framed squares fall in usable parts of the \ion{He}{ii} spectra (see Figure~\ref{Fig:HeSpectra1}) and can be included in our statistical analysis. An arrow indicates the \citet{Jakobsen2003} quasar.
}
\label{Fig:Sample_Summary}
\end{figure}

After restricting the combined quasar sample to have science-grade \ion{He}{ii} $\lya$ coverage along the background sightline (\S~\ref{sec:HeII_Sightlines}) and $\mathrm{\Gqso  > 0.5 \times 10^{-15}\:s^{-1}}$ we end up with 66 foreground quasars, summarized in Table~\ref{Tab:Table1}. Here we have included quasars discovered in previous dedicated surveys in these fields \citep{Jakobsen2003, Steidel2003, Hennawi2006, Worseck2006, Worseck2007, Syphers2013}. Our survey yields in total 131 new quasars in the projected vicinity of the 22 \ion{He}{ii}-transparent quasars (Table~\ref{Tab:Objects}), 27 of which have $\mathrm{\Gqso  > 0.5 \times 10^{-15}\:s^{-1}}$ and fall in regions with science-grade \ion{He}{ii} $\lya$ absorption.

Our final foreground quasar sample is illustrated in Figure~\ref{Fig:Sample_Summary}. With our deep survey (yellow symbols) we are able to detect quasars $\mathrm{\approx2\,mag}$ fainter than SDSS, and preferentially discover quasars close to the background sightline ($D \lesssim 8 \pMpc$) with still high photoionization rates ($\mathrm{ \Gqso > 2 \times 10^{-15}\,{s}^{-1}}$). Figure~\ref{Fig:Sample_Summary} also shows that with our wide survey (blue and orange symbols) we find numerous quasars that have not been discovered by SDSS and BOSS.
This substantially expands our sample in the region of interest, meaning foreground quasars with $\mathrm{ \Gqso > 0.5 \times 10^{-15}\,{s}^{-1}}$. In fact, the two usable quasars with the highest photoionization rates in our sample ($\mathrm{\Gqso=18.2}$ and $\mathrm{16.9 \times 10^{-15}\,{s}^{-1}}$, see \S~\ref{Sec:Highest_Gamma_Quasars} for details) are from our wide survey. Such bright ($r\sim 19$) quasars outside the field of view of our deep survey are particularly important to constrain the quasar lifetime.

\begin{figure*}[t]
\includegraphics[width=\textwidth]{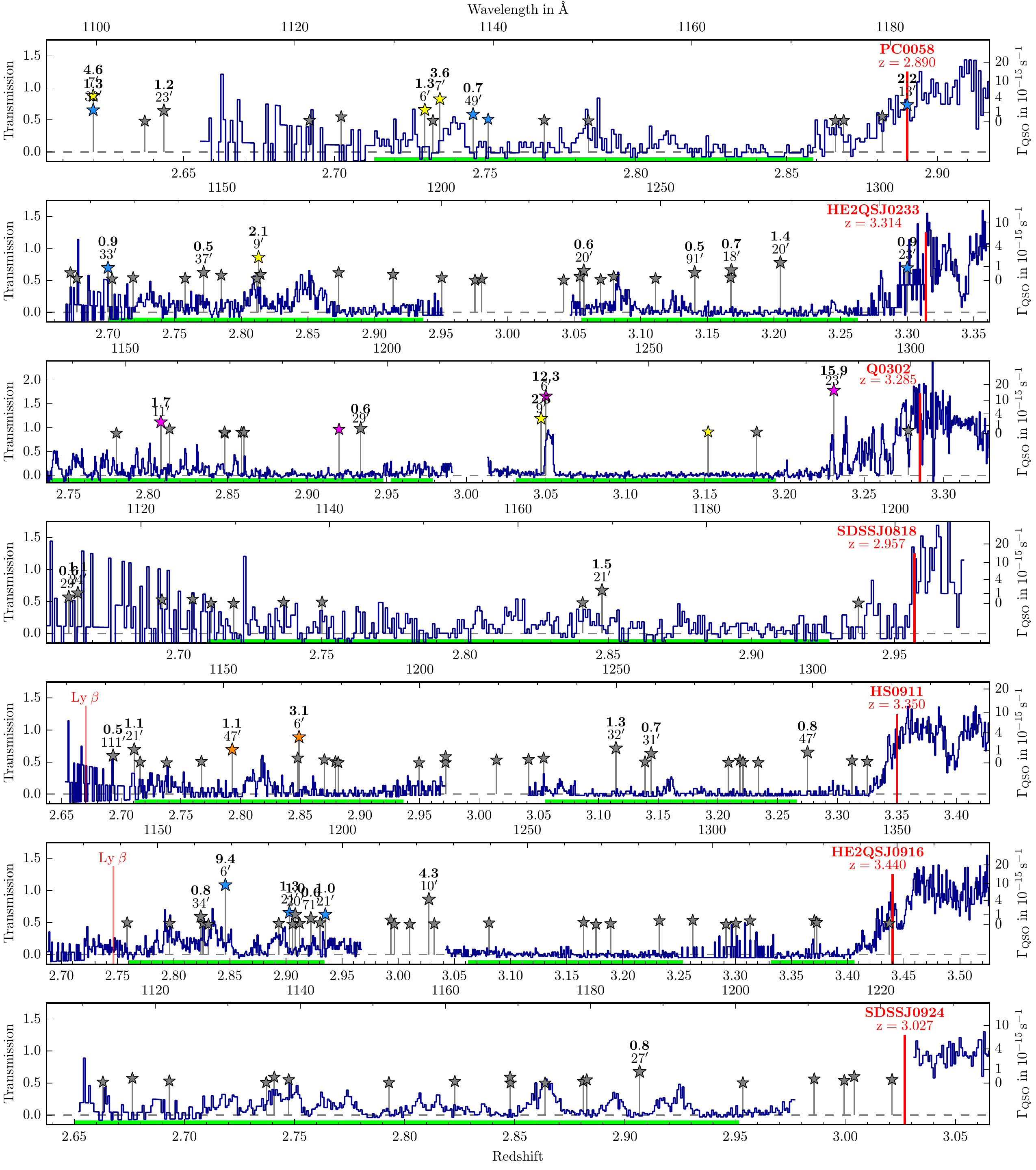}
\caption{Gallery of the 22 \ion{He}{ii} sightlines and the foreground quasar sample. The \textit{HST}/COS G130M spectrum of Q\,0302$-$003 has been binned to $\mathrm{\approx0.15\AA}$ per pixel for visualization purposes. Overplotted are positions of foreground quasars which are vertically displaced as indicated by the right axis to visualize their estimated \ion{He}{ii} photoionization rate at the background sightline. For foreground quasars with high impact, $\Gqso$ is also given as a label in units of $10^{-15}\:\mathrm{s}^{-1}$ (bold top number) and the separation from the background sightline in arcminutes (bottom). Foreground quasars with $\Gqso < 0.1 \times 10^{-15}\:\mathrm{s}^{-1}$ are omitted. Colors denote quasars discovered by us with different instruments (yellow: VLT/VIMOS; orange: CAHA/TWIN; blue: ESO NTT/EFOSC2; purple: Keck/LRIS), quasars from SDSS and BOSS (gray), and quasars from previous dedicated surveys (black). A green strip on the bottom edge of the plots indicates the regions of the \ion{He}{ii} spectra we use in the statistical analysis.}
\label{Fig:HeSpectra1}
\end{figure*}

\begin{figure*}
\addtocounter{figure}{-1}
\includegraphics[width=\textwidth]{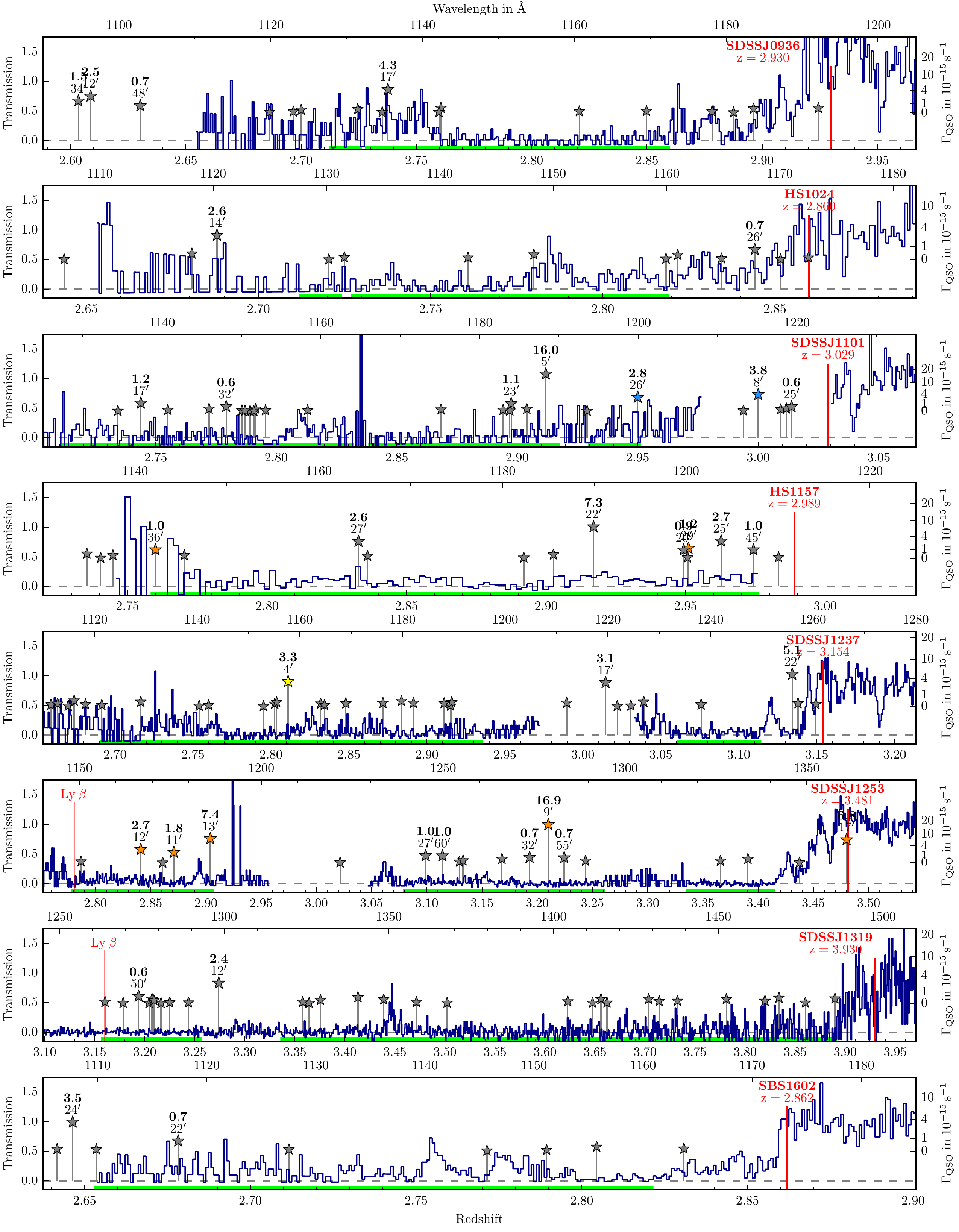}
\caption{\ion{He}{ii} sightlines continued.}
\label{Fig:HeSpectra2}
\end{figure*}

\begin{figure*}
\addtocounter{figure}{-1}
\includegraphics[width=\textwidth]{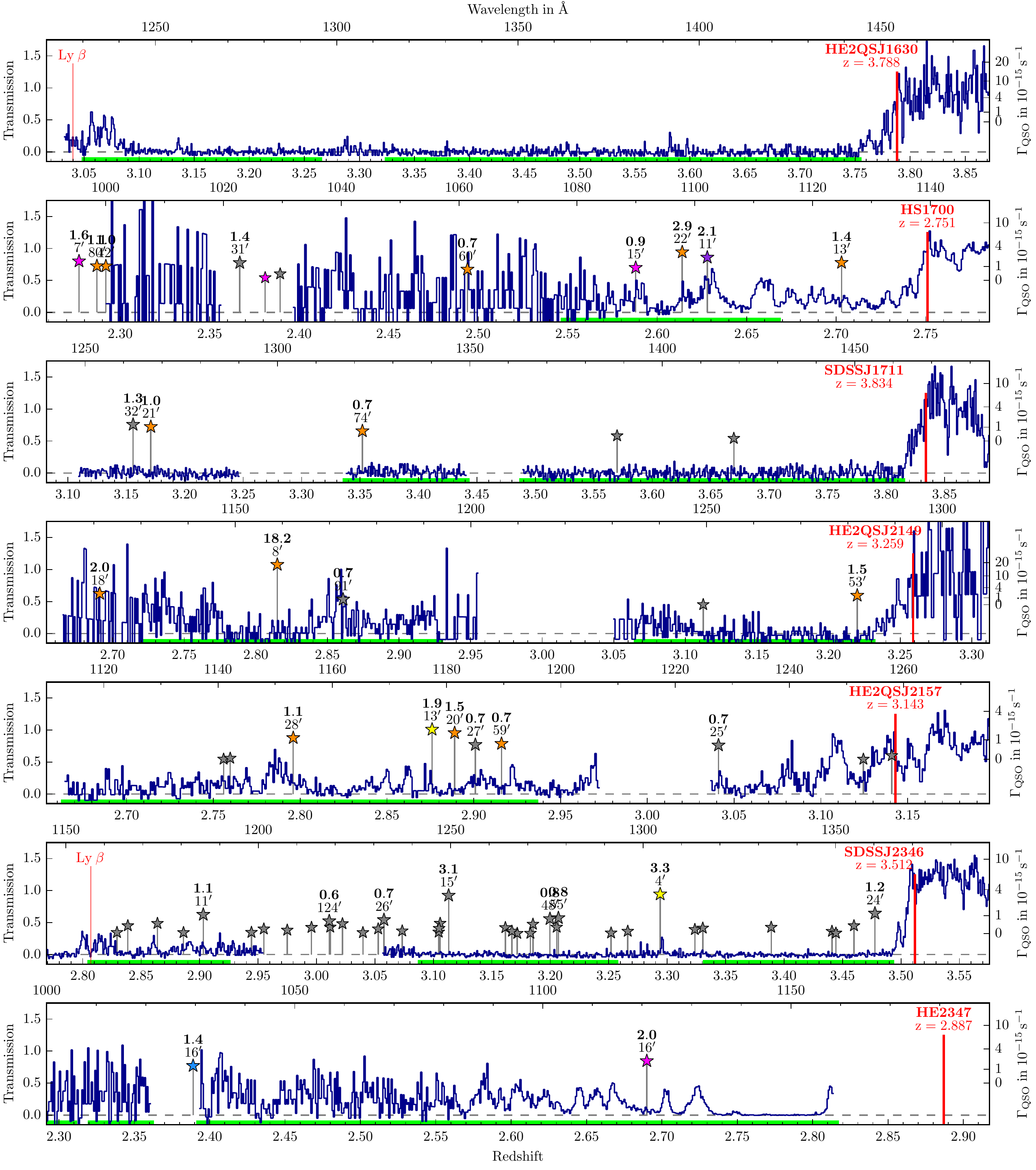}
\caption{\ion{He}{ii} sightlines continued. The vicinity of HE2QS\,J1630$+$0435 was targeted as part of our wide survey but no relevant foreground quasar was found. Although this sightline does not add any transverse proximity effect data, we include it in the statistical analysis (in particular the Monte Carlo analysis \S~\ref{Sec:Significance}), and therefore show it in this figure.}
\label{Fig:HeSpectra3}
\vspace{2cm}
\end{figure*}

Spectra of all \ion{He}{ii} sightlines are shown in Figure~\ref{Fig:HeSpectra1} together with their respective foreground quasars.
Gaps in the spectra are due to masking of geocoronal residuals (\S~\ref{sec:HeII_Sightlines}).
The regions that pass our quality criteria are indicated with a light green stripe at the bottom of the plots. 
Overplotted are the positions of foreground quasars.

\section{Search for the Transverse Proximity Effect in Individual Sightlines}
\label{Sec:Special_Objects}
Although the main aim of this work is a statistical analysis of the \ion{He}{ii} transverse proximity effect, we will briefly discuss a few special objects to build intuition about the proximity effect signal.

\subsection{HS\,1700$+$6416}
Near HS\,1700$+$6416 four foreground quasars were discovered by \citet{Syphers2013}, one of which had an uncertain redshift assignment ($z=2.625$) due to a single detected emission line. With our Keck/LRIS spectrum we confirmed this quasar at a redshift $z = 2.628\pm0.003$, matching a broad \ion{He}{ii} transmission peak in the COS spectrum. Our CAHA survey discovered three additional foreground quasars at $z>2.3$, one of which lines up with a narrow \ion{He}{ii} spike at $z=2.614$.

However, at these low redshifts where \ion{He}{ii} reionization is likely to be already completed, it is unclear whether these features are caused by the additional ionizing radiation of foreground quasars or arise due to density fluctuations. Given the generally high \ion{He}{ii} transmission at $z<2.7$, a random association of a foreground quasar with such a region is much more likely to occur than at higher redshifts. Chance alignments of the two foreground quasars with the HS\,1700$+$6416 transmission spikes thus cannot be excluded.

\subsection{Q\,0302$-$003}
\label{Sec:Q0302}
The sightline towards Q\,0302$-$003 presents the prototypical case for the \ion{He}{ii} transverse proximity effect.

At $z<3.19$, i.e.\ outside the large line-of-sight proximity zone of the background quasar ($\approx 80\cMpc$), the \ion{He}{ii} spectrum shows almost no transmission down to redshifts $2.9$. The exception is a striking transmission feature at $z=3.05$ \citep{Heap2000} with a width of $\mathrm{\approx 450\,km\,s^{-1}}$ ($5.7\cMpc$) and a peak transmission of 80\%
\footnote{Only for Q\,0302$-$003 we use a high-resolution COS G130M spectrum. The spike would be marginally resolved at G140L resolution and show a lower peak transmission but would still be identified as an outstanding spectral feature.}.
\citet{Jakobsen2003} found an $r=20.6\mag$ foreground quasar with a corresponding redshift $z=3.050\pm0.003$ $6.5\arcmin$ away from the sightline and argued that this results from the \ion{He}{ii} transverse proximity effect.

Our survey uncovered a second quasar at a very similar redshift of $z=3.047$, but further away from the background sightline ($8.5\arcmin$) and $0.7\mag$ fainter than the \citet{Jakobsen2003} quasar (Figure~\ref{Fig:Zooms}).
Based on its higher photoionization rate ($\Gqso = 12 \times 10^{-15}\,\mathrm{s}^{-1}$ vs.\ $\Gqso = 2.5 \times 10^{-15}\,\mathrm{s}^{-1}$) the \citet{Jakobsen2003} quasar seems to be the dominant source of \ion{He}{ii}-ionizing photons, but given the uncertain impact of obscuration, lifetime and quasar SED effects on the resulting photoionization rate from either quasar (\S~\ref{Sec:Gamma_QSO}), we cannot make any judgment about their relative contribution. The second quasar could still have a significant impact and it might be that the prominent \ion{He}{ii} transmission spike is the result of the combined ionizing power of both quasars. In any case, it demonstrates that a one-to-one assignment of transmission spikes to foreground quasars is an oversimplified approach that does not capture the full complexity of the \ion{He}{ii} transverse proximity effect.

\subsection{Foreground Quasars with the Highest Observed Photoionization Rates}
\label{Sec:Highest_Gamma_Quasars}

\begin{figure*}
\includegraphics[width=\linewidth]{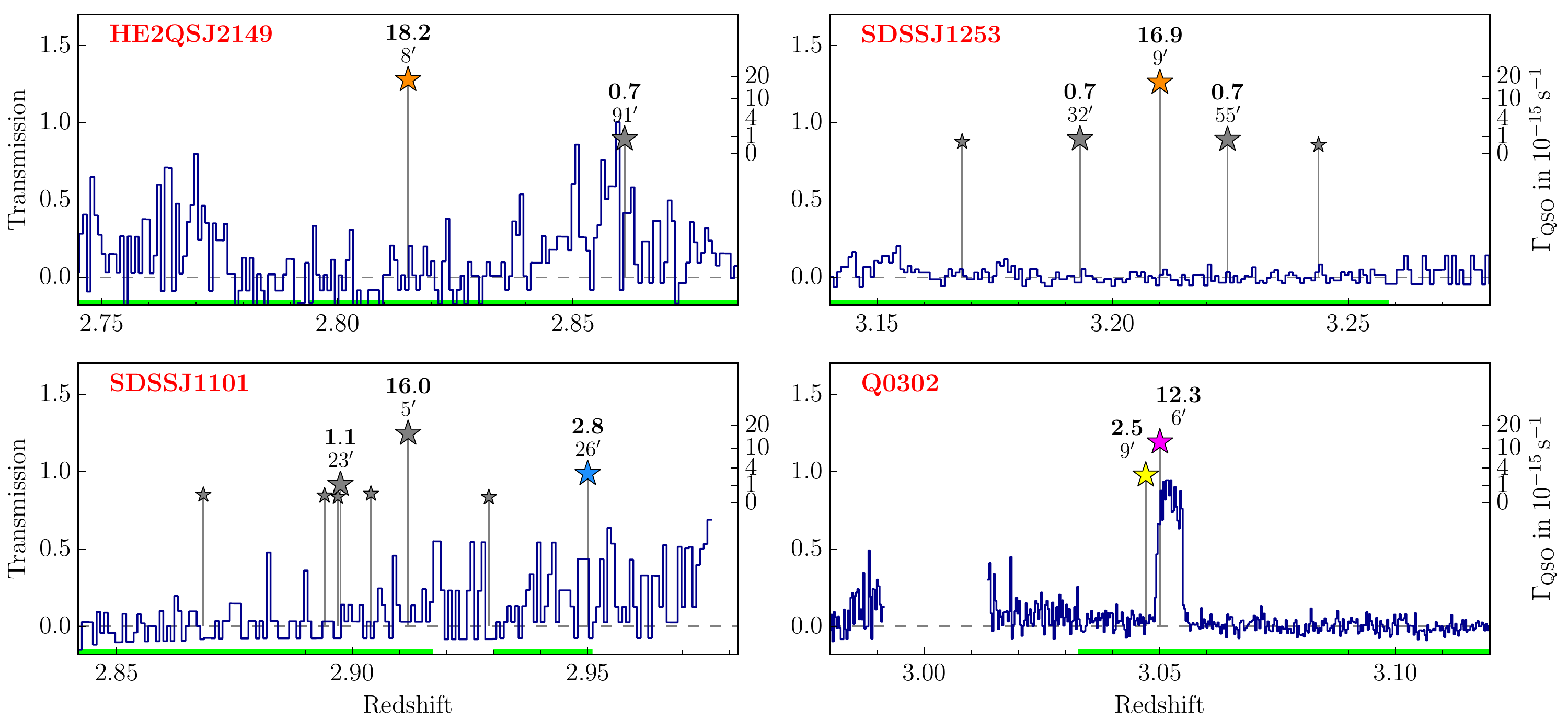}
\caption{Enlarged versions of Figure~\ref{Fig:HeSpectra1} for the regions around the four foreground quasars with the highest $\Gqso$ (labeled in bold). Only for the \citet{Jakobsen2003} quasar near the Q\,0302$-$003 sightline a \ion{He}{ii} transmission spike is observed. For the other three quasars, despite their higher \ion{He}{ii} photoionization rates, we observe strong absorption at the foreground quasar positions. All four panels have the same scale in the sense that they all show a pathlength of $\Delta z = 0.14$.}
\label{Fig:Zooms}
\end{figure*}

Our sample contains three new foreground quasars for which we infer a photoionization rate substantially higher than the \citet{Jakobsen2003} quasar. The respective regions in the \ion{He}{ii} spectra are shown in Figure~\ref{Fig:Zooms}.
Near the sightline toward SDSS\,J1253$+$6817 we have discovered a $z=3.20$ quasar with an estimated $\Gqso = 16.9 \times 10^{-15}\,\mathrm{s}^{-1}$, 40\% larger than for the \citet{Jakobsen2003} quasar. Despite this, there is no transmission spike observed even remotely comparable to the one in the Q\,0302$-$003 sightline. The spectrum in this region is consistent with zero transmission ($\tau_\mathrm{eff} > 4$). Despite the somewhat larger separation from the background sightline and the slightly higher redshift, this is a very surprising result.
A similar situation is observed for HE2QS\,J2149$-$0859, where we find a foreground quasar at $z=2.82$ with $\Gqso = 18.2 \times 10^{-15}\,\mathrm{s}^{-1}$ and no enhancement in transmission in the \ion{He}{ii} sightline. Another quasar with high $\Gqso = 16.0 \times 10^{-15}\,\mathrm{s}^{-1}$ but no visible impact on the \ion{He}{ii} transmission is located near the SDSS\,J1101$+$1053 sightline at $z=2.91$ .

We have discovered three additional quasars for which the inferred \ion{He}{ii} photoionization rate is in excess of the \citet{Jakobsen2003} system, and an order of magnitude above the estimated UV background. Whereas the precedent set by \citet{Jakobsen2003} might lead one to expect
strong transmission spikes associated with all three of these new foreground  quasars, this is clearly not the case. These new objects illustrate that there is no simple deterministic relationship between transmission spikes and our inferred \ion{He}{ii} photoionization rate. Instead they point to a large scatter in the transverse proximity effect which probably results from dependencies on other parameters  (e.g.\ lifetime, obscuration, equilibration time, IGM absorption, IGM density fluctuations) which are not captured by our simple isotropic $\Gqso$ model (see \S~\ref{Sec:Gamma_QSO}). This highlights that a statistical analysis on a large sample of foreground quasars is required to overcome the large sightline-to-sightline variation encountered in the analysis of individual associations, which is the main aim of this work.

\section{Statistical Data Analysis of the \ion{He}{ii} Transverse Proximity Effect}
\label{Sec:StatisticalAnalysis}

\subsection{Average \ion{He}{ii} Transmission near Foreground Quasars}
\label{Sec:Stacking}

Our strategy is to stack the \ion{He}{ii} $\lya$ spectra centered on the redshifts of the foreground quasars. This gives us an average \ion{He}{ii} transmission profile along the background sightlines which can then be tested for a local enhancement in the vicinity of the foreground quasars.

From the masked \ion{He}{ii} spectra (\S~\ref{sec:HeII_Sightlines}, Figure~\ref{Fig:HeSpectra1}) we extract regions of $\pm120\cMpc$ (corresponding to $\Delta z = 0.125$ or $\mathrm{9400\,km\,s^{-1}}$ at $z = 3$) centered on the redshifts of the foreground quasars and rebin them to a common pixel scale of $5\cMpc$ ($\mathrm{390\,km\,s^{-1}}$ or $\Delta z = 0.005$ at $z=3$), much coarser than the typical resolution of our \ion{He}{ii} spectra ($R\approx2000$). From this set of extracted and rebinned spectra we select a subsample for which the individual foreground quasars exceed a threshold in the inferred \ion{He}{ii} photoionization rate (\S~\ref{Sec:Gamma_QSO}). We then compute for each spatial bin the mean transmission by averaging over all spectra in the subsample.

For each bin we estimate uncertainties using the bootstrap resampling technique and compute the 15\%--85\% percentile interval. The bootstrap errors should give a combined estimate on the measurement error and the statistical fluctuations in the transverse proximity effect signal.
However, bootstrapping is probably not converged for very small samples with $N \lesssim 5$ quasars.
We therefore also propagate the formal photon-counting error of the initial spectra and take the maximum of propagated and bootstrap error. For bins with sufficient samples the bootstrap errors give the larger and more physical uncertainty. 
In cases where we have very few samples ($\lesssim 3$) the errors will be underestimated. However, these errors are only used
for qualitative visualization purposes to provide an estimate of the error in our stack.
Our analysis and estimate of the significance does not depend on these error estimates but is instead based on Monte Carlo simulations (see \S~\ref{Sec:Significance}).

\begin{figure}
\includegraphics[width=\linewidth]{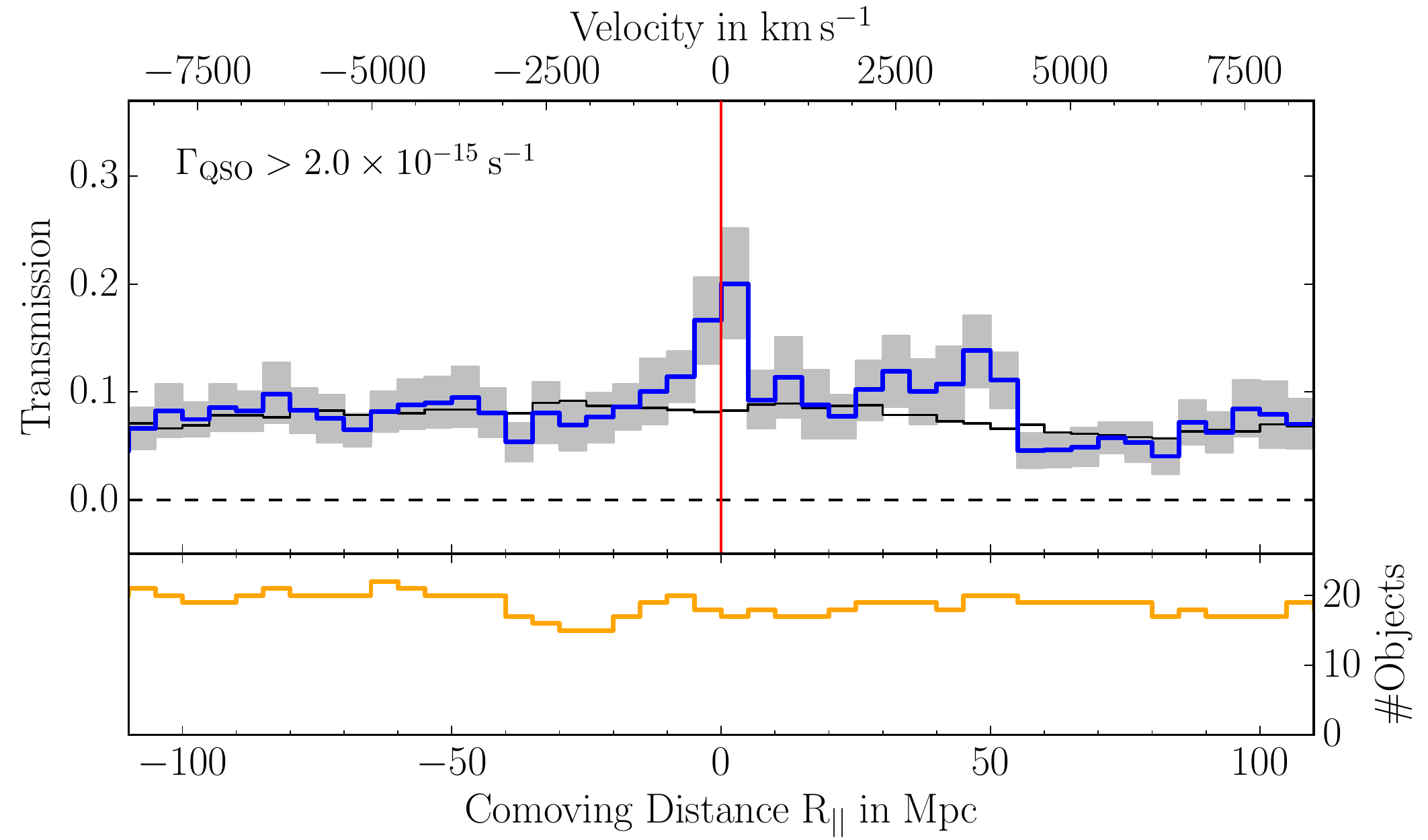}
\caption{
Average \ion{He}{ii} $\lya$ transmission profile close to foreground quasars with $\Gqso > 2.0 \times 10^{-15}\,\mathrm{s}^{-1}$, 
derived by stacking the $5\cMpc$-binned \ion{He}{ii} spectra centered on the foreground quasar redshifts.
The vertical red line marks the position of the foreground quasars. 
Conversion between comoving distance and velocity assumes the median redshift of the sample $z=2.89$.
Gray shaded areas give a bootstrap error estimate. The number of contributing spectra per bin is given in the bottom panel. An estimate for the mean IGM \ion{He}{ii} transmission stacked in the same way as the \ion{He}{ii} spectra is shown as a thin black line in the top panel. 
The averaged \ion{He}{ii} transmission profile shows a clear enhancement within $|R_\parallel| < 15\cMpc$. We estimate its significance to be $3.1\,\sigma$, based on a Monte Carlo analysis.
}\label{Fig:FUV_Stack_20_00_00-99}
\end{figure}

A stack including all foreground quasars with $\Gqso > 2.0 \times 10^{-15}\,\mathrm{s}^{-1}$ is shown in Figure~\ref{Fig:FUV_Stack_20_00_00-99}. The top panel shows the stacked \ion{He}{ii} spectra centered on the foreground quasars. The lower x-axis gives the distance in comoving Mpc from the points on the sightlines that are closest to the foreground quasars, denoted as $R_\mathrm{\parallel}$. The upper x-axis converts this distance into an approximate velocity assuming Hubble flow and using the median redshift of the sample $z=2.89$ for the conversion. Bootstrap errors (15\%--85\% percentile range) are shown as the gray shaded area. Since we have to mask the \ion{He}{ii} spectra according to the conditions described in \S~\ref{sec:HeII_Sightlines}, the spectra do not necessarily have continuous coverage. This leads to some variation in the number of spectra contributing per bin which we therefore explicitly show in the bottom panel below the stack.

\begin{figure}
\begin{center}
\includegraphics[width=\linewidth]{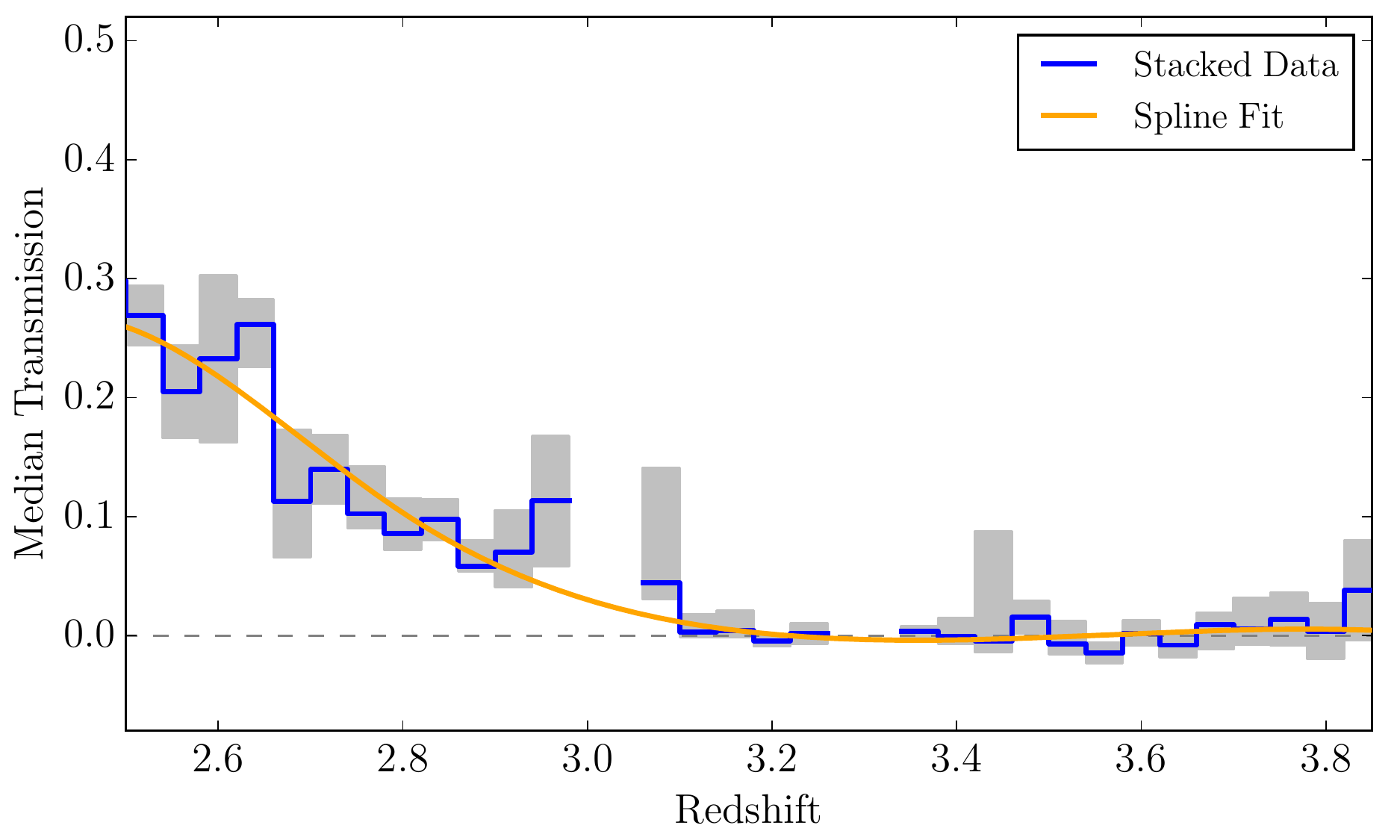}
\caption{Redshift evolution of the median \ion{He}{ii} $\lya$ transmission, derived by binning the \ion{He}{ii} spectra to a common grid of width $\Delta z = 0.04$ and taking for each bin the median transmission values (blue line). Gray shaded regions represent bootstrap errors.
Gaps in the data at $z=3$ and $z=3.3$ are due to geocoronal contamination. The orange curve shows our spline fit.
}
\label{Fig:Average_Transmision}
\end{center}
\end{figure}

For the sake of illustration we also compute the expected stacked signal at random positions in the IGM, which is shown as the thin black line in Figure~\ref{Fig:FUV_Stack_20_00_00-99}. We binned the \ion{He}{ii} spectra to a common grid of width $\Delta z = 0.04$ and computed for each bin the median transmission over all contributing spectra. Through these points we then fit a cubic spline as our model of the median IGM \ion{He}{ii} transmission. The result is shown in Figure~\ref{Fig:Average_Transmision}, which clearly shows the high \ion{He}{ii} opacity of the IGM above $z>3.2$ and the rapid increase in transmission below that redshift. To compute the thin black line in Figure~\ref{Fig:FUV_Stack_20_00_00-99}, we evaluate this spline fit at the redshifts corresponding to the pixels in the selected background sightlines and mask and stack these values in exactly the same way as the original data.

At line-of-sight distances of $|R_\parallel| > 30\cMpc$ one observes a fairly flat and constant transmission at the 8\% level,
consistent with the median IGM transmission. At $|R_\parallel|<15\cMpc$ we see a clear enhancement to 20\% transmission that is centered on the
position of the foreground quasars and falls off gradually to larger distances. As $\approx 20$ sightlines contribute per $5\cMpc$ bin, this enhancement is a clear indication for the presence of a \ion{He}{ii} transverse proximity effect in the average IGM around quasars.

The shape of the observed transmission profile seems to be asymmetric, having the peak slightly behind the foreground quasar,
falling off gradually to lower redshifts ($R_\parallel < 0$) and more steeply to higher redshifts ($R_\parallel > 0$).
However, quasar redshift errors are large enough to significantly contribute to the observed shape of the transmission profile ($\sigma_z\approx \mathrm{1000\:km\:s^{-1}}$ corresponding to $13\cMpc$ at the median redshift $z=2.89$). We therefore cannot make any statement about its intrinsic shape. Our stack also shows some enhanced transmission between $+20\cMpc$ and $+50\cMpc$, i.e.\ behind the foreground quasar. It is not as pronounced as the central peak and is marginally consistent with the median IGM transmission in most of the bins. In the following we will quantify the statistical significance of the central transmission feature as well as the transmission enhancement around $+40\cMpc$.

\subsection{Monte Carlo Significance Estimate}
\label{Sec:Significance}

To formally estimate the significance of the increased \ion{He}{ii} transmission in the vicinity of foreground quasars we perform a Monte Carlo analysis. The first step is to quantify the strength of the enhanced transmission. For this we require a measure for the amount of excess transmission. We define a kernel of $\pm15\cMpc$, i.e.\ slightly larger than the typical $13\cMpc$ position uncertainty of foreground quasars, and compute the average transmission of the stack inside this window. The average transmission within $|R_\parallel|<15\cMpc$ can then be compared to the average transmission in a control window ($15\cMpc<|R_\parallel|<120\cMpc$). The control window should yield a fair estimate of the average transmission since the bins sufficiently far away are unlikely to be correlated with the foreground quasar. They therefore represent the average IGM at approximately the same redshift and probed by the same sightlines which naturally incorporates the redshift distribution of the stacked subsample. The large size of our control window of $210\cMpc$ ensures that we do not add unnecessary noise to our statistic. We define the transmission enhancement as the difference of the \ion{He}{ii} transmission ($T$) within the central $\pm15\cMpc$ and the transmission outside of this and denote it as
\begin{equation}
\xiTPE = \langle T_{|R_\parallel|<15\cMpc}\rangle - \langle T_{15\cMpc<|R_\parallel|<120\cMpc}\rangle
\label{Eq:xi}.
\end{equation}

\begin{figure}
\includegraphics[width=\linewidth]{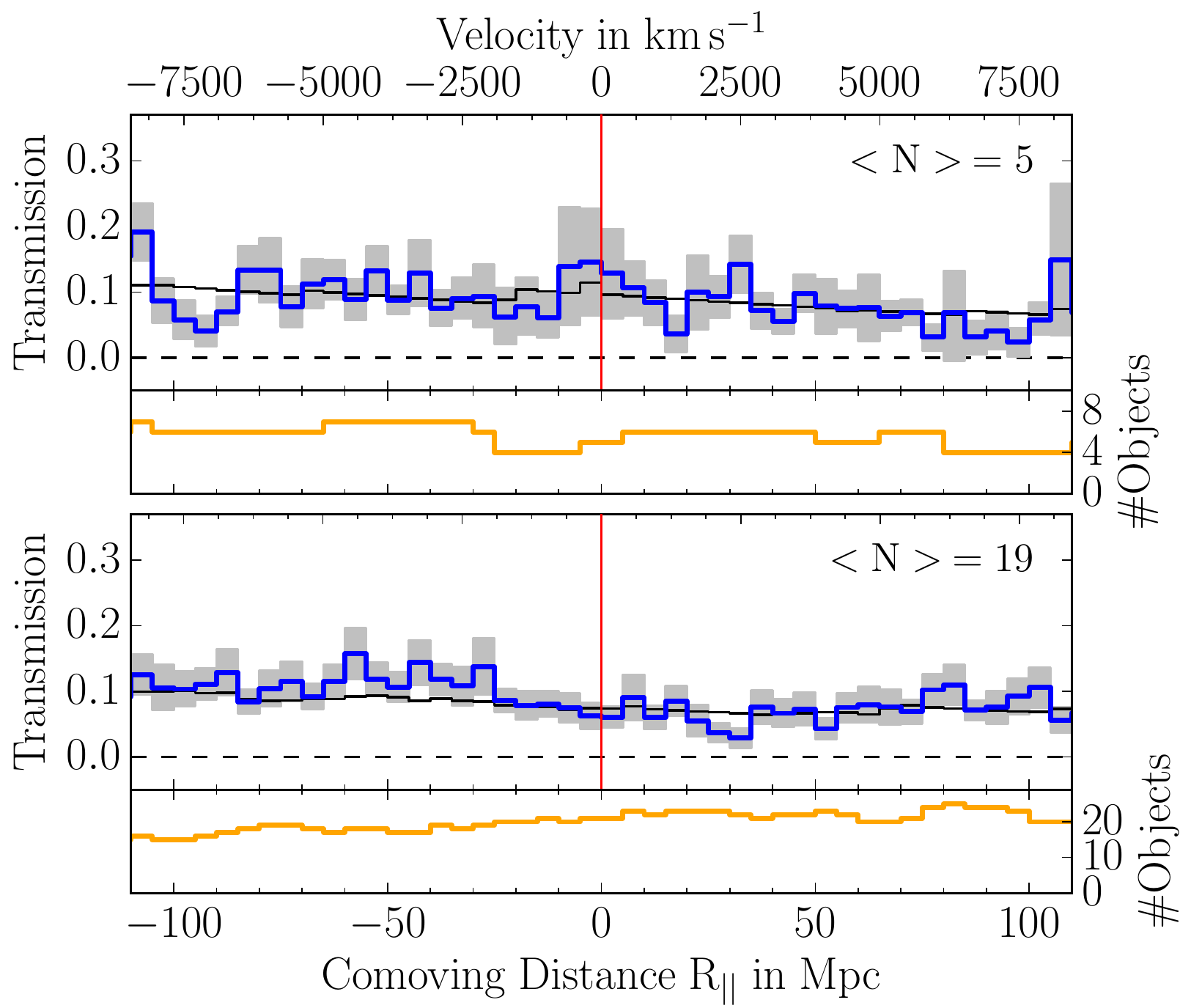}
\caption{Examples of mock stacks used for the Monte Carlo analysis, created by stacking the \ion{He}{ii} spectra on random redshifts. The top (bottom) panel shows a stack with 5 ($\simeq 19$) contributing spectra per bin.}
\label{Fig:Random_Stacks}
\end{figure}

To evaluate the probability distribution of the transmission enhancement at \emph{random} locations in the IGM we create a large number of mock stacks.We take \ion{He}{ii} spectra centered on random redshifts drawn from a redshift distribution matching the redshift distribution of foreground quasars in the science stack, and keep adding such spectra to the random stacks until the average number of contributing spectra per spectral bin equals the one in the science stack.
We stress that due to the the limited number of \ion{He}{ii} sightlines and discontinuous coverage the number of contributing spectra can only be matched on average but not necessarily for each bin of the stack. Examples of mock stacks are shown in Figure~\ref{Fig:Random_Stacks}. We then count for how many mock stacks the transmission enhancement $\xiTPE$ exceeds the measured value in the science stack. We create mock stacks until 500 of them fulfill this condition to ensure proper sampling of the tail of the distribution.

\begin{figure}
\includegraphics[width=\linewidth]{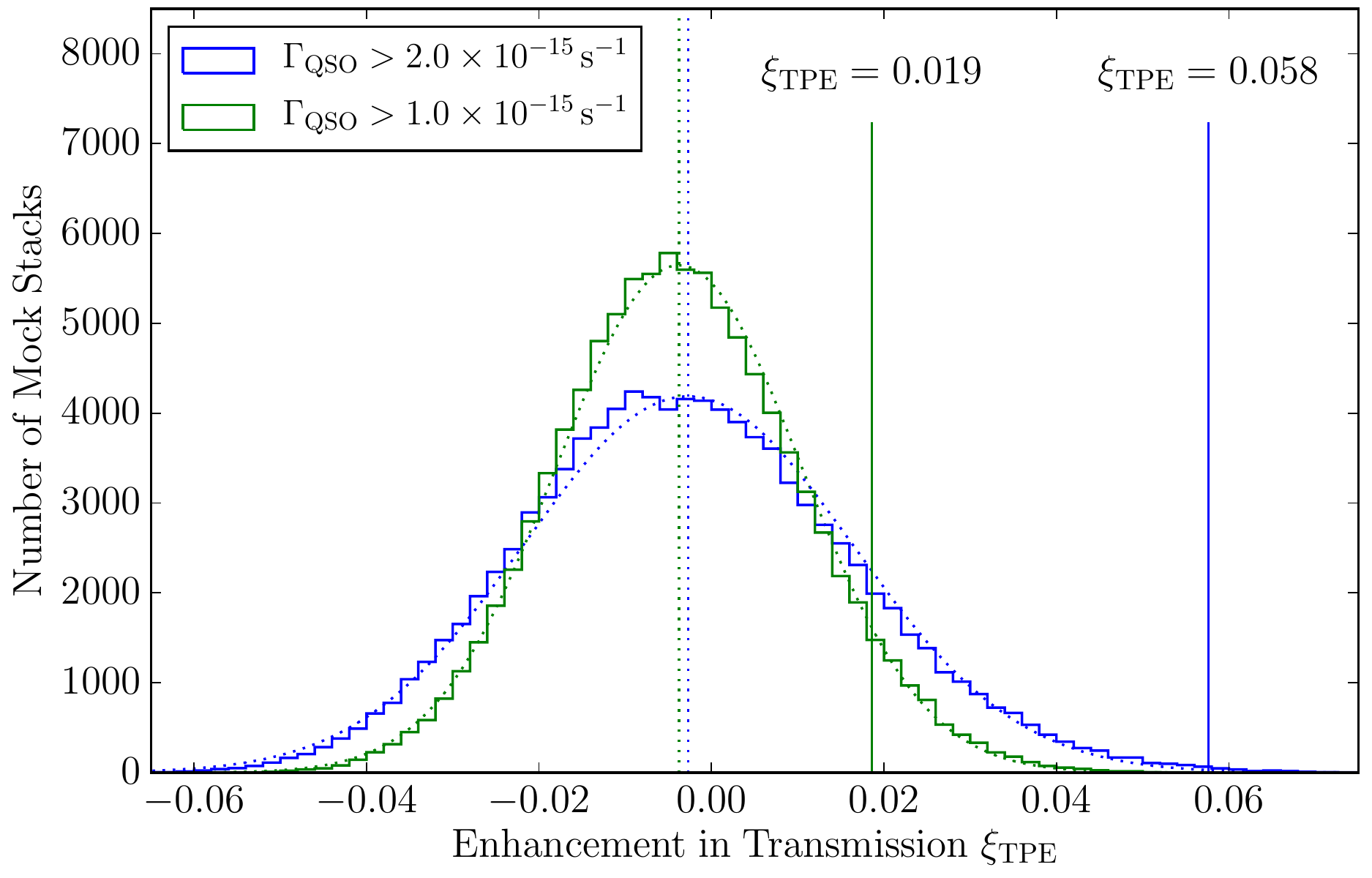}
\caption{Distribution of the transmission enhancement in $10^5$ mock stacks for two different cuts in $\Gqso$. 
The $\Gqso > 2 \times 10^{-15}\,\mathrm{s}^{-1}$ stacks contain 19 spectra per bin and therefore show a wider spread in $\xiTPE$ than the $\Gqso > 1 \times 10^{-15}\,\mathrm{s}^{-1}$ stacks which have 34 spectra per bin. 
The dotted lines give the mean and Gaussian approximation for the distributions. 
The $\xiTPE$ values measured in the science stacks are indicated with solid vertical lines. For $\Gqso > 1 \times 10^{-15}\,\mathrm{s}^{-1}$ ($\Gqso > 2 \times 10^{-15}\,\mathrm{s}^{-1}$) $5.3$\% ($<0.1$\%) of the mock stacks exceed the measured $\xiTPE=0.019$ ($\xiTPE=0.058$).}
\label{Fig:Hist_Mock_Stacks}
\end{figure}

The transmission enhancement in the science stack with $\Gqso > 2 \times 10^{-15}\,\mathrm{s}^{-1}$ (Figure~\ref{Fig:FUV_Stack_20_00_00-99}) is $\xiTPE=0.058$. The blue histogram in Figure~\ref{Fig:Hist_Mock_Stacks} shows the $\xiTPE$ distribution of random stacks matched to this science stack in terms of redshift distribution and number of contributing spectra per bin. We find that in $0.1$\% of the random stacks the transmission enhancement exceeds the measured $\xiTPE=0.058$. The $\xiTPE$ distribution is well approximated by a Gaussian (blue dotted curve in Figure~\ref{Fig:Hist_Mock_Stacks}), so we detect a transmission enhancement at a statistical significance of $3.1\sigma$. In the following subsections we will create stacks with modified selection criteria. To illustrate how this influences the corresponding Monte Carlo analyses,
we consider a science stack of all foreground quasars with a lower threshold value $\Gqso > 1\times 10^{-15}\,\mathrm{s}^{-1}$ in which we measure $\xiTPE = 0.019$. For $\Gqso > 1 \times 10^{-15}\,\mathrm{s}^{-1}$ the $\xiTPE$ distribution (green) is narrower than for $\Gqso > 2 \times 10^{-15}\,\mathrm{s}^{-1}$ (blue) due to the larger number of contributing spectra per bin (34 instead of 19). Still, $5.3$\% of the random stacks have $\xiTPE>0.019$, corresponding to a significance of $1.6\sigma$ (Figure~\ref{Fig:Hist_Mock_Stacks}).

We confirm that the previously known transmission spike and foreground quasar in the Q\,0302$-$003 sightline \citep{Heap2000, Jakobsen2003} does not dominate our signal. Excluding the region of the transmission spike at $z=3.05$ together with the two nearby foreground quasars and repeating our Monte Carlo analysis, we still end up with a chance probability of $1.5$\% for the measured enhanced transmission, corresponding to a $2.2\sigma$ detection. We show later in \S~\ref{Sec:Single_QSO_Stat} that while these two foreground quasars contribute significantly to our transverse proximity signal, they are however not the ones with the strongest transmission enhancements. 

We further use our Monte Carlo scheme to evaluate the significance of the excess transmission seen in the stack (Figure~\ref{Fig:FUV_Stack_20_00_00-99}) at $R_\parallel \approx +40\cMpc$. For this we define a window $+30\cMpc < R_\parallel < +55\cMpc$ and compute $\xiTPE$ for this range, while defining control windows at $R_\parallel < -15\cMpc$ and $R_\parallel > +55\cMpc$ to exclude the central transmission peak. We find a by-chance probability of $1.9$\% corresponding to a significance level of $2.1\sigma$. Given that we specifically chose the limits of the window to maximize the impact of the secondary peak, it could still be consistent with a statistical fluctuation. It is, however, difficult do draw definitive conclusions without additional data.

\subsection{Dependence on $\Gqso$}
\label{Sec:Gamma_Dependence}

\begin{figure}
\includegraphics[width=\linewidth]{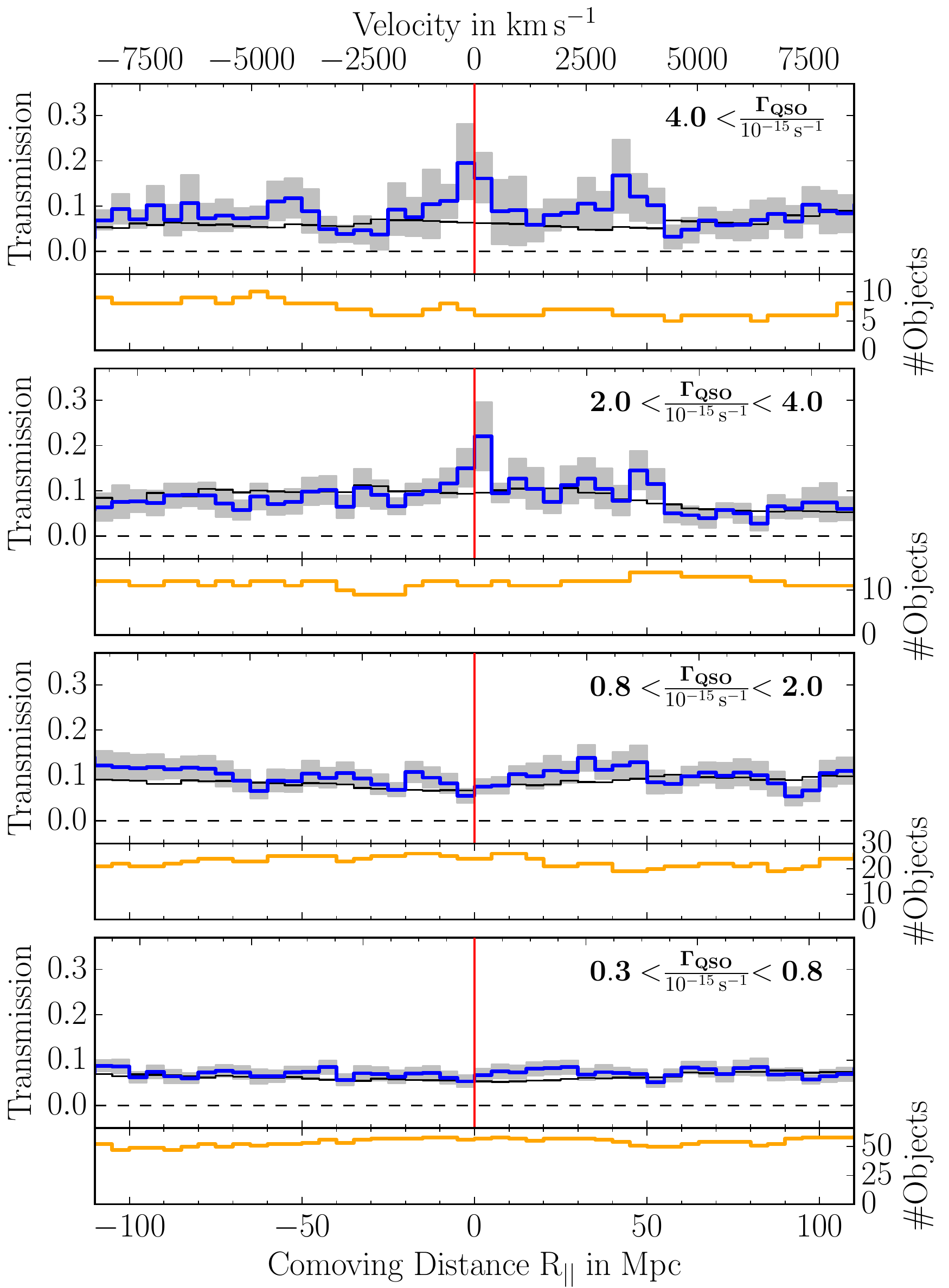}
\caption{
Average \ion{He}{ii} $\lya$ transmission profiles for foreground quasars with different ranges in $\Gqso$ (labeled). Small panels below the stacks show the number of contributing spectra per $5\cMpc$ bin. The dependence of the transmission enhancement on $\Gqso$ shows that the transmission enhancement is due to the transverse proximity effect.
}
\label{Fig:Gamma_Dependence}
\end{figure}

So far we have focused on stacks with a photoionization rate threshold of $\Gqso > 2 \times 10^{-15}\,\mathrm{s}^{-1}$ that is similar to or higher than the estimated UV background photoionization rate \citep{HaardtMadau2012, Khrykin2016} and gives the strongest signal. It is, however, informative to vary the range of $\Gqso$. Figure~\ref{Fig:Gamma_Dependence} shows spectral stacks centered on foreground quasars in four consecutive ranges of $\Gqso$. The upper panel includes foreground quasars with an estimated photoionization rate of $\Gqso > 4 \times 10^{-15}\,\mathrm{s}^{-1}$, which should result in the strongest impact on the background sightline. The next panel from the top includes quasars with intermediate impact ($2 \times 10^{-15}\,\mathrm{s}^{-1} < \Gqso < 4 \times 10^{-15}\,\mathrm{s}^{-1}$), and the lower two panels are for quasars with an estimated low impact.

The strength of the excess transmission clearly depends on the range in $\Gqso$. For foreground quasars with $\Gqso > 4 \times 10^{-15}\,\mathrm{s}^{-1}$ we detect a strong transmission enhancement of $\xiTPE=0.049$ despite the small number of $\simeq 7$ sightlines per $5\cMpc$ bin. The amplitude of the peak is even higher for the quasars with $2 \times 10^{-15}\,\mathrm{s}^{-1} < \Gqso < 4 \times 10^{-15}\,\mathrm{s}^{-1}$ ($\xiTPE = 0.063$, second panel from top). For these two cases our Monte Carlo scheme yields a significance of $1.8\sigma$ and $2.1\sigma$ (probability $p=0.035$ and $p=0.017$), respectively. The foreground quasars included in these two stacks should dominate over the UV background of $\Gamma^\mathrm{HeII}_\mathrm{UVB}\approx 10^{-15}\,\mathrm{s}^{-1}$ \citep{HaardtMadau2012, Khrykin2016}, and we interpret the detected excess transmission as their transverse proximity effect.

For foreground quasars with lower estimated photoionization rates (lower two panels in Figure~\ref{Fig:Gamma_Dependence}) any indication of the transverse proximity effect vanishes, in agreement with our expectation. We find $\xiTPE=-0.019$ and $\xiTPE=0.0$, respectively. The increased absorption in the panel third from top has a significance of $0.85\,\sigma$ ($p=0.2$) and is therefore consistent with the average transmission.

From the sequence of decreasing excess transmission with decreasing $\Gqso$ in Figure~\ref{Fig:Gamma_Dependence} we conclude that the transmission peaks are due to the transverse proximity effect of the foreground quasars. Additionally, this sequence confirms that $\Gqso$ is a useful estimator for the strength of the transverse proximity effect in an ensemble of foreground quasars, despite the assumptions in \S~\ref{Sec:Gamma_QSO}. However, this is only true when averaging over a sample. As shown in \S~\ref{Sec:Special_Objects} and later in \S~\ref{Sec:Single_QSO_Stat}, there is a large object-to-object variance.

\subsection{Redshift Evolution}
\label{Sec:RedshiftEvolution}

\begin{figure*}
\begin{center}
\includegraphics[width=.325\linewidth]{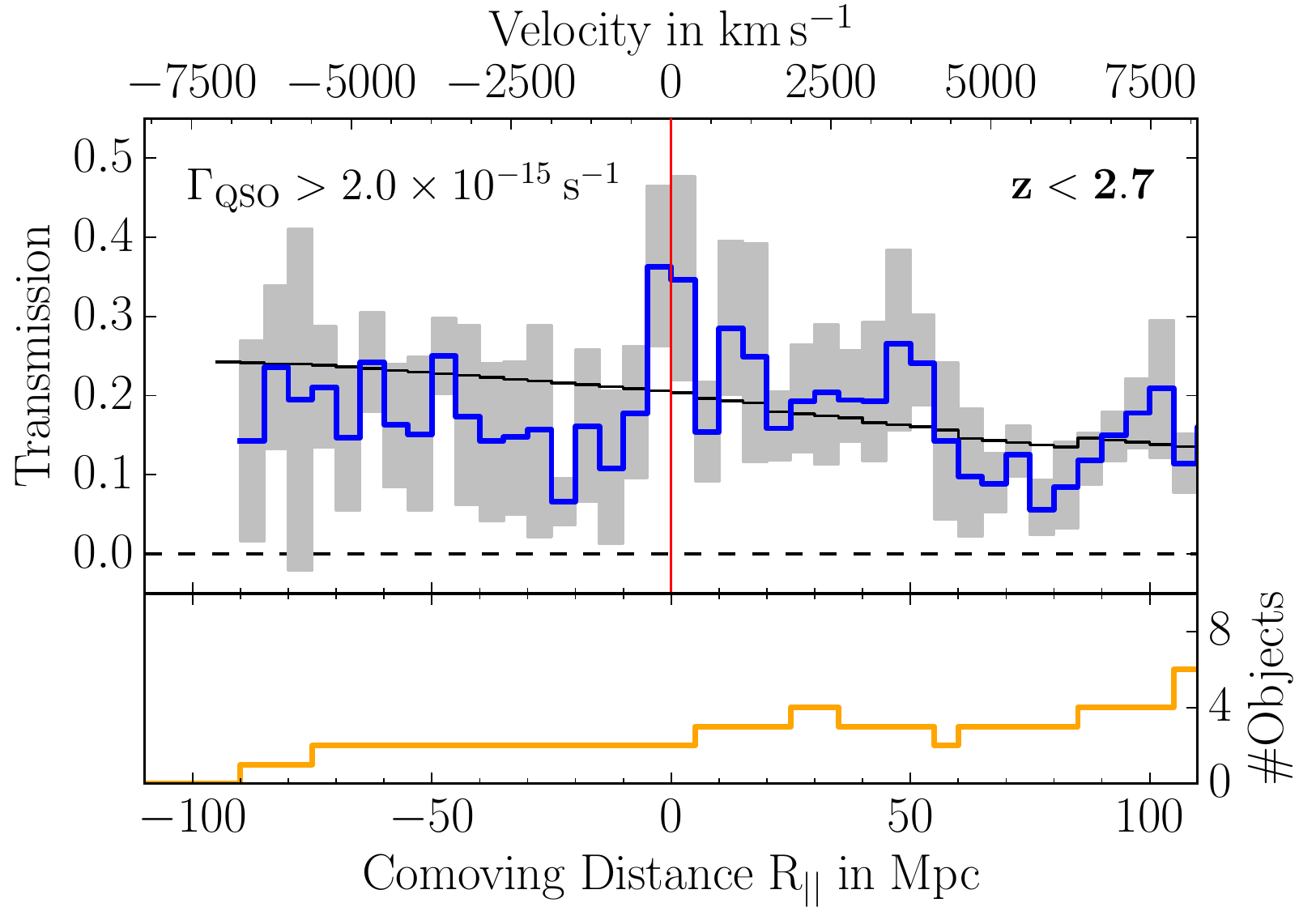}
\includegraphics[width=.325\linewidth]{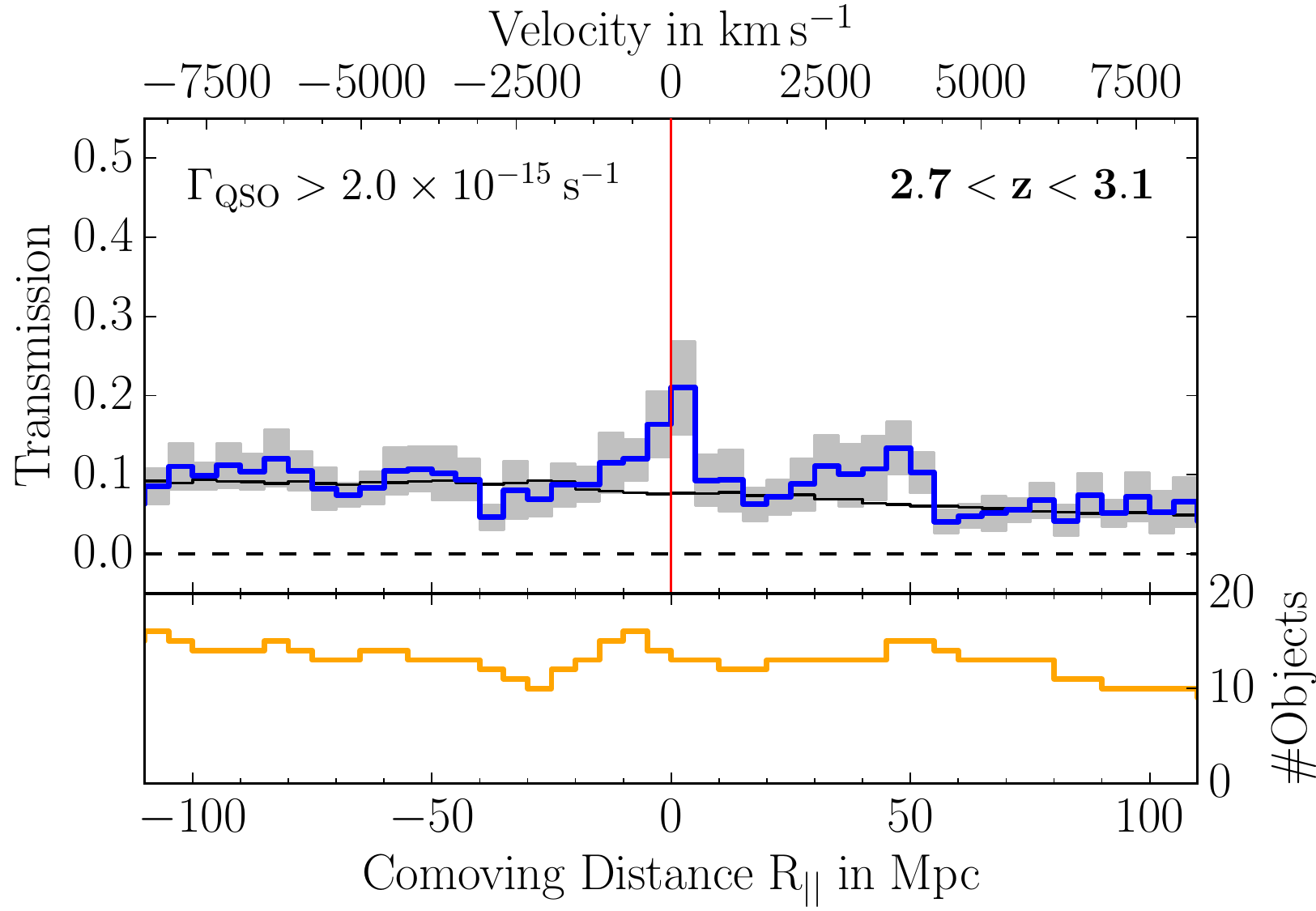}
\includegraphics[width=.325\linewidth]{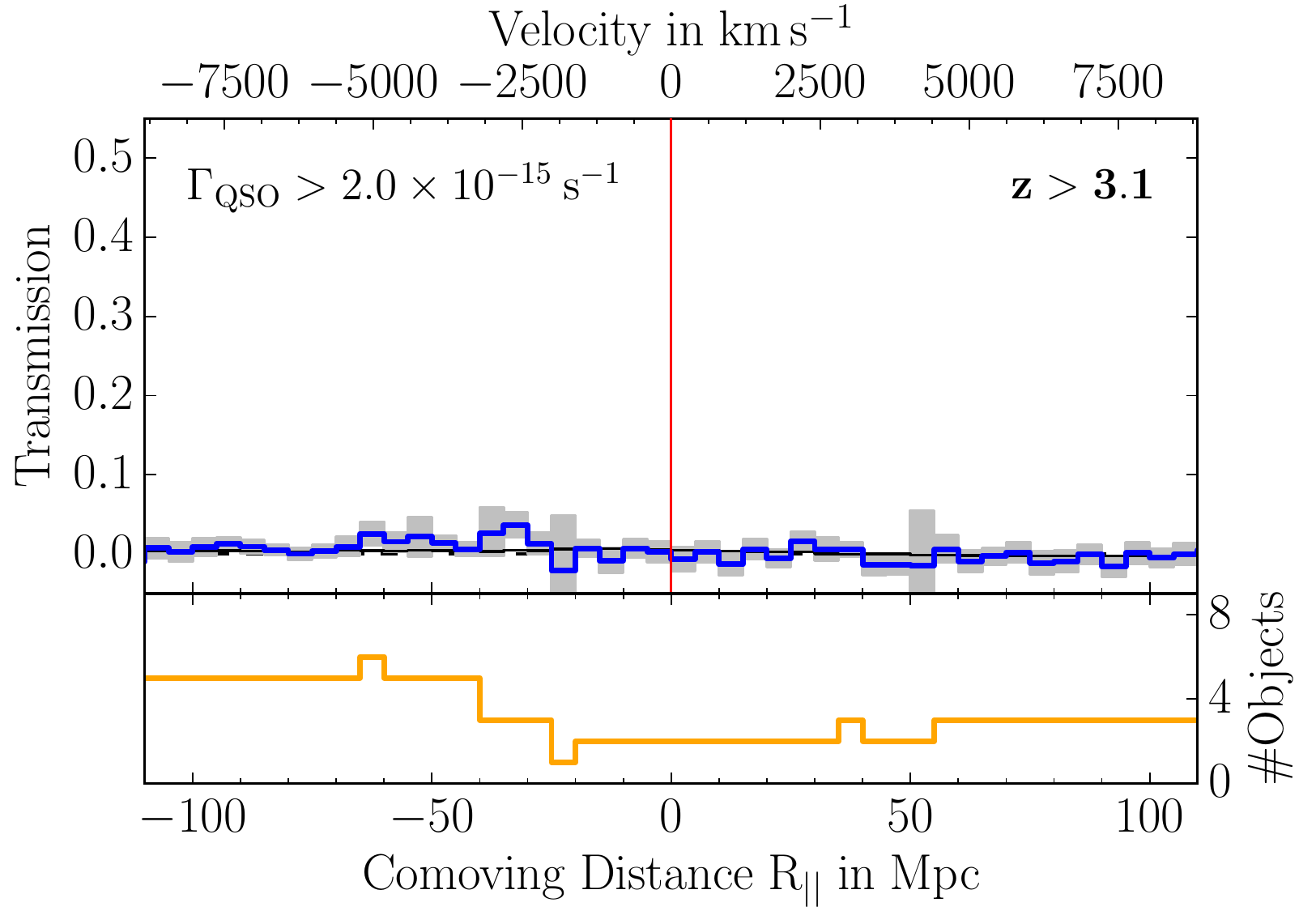}
\caption{
Redshift evolution of the average \ion{He}{ii} $\lya$ transmission near foreground quasars with $\Gqso > 2.0 \times 10^{-15}\,\mathrm{s}^{-1}$ in three redshift ranges (labeled). The median redshifts are $z = 2.63$, $2.90$ and $3.33$, respectively. In every stack the mean transmission evolves with redshift as the bins on the left side of the plots have on average a lower redshift than the bins on the right side.
}
\label{Fig:redshift_evolution}
\end{center}
\end{figure*}

The foreground quasars with $\Gqso > 2.0 \times 10^{-15}\,\mathrm{s}^{-1}$ have a median redshift of $z=2.89$ over a range $2.5 \lesssim z \lesssim 3.5$. We expect helium reionization to finish at these redshifts \citep{Madau1994, Reimers1997, MiraldaEscude2000, McQuinn2009, HaardtMadau2012, Compostella2013, Compostella2014, Worseck2016} and it is thus expected that the average properties of the IGM are different around our high-redshift quasars compared to the ones at low redshift. The evolution in mean transmission was already illustrated in Figure~\ref{Fig:Average_Transmision} and it can also be seen in our stacks. In Figure~\ref{Fig:redshift_evolution} we show stacks of the \ion{He}{ii} transmission near foreground quasars with $\Gqso > 2.0 \times 10^{-15}\,\mathrm{s}^{-1}$ in three redshift ranges ($z<2.7$, $2.7<z<3.1$, $z > 3.1$). 

The $z>3.1$ sample is consistent with zero transmission in most of the bins including the positions close to the foreground quasars. We see no indication for a transverse proximity effect ($\xiTPE\approx0$), but note that only $\approx3$ quasars contribute to this stack. For intermediate redshifts ($2.7<z<3.1$) we find a detectable non-zero average \ion{He}{ii} transmission in the IGM that is enhanced in the vicinity of foreground quasars ($\xiTPE=0.055$). The contrast between the transmission in the quasar vicinity and the average transmission in the IGM far from quasars is largest for this redshift interval, and we find a significance of $1.9\sigma$ ($p=0.026$). At $z < 2.7$ the mean transmission rises quickly. We find a transmission enhancement of $\xiTPE =0.081$, but fluctuations in the $\lya$ forest and the relatively high mean transmission reduce its significance. Also, the COS sensitivity drops at $\lambda<1110$\,\AA\ ($z<2.65$) and the few available spectra are very noisy. Therefore we find a significance of only $1.2\sigma$ ($p=0.12$).

\subsection{Constraining the Quasar Episodic Lifetime with the Transverse Proximity Effect}
\label{Sec:Lifetime}

\begin{figure}
\includegraphics[width=\linewidth]{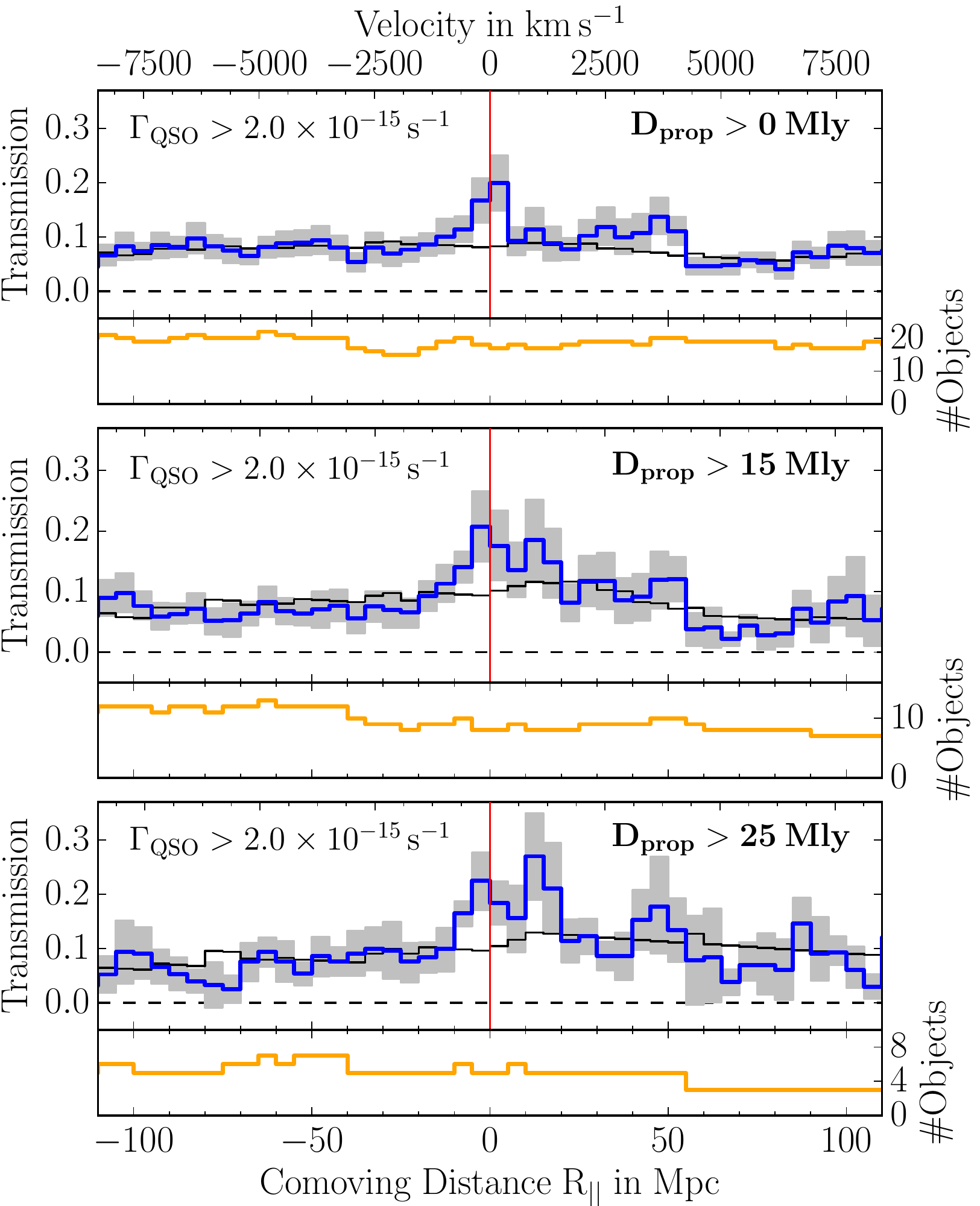}
\caption{Constraint on the quasar lifetime from the \ion{He}{ii} transverse proximity effect. The top panel shows a stack including all foreground quasars (i.e.\ Figure~\ref{Fig:FUV_Stack_20_00_00-99}) while the two others only include quasars above a minimum separation of $D_\mathrm{prop} > 15\,\Mly$ and $\mathrm{> 25\,\Mly}$. The transverse proximity effect persists in these stacks with a significance of $3.2\sigma$ and $2.6\sigma$, respectively. Since we simultaneously observe quasars and enhanced ionization at the background sightline, the quasars must shine for longer than the transverse light crossing time, implying a quasar lifetime of $\tQ > 25\Myr$.   
}
\label{Fig:Distance_Dependence}
\end{figure}

The detection of the \ion{He}{ii} transverse proximity effect presented in the previous subsections sets the stage for constraining quasar properties, in particular the episodic quasar lifetime $\tQ$. Conceptually, this requires only a few further assumptions. If we attribute enhanced \ion{He}{ii} transmission to the transverse proximity effect of a nearby foreground quasar at the same redshift, we see the quasar and this IGM parcel at the same lookback time. However, to ionize the gas, the quasar had to emit \ion{He}{ii}-ionizing photons at least for the transverse light crossing time between the foreground quasar and the sightline \citep[e.g.][]{Jakobsen2003, Worseck2006, Furlanetto2011}. If we assume a simple lightbulb model in which the quasar turns on, shines with constant luminosity for some time and turns off again, it had to shine for at least the light crossing time to allow simultaneous observation of the quasar and the additional ionization at the background sightline. We can therefore infer a geometric limit for how long the quasars already had to be active ($\tTO$) which sets a lower limit on the episodic lifetime of quasars ($t_\mathrm{Q}$). 

To do so, we create stacks for quasars above a minimum transverse separation from the background sightline. If the transverse proximity effect persists in the stack, the quasars on average have to shine for at least the light crossing time corresponding to that distance. This is presented in Figure~\ref{Fig:Distance_Dependence}. We select quasars with $\Gqso > 2 \times 10^{-15}\,\mathrm{s}^{-1}$ and apply cuts on the proper distance to the background sightline of $D_\mathrm{prop} > 15\,\Mly$ and $> 25\,\Mly$.

For $D_\mathrm{prop} > 15\,\Mly$ we detect enhanced transmission ($\xiTPE=0.097$), and our Monte Carlo simulations with this cut in $D_\mathrm{prop}$ yield a significance level of $3.2\sigma$ or a by-chance probability $p=0.0006$. The slightly higher significance compared to the stack including all foreground quasars ($3.1\sigma$) may be because at larger distances more luminous quasars are required to meet our threshold in $\Gqso$. These more luminous quasars probably have larger proximity zones \citep{Khrykin2016} possibly explaining the slightly higher significance. However, this could also be due to stochasticity in the data. For the stack including only foreground quasars with $D_\mathrm{prop} > 25\,\Mly$ the enhanced transmission persists ($\xiTPE=0.116$) at $2.6\sigma$ significance ($p=0.0053$). However, at $D_\mathrm{prop} > 25\,\Mly$ only $\approx5$ quasars are bright enough to yield a photoionization rate $\Gqso > 2 \times 10^{-15}\,\mathrm{s}^{-1}$. Note that our measurements at large separations do not include the prominent $z=3.05$ \ion{He}{ii} transmission spike in the Q\,0302$-$003 sightline, as both foreground quasars are at $D_\mathrm{prop}<13\,\Mly$ (Figure~\ref{Fig:Zooms}).

\begin{figure}[h]
 \includegraphics[width=\linewidth]{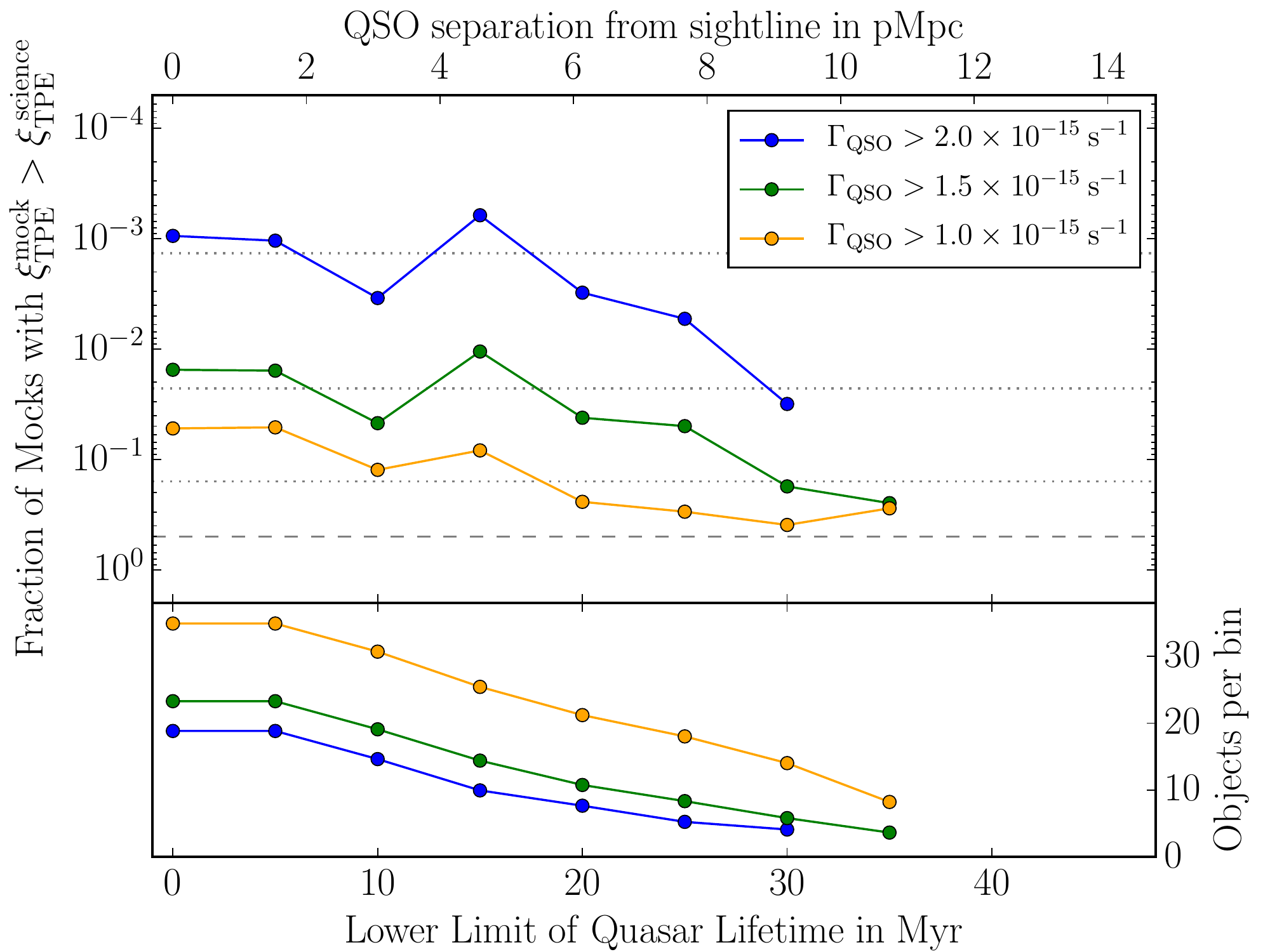}
 \caption{Significance estimate of the lower limit on the quasar lifetime from our Monte-Carlo analysis. The top panel gives the probability that the transmission enhancement $\xiTPE$ in our mock stacks exceeds the value that is measured in the science stack. This is given as function of the cuts on separation and $\Gqso$.  In absence of the transverse proximity effect ($\xiTPE=0$) we expect 50\% of the random stacks to exceed the measured transmission (dashed line). The probabilities corresponding to 1, 2 and $3\sigma$ detections are indicated by dotted lines. The bottom panel gives the average number of spectra contributing per bin in the stacks.}
 \label{Fig:Significance}
\end{figure}

We also test stacks with lower cuts on $\Gqso$ that yield consistent results, however at an overall lower confidence level. A detailed overview of the significance estimates for various QSO separations and cuts on $\Gqso$ is given in Figure~\ref{Fig:Significance}. For all parameter combinations we first create the science stack and then run a Monte Carlo analysis matched to that sample. Figure~\ref{Fig:Significance} clearly illustrates the presence of a transverse proximity effect out to substantial distances. The corresponding lifetime constraints depend on the minimum requirements (sample size, $\Gqso$ cut, $\sigma$ limit). We obtain a $2.6\sigma$ detection (which we call significant) for $\Gqso > 2 \times 10^{-15}\,\mathrm{s}^{-1}$ and $D_\mathrm{prop} > 25\Mly$ (i.e.\ $8\pMpc$), resulting in a lower limit on the quasar lifetime of $25\Myr$.

\section{Quantifying the Transverse Proximity Effect for Individual Quasars}
\label{Sec:Single_QSO_Stat}

\begin{figure*}
\includegraphics[width=\linewidth]{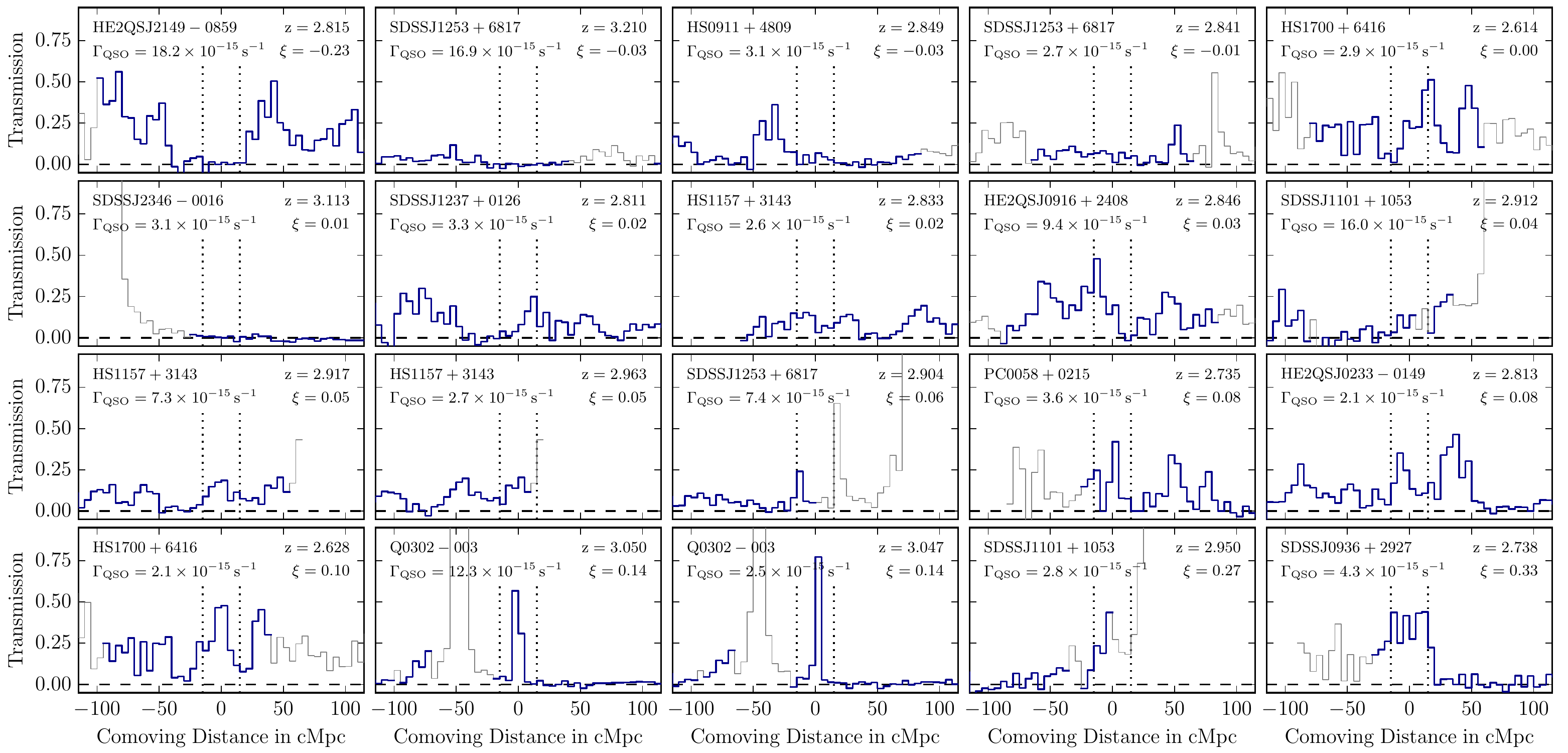}
\caption{\ion{He}{ii} spectra in the vicinity of all foreground quasars with
$\Gqso>2\times 10^{-15}\,\mathrm{s}^{-1}$. Masked and therefore ignored parts
of the spectra are shown in gray. Foreground quasar redshifts, estimated
$\Gqso$ values and \ion{He}{ii} sightlines are labeled.
The transmission enhancement $\xiTPE$ is measured by taking the difference
of the average transmission within the central $\pm15\cMpc$ (dotted lines)
and outside of this. Panels are ordered by $\xiTPE$ from top left to bottom right.
The transmission spike at $z = 3.05$ in the Q\,0302$-$003 sightline appears twice
since there are two associated foreground quasars. Due to their slightly different redshifts,
the centered and rebinned \ion{He}{ii} spectra look differently,
but their $\xiTPE$ values are identical.
}
\label{Fig:FUV_Tiles}
\end{figure*}

\begin{figure}
\includegraphics[width=\linewidth]{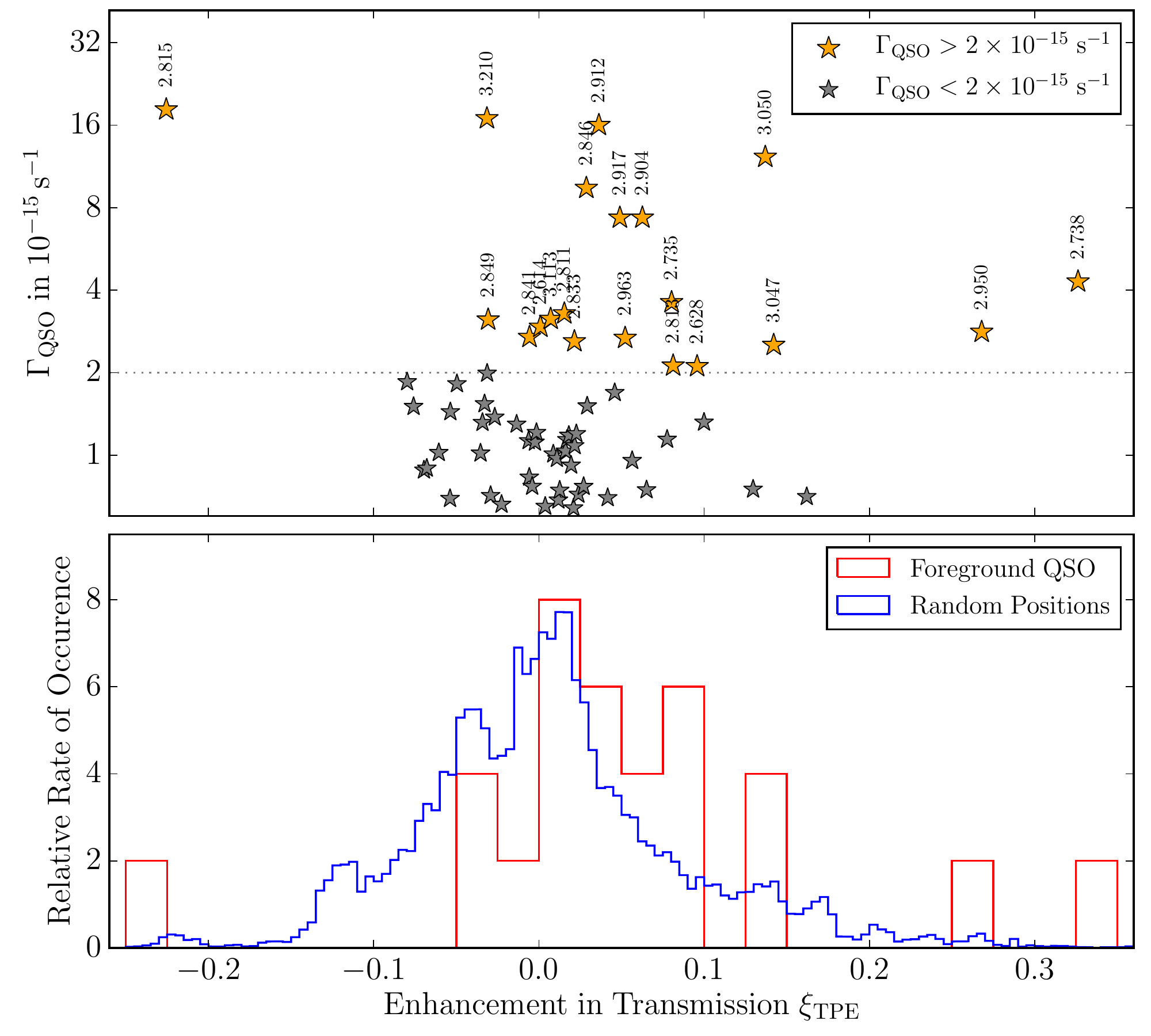}
\caption{\emph{Top panel: } 
Ionization rate $\Gqso$ compared to the transmission enhancement $\xiTPE$ for our sample of foreground quasars. Quasars with $\mathrm{\Gqso>2 \times 10^{-15}\,s^{-1}}$ are shown in orange, quasars with $\mathrm{0.6 \times 10^{-15}\,s^{-1} < \Gqso < 2 \times 10^{-15}\,s^{-1}}$ in gray. For identification, the quasars are labeled with their redshift.
\emph{Bottom panel: } Histogram of $\xiTPE$ for quasars with $\mathrm{\Gqso>2 \times 10^{-15}\,s^{-1}}$ (red) and for $10^6$ random positions in the \ion{He}{ii} spectra with a matched redshift distribution (blue). Both histograms are normalized to unity. The $\xiTPE$ distribution of foreground quasars with $\mathrm{\Gqso > 2 \times 10^{-15}\,s^{-1}}$ is inconsistent with being drawn from the random distribution ($p=4.7$\%).
}
\label{Fig:SingleObjectStatistic}
\end{figure}

\begin{figure}[b]
\includegraphics[width=\linewidth]{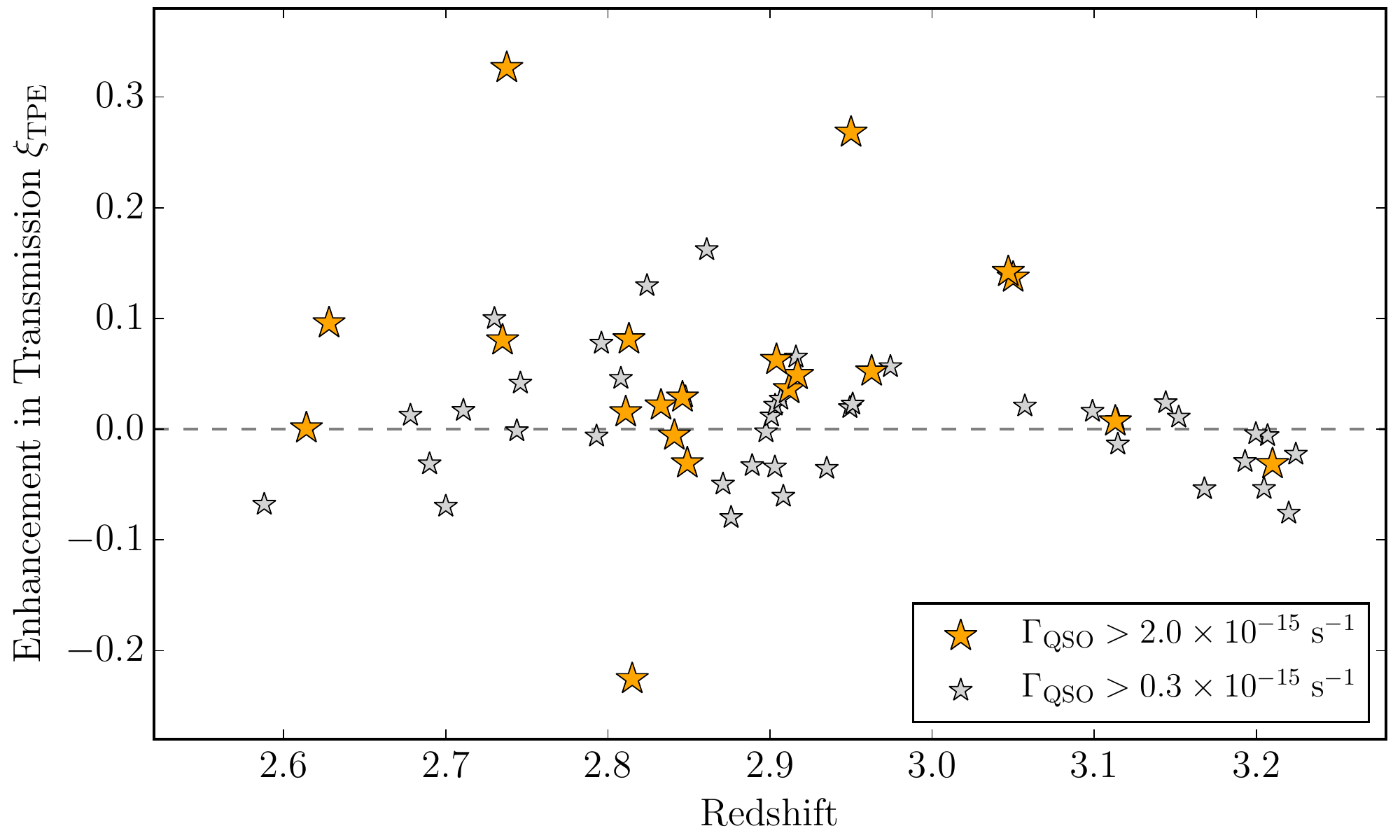}
\caption{
Distribution of the transmission enhancement with respect to redshift. 
No correlation with redshift is observed.
}
\label{Fig:Enhancement_Redshift}
\end{figure}

Given the apparently discrepant results that foreground quasars with high estimated \ion{He}{ii} photoionization rates often show no signs of a transverse proximity effect (\S\ref{Sec:Special_Objects}) whereas the effect is clearly detected in stacked spectra (\S\ref{Sec:StatisticalAnalysis}), we apply the $\xiTPE$ statistic (Equation~\ref{Eq:xi}) to individual sightlines and investigate the variance in $\xiTPE$ among our foreground quasar sample. This necessarily results in a noisy statistic since we do not average down fluctuations caused by e.g.\ the cosmic density structure or other sources of stochasticity in the transverse proximity effect, but it allows us to investigate the effect of individual foreground quasars.

Figure~\ref{Fig:FUV_Tiles} shows a zoom-in of the \ion{He}{ii} spectra in the vicinity of each foreground quasar contributing to the stack shown in Figure~\ref{Fig:FUV_Stack_20_00_00-99}. The regions are again chosen to extend for $\pm 120\cMpc$ and binned to $5\cMpc$, identical to the procedure adopted for stacking. The objects are sorted by transmission enhancement $\xiTPE$ in increasing order from top left to bottom right. As already mentioned in Section~\ref{Sec:Special_Objects}, there is a large diversity in the appearance of the \ion{He}{ii} spectra close to the foreground quasars. For some there is a substantial enhancement in transmission, e.g. for the quasars in the bottom row of Figure~\ref{Fig:FUV_Tiles} including the $z=3.05$ quasar near the Q\,0302$-$003 sightline with $\xiTPE=0.14$. For others we see no significant transmission enhancement ($\xiTPE \approx 0$). For the two highest redshift foreground quasars at $z > 3.1$, the \ion{He}{ii} spectra show no transmission at all ($\xiTPE\approx0$) and for one $z=2.815$ quasar close to the HE2QS\,J2149$-$0859 sightline we even find substantially lower transmission at the quasar location than in the region around it ($\xiTPE = -0.23$, upper left corner of Figure~\ref{Fig:FUV_Tiles}).

Although the transmission spike in the Q\,0302$-$003 sightline is the most
prominent feature in our sample, it does not have the highest $\xiTPE$ (Equation~\ref{Eq:xi}).
The reason is that, although the transmission spike is very strong, it is also rather narrow ($\mathrm{\approx 450\,km\,s^{-1}}$ or $5.7\cMpc$)%
\footnote{The Q\,0302$-$003 sightline is the only one for which we use the high-resolution COS G130M spectrum and therefore clearly resolve the transmission spike. However, since the average transmission within $\pm15\cMpc$ used by our $\xiTPE$ statistic is not affected by the resolution we do not smooth the spectrum to G140L resolution.},
far narrower than the $\pm15\cMpc$ window we average over, chosen to be slightly larger than the typical redshift uncertainty of $\mathrm{1000\,km\,s^{-1}}$. The average transmission within this window is therefore not extreme and for two other quasars we find substantially higher enhancements. Near the foreground quasar at $z=2.738$ the SDSS\,J0936$+$2927 sightline exhibits transmission only at the 35\% level but extending over the full width of the window, resulting in the highest measured transmission enhancement in our sample ($\xiTPE = 0.33$, bottom right panel in Figure~\ref{Fig:FUV_Tiles}). However, note that at $z=2.7$ it is unclear how much of this large transmission can be attributed to density
fluctuations in the post-reionization IGM. Indeed, we adopt the stacking technique to average down these large fluctuations.

In Figure~\ref{Fig:SingleObjectStatistic} we show the distribution of $\xiTPE$ and plot it against $\Gqso$. First, we find a large spread in $\xiTPE$ ($-0.23\le\xiTPE\le 0.33$) for quasars with $\mathrm{\Gqso > 2 \times 10^{-15}\,s^{-1}}$. Second, there is no obvious trend with $\Gqso$ or redshift (Figure~\ref{Fig:Enhancement_Redshift}). The $z=2.815$ quasar near the HE2QS\,J2149$-$0859 sightline has the lowest value $\xiTPE = -0.23$, but the highest photoionization rate in our sample ($\mathrm{\Gqso = 18.2 \times 10^{-15}\,s^{-1}}$, \S~\ref{Sec:Highest_Gamma_Quasars}). In contrast, the quasar at $z = 3.05$ near the Q\,0302$-$003 sightline has a photoionization rate 30\% lower ($\mathrm{\Gqso = 12.3 \times 10^{-15}\,s^{-1}}$) and shows a strong enhancement in transmission ($\xiTPE=0.14$). The $\xiTPE$ distribution has a mean value of $\overline{\xiTPE} = 0.056$ which is consistent with $\xi_\mathrm{TPE}^\mathrm{stack}=0.058$ in our stack (Figure~\ref{Fig:FUV_Stack_20_00_00-99}). For comparison we also show foreground quasars with $0.6 \times \mathrm{10^{-15}\,s^{-1}< \Gqso < 2 \times 10^{-15}\,s^{-1}}$. These have $\overline{\xiTPE} = 0.004$, indicating an insignificant transverse proximity effect, as expected given their low photoionization rates.

The bottom panel of Figure~\ref{Fig:SingleObjectStatistic} shows a histogram
of $\xiTPE$ for the sample of foreground quasars with $\mathrm{\Gqso > 2 \times 10^{-15}\,s^{-1}}$ (red).
The blue histogram is based on our $\xiTPE$ statistic applied to $10^6$ random positions along the \ion{He}{ii} sightlines with a redshift distribution matched to the one of the foreground quasars.
We use a Kolmogorov-Smirnov test to investigate if the foreground quasar distribution is consistent with being drawn from the distribution of random positions.
For the foreground quasars with $\mathrm{0.6 \times 10^{-15} < \Gqso < 2 \times 10^{-15}\,s^{-1}}$ this is the case ($p=0.413$), however the sample with $\mathrm{\Gqso > 2 \times 10^{-15}\,s^{-1}}$ is inconsistent with the random distribution ($p=0.047$).
This is additional proof for the presence of the \ion{He}{ii} transverse proximity effect in our sample of foreground quasars, based on the full distribution of $\xiTPE$ rather than just its mean value as in the stack.

\section{Discussion}
\label{Sec:Discussion}

\subsection{Interpretation and Limitations of our Lifetime Constraint}
In \S~\ref{Sec:StatisticalAnalysis} we showed statistical evidence for a transverse proximity effect in our foreground quasar sample and derived a lower limit on the episodic quasar lifetime. There are several reasons why the intrinsic episodic quasar lifetime might in fact be larger than our limit.
When creating stacks with a lower cut on the separation from the background sightline, we run out of bright foreground quasars with $\Gqso > 2 \times 10^{-15}\,\mathrm{s}^{-1}$ at $D_\mathrm{prop}>30\Mly$. Lower photoionization rates are comparable to the UV background and apparently not sufficient to cause a detectable effect (Figures~\ref{Fig:Gamma_Dependence} and \ref{Fig:Significance}). With the given foreground quasar sample our geometrical method can therefore not probe lifetimes longer than this.
In addition, it might be that IGM absorption limits the extent of the proximity zone and not the finite lifetime of the quasars \citep{Khrykin2016}.
We estimate $\Gqso$ ignoring IGM absorption (assuming $\lambda_\mathrm{mfp} = \infty $ in Equation~\ref{Eq:Q_QSO}),
however a mean free path of $\lambda_\mathrm{mfp} \approx 50\cMpc$ -- consistent with \citet{Davies2014} -- would cause a reduction of $\Gqso$ by 50\% at $10\pMpc$ or $32\Mly$ separation. IGM absorption therefore limits the impact of distant quasars. Quasars at $D_\mathrm{prop} = 10\pMpc$ would have to be twice as luminous to still exceed our threshold of $\Gqso > 2 \times 10^{-15}\,\mathrm{s}^{-1}$ at the background sightline%
\footnote{This threshold was derived in \S~\ref{Sec:Gamma_Dependence} from all available quasars and is dominated by small separations.}. 
Such objects do no exist within our sample.
On the other hand, our measurement of the transverse proximity effect constrains the mean free path to \ion{He}{ii}-ionizing photons. 
From our detection of the effect at $D_\mathrm{prop} = 25\Mly$ we conclude that the mean free path cannot be much shorter than this.

Under the assumption of a lightbulb model for the quasar lightcurve,  our lifetime measurement is sensitive to the age of the quasars $\tTO$. This is always shorter than the episodic lifetime $\tQ$ and the average age of a population with a flat age distribution is half its lifetime, i.e.\ $\overline{\tTO} =\tQ/2$. As such, we place a lower limit on $\tTO > 25\Myr$, which could in principle indicate $\tQ>50\Myr$. 
However, due to our limited sample size ($N\approx5$) we can not be sure that we have converged to this average and conservatively only constrain $t_Q > 25\Myr$.

Furthermore, we have to consider the time the IGM requires to adjust to a new photoionization equilibrium.
The equilibration timescale is the inverse of the ionization rate and is approximately $10^{7}\,\mathrm{yr}$ \citep{Khrykin2016}%
\footnote{Recombination can be neglected since the recombination timescale for helium is comparable to the Hubble time}.
This has to be added to the light crossing time to get a more precise estimate on the lifetime. However, the local equilibration time depends on the amplitude of the UV field which is heavily influenced by the quasar within its proximity zone and we consider the exact effects of this too uncertain to include it in our lifetime estimate.
Additionally, equilibration effects render our method insensitive to quasar variability on scales shorter than the equilibration timescale.
However, as shown in Section~\ref{Sec:Gamma_Dependence} $\Gqso$ is, at least when applied to a
population, a reasonable proxy for the impact on the background sightline. The quasar luminosity is therefore not allowed to differ tremendously over the timescales probed by our analysis and the actual quasar lightcurve -- even if it deviates from the lightbulb model -- has to be consistent with sustained activity over $25\Myr$. 

\subsection{Absence of Transmission Spikes for Large Photoionization Rate Enhancements}
\label{Sec:NoSpikes}

Our statistical sample shows that the presence of a foreground quasar close to the background sightline does not necessarily imply a prominent \ion{He}{ii} transmission spike, even when the expected photoionization rate should be greatly enhanced. Instead, out of the four foreground quasars with the highest $\Gqso$, only the one near the Q\,0302$-$003 sightline is associated with an obvious transmission spike (Figure~\ref{Fig:Zooms}).
So far we can only speculate why the other three sightlines do not show similar transmission spikes.

Possibly, these quasars might be active for too short a period ($\tTO \lesssim 10\Myr$) to allow their radiation to reach the background sightline. However, these quasars are not substantially farther away from their background sightline or even closer than the quasar near Q\,0302$-$003 ($14.7$, $13.5$ and $7.3\Mly$ compared to $9.7\Mly$) and all have a higher $\Gqso$. From our stacking analysis we find evidence for continued quasar activity over $25\Myr$ (Figure~\ref{Fig:Significance}). If this is representative of the quasar population, it would be surprising if the three quasars with the highest $\Gqso$ are all very young.

Another possible explanation could be that due to variations in the quasar SED, in particular the poorly constrained part between $1$ and $4\,\mathrm{Ry}$, or because of substantial fluctuations in $\lambda_\mathrm{mfp}$ the ionizing rate at the background sightline is actually lower than our estimate. 
While this might be the case for individual quasars, our statistical detection of the transverse proximity effect for foreground quasars with
$\Gqso > 2 \times 10^{-15}\,\mathrm{s}^{-1}$ (Figure~\ref{Fig:Gamma_Dependence}) argues against a
a scenario where our estimates of $\Gqso$ are systematically off.  
Any fluctuation in the SED or $\lambda_\mathrm{mfp}$ would have to be very substantial to bring the $8\times$ higher estimated $\Gqso$ of our three strongest foreground quasars to a level for which we expect to see no effect.

Quasar obscuration coudl certainly be important, which we have so far ignored when calculating $\Gqso$ (see \S~\ref{Sec:Gamma_QSO}).  The effect of anisotropic emission on the appearance of a particular \ion{He}{ii} spectrum clearly needs further exploration.  Since quasar orientation is random and only constrained to be unobscured towards Earth, one can in any case only expect a probabilistic answer \citep[see e.g.][]{Furlanetto2011}. Determining how realistic it might be that three out of four foreground quasars do not illuminate the background sightline is an important question for future work.

Despite the three discussed foreground quasars having no visible impact on the background sightlines, the quasar at $z=3.05$ near Q\,0302$-$003 might show a particularly strong signature. Comparison of the \ion{H}{i} and \ion{He}{ii} spectrum of the Q\,0302$-$003 sightline indicates a substantial modulation of the \ion{He}{ii} transmission by the IGM density field, with the \ion{He}{ii} transmission spike being located in a region of high \ion{H}{i} transmission \citep{Worseck2006, Syphers2014}.
One could speculate that a favorable association of one (or even two) foreground quasars with a low-density region in the IGM allows for an unusually strong \ion{He}{ii} transmission spike. Analyzing high-resolution \ion{H}{i} spectra for the other \ion{He}{ii} sightlines might shed light on this point.

We conclude that a high photoionization rate ($\Gqso > 2 \times 10^{-15}\,\mathrm{s}^{-1}$, Figure~\ref{Fig:Gamma_Dependence}) is a necessary condition to cause a significant transverse proximity effect, but is itself not sufficient.
Apparently, there are other factors that govern the presence of excess transmission of which we have discussed a few.
Additional data and better statistics will certainly help to constrain this, but we inevitably require further modeling with a realistic implementation of lifetime, obscuration and $\lambda_\mathrm{mfp}$ effects that accounts for the stochasticity in these quantities.

\subsection{Toy Model for the Ionization Rate along the Sightline}
\label{Sec:GammaModel}

\begin{figure}
 \includegraphics[width=\linewidth]{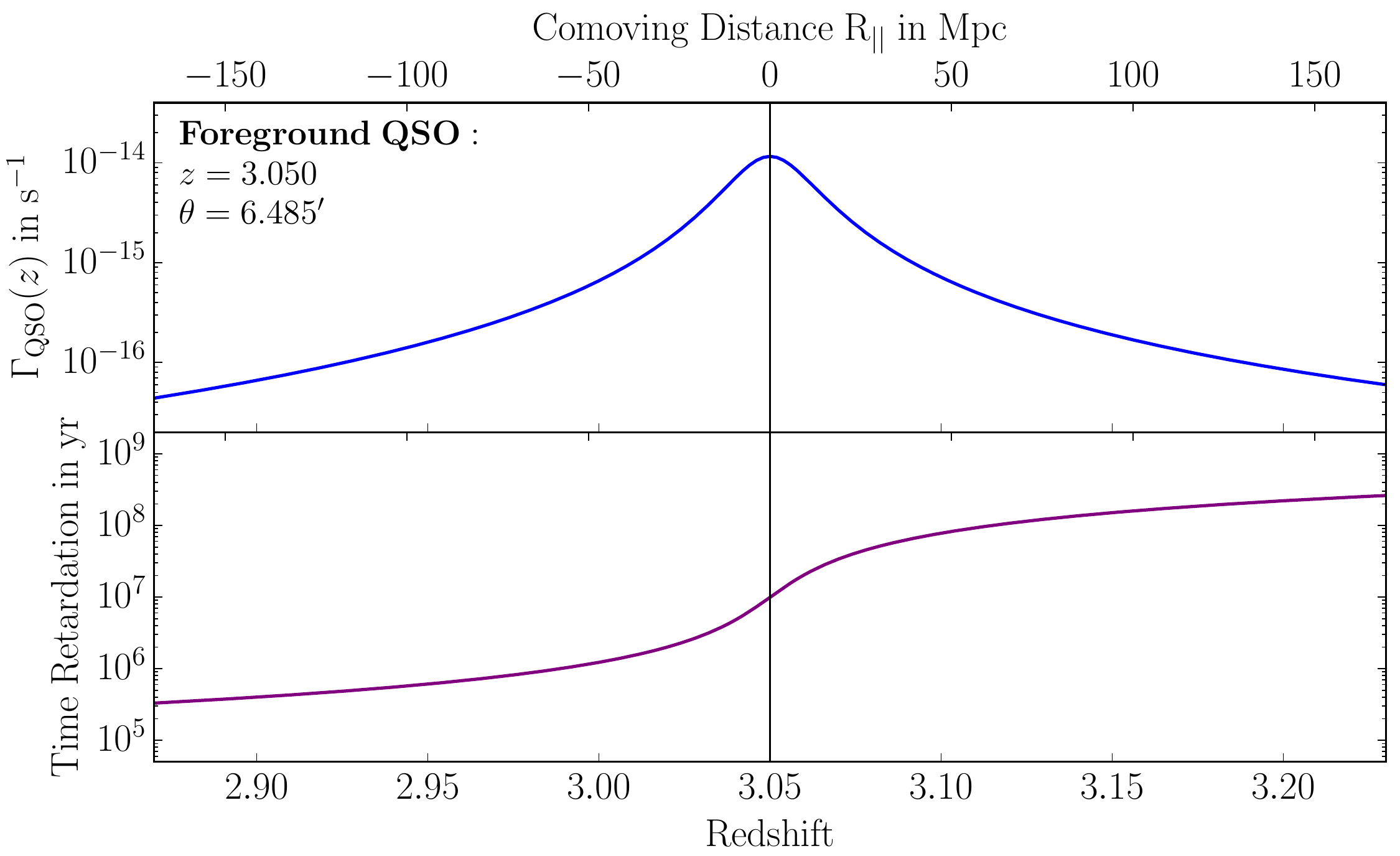}
 \caption{Photoionization rate $\Gqso(z)$ and time retardation $\Delta t(z)$ along the Q\,0302$-$003 sightline for the brighter of the two foreground quasars at $z=3.05$. Note the monotonic increase of $\Delta t(z)$ toward higher redshifts.}
 \label{Fig:TimeRetardation}
\end{figure}

\begin{figure*}
 \includegraphics[width=\linewidth]{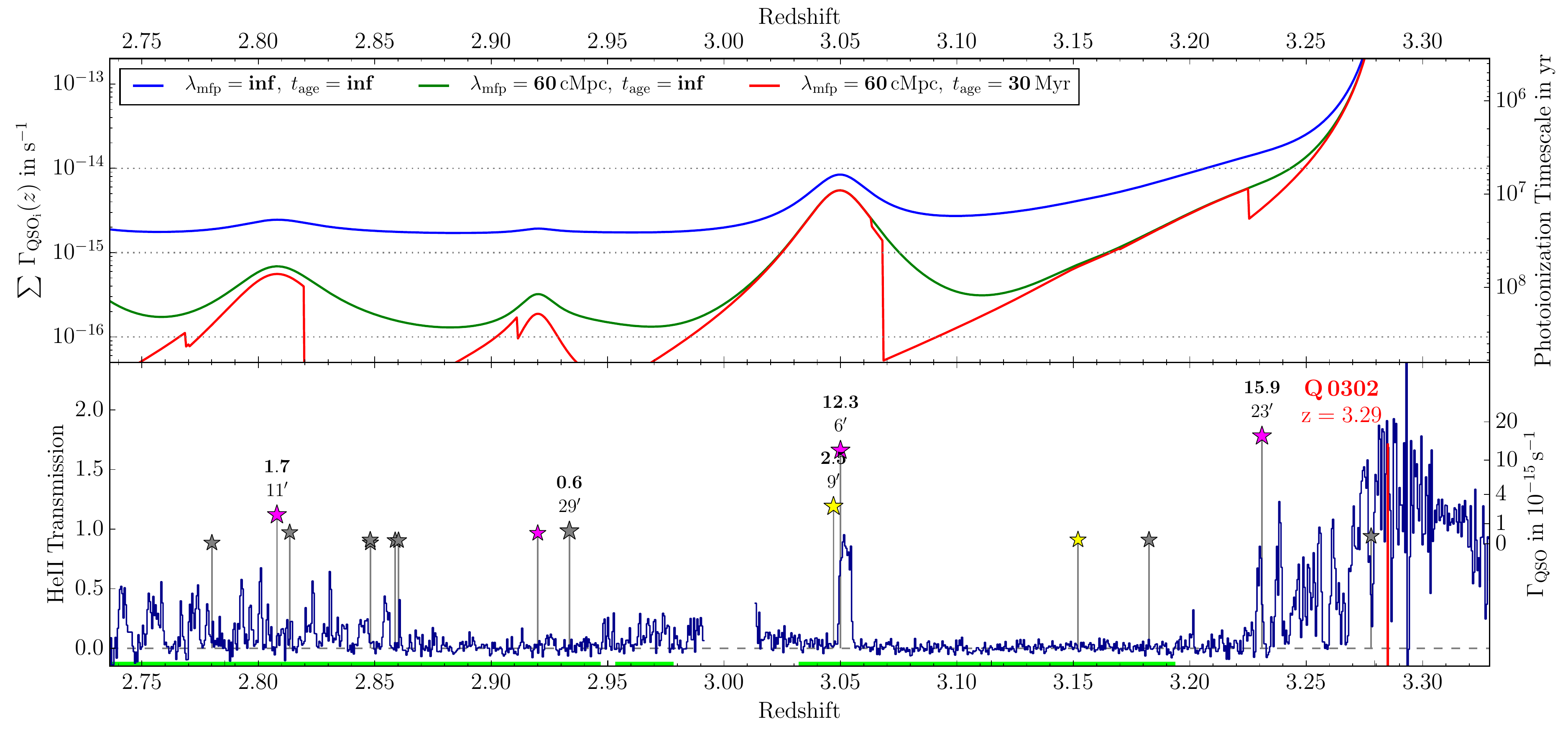}
\caption{Models for the varying \ion{He}{ii}-ionizing radiation field along the \ion{He}{ii} sightline of Q\,0302$-$003 (top) and its \ion{He}{ii} transmission spectrum indicating the locations of the quasars (bottom).
We synthesize the photoionization rate $\Gsl(z)$ based on the background quasar and the population of known foreground quasars by summing up their individual contributions. 
For computation of the $\Gsl(z)$ model shown in blue we assume isotropic emission, infinite quasar lifetime and a transparent IGM. We show the effects of IGM absorption with a mean free path of $\lambda_\mathrm{mfp} = 60\,\cMpc$ (green curve) and in addition a finite quasar age of $\tTO=30\,\mathrm{Myr}$ (red curve). 
Assuming a finite age causes a strong asymmetry in the radiation field around individual foreground quasars since these only illuminate regions in the background sightline for which the time retardation is smaller than the quasar age ($\Delta t(z) < \tTO$).
}
 \label{Fig:GammaModel}
\end{figure*}

Here, we present a toy model of the UV radiation field along a background sightline to illustrate effects that are relevant to the \ion{He}{ii} transverse proximity effect. A complete model requires 3D numerical radiative transfer calculations and is beyond the scope of this paper. For now however, we want to build some intuition for the processes involved and highlight key aspects that would be part of a future modeling attempt.

The light crossing time between a foreground quasar and the background sightline is the governing effect we use to constrain the quasar episodic lifetime. While the light from the foreground quasar travels to Earth on the direct path, the distance to any point on the background sightline and from there to the observer on Earth is longer. This difference in pathlength and therefore light travel time is the time retardation $\Delta t(z)$.
In the bottom panel of Figure~\ref{Fig:TimeRetardation} we show as an example the time retardation for any point along the sightline of Q\,0302$-$003 with respect to the foreground quasar at $z=3.05$ at a separation of $6.5\arcmin$ (for formal derivations and similar models see e.g.
\citealt{Liske2001, Smette2002, Adelberger2005, Worseck2008}).
Evaluated at the redshift of the foreground quasar, the time retardation equals the transverse light crossing time used so far in our analysis.
Toward lower redshifts $\Delta t(z)$ decreases since the additional pathlength of the photons is shorter, whereas for gas behind the foreground quasar (at higher redshifts) the time retardation rapidly increases.  
Taking a quasar that shines for a given time $\tTO$, gas at lower redshift, when probed by the photons from the background quasar, has already been exposed to the quasar's radiation for longer times than gas at higher redshifts.
For locations with $\Delta t(z) > \tTO$ the gas has not yet been exposed to the quasar radiation at all. 
The photoionization rate along the background sightline $\Gqso(z)$ (top panel in Figure~\ref{Fig:TimeRetardation}) can be calculated as described in \S~\ref{Sec:Gamma_QSO}, but using the luminosity distance to a given point on the background sightline instead of the transverse distance in Equation~\ref{Eq:GeometricDilution}.

Given that we can model the time retardation and photoionization rate caused by a single quasar as described above, we can synthesize the \ion{He}{ii}-ionizing radiation field along a sightline by summing the contributions of all foreground quasars as
\begin{equation}
\Gsl(z) = \sum_i \: \Gamma^\mathrm{HeII}_{\mathrm{QSO}_i}(z) \:.
\end{equation}
Figure~\ref{Fig:GammaModel} shows this an example of this model for the sightline toward Q\,0302$-$003.
The bottom panel displays again the \ion{He}{ii} transmission spectrum and the location of the foreground quasars, while the top panel shows several models of $\Gsl(z)$.
For the model shown in blue we assume isotropic emission and infinite quasar lifetime for all foreground quasars and no IGM absorption ($\lambda_\mathrm{mfp}=\infty$).
The right y-axis of Figure~\ref{Fig:GammaModel} indicates the \ion{He}{ii} equilibration timescale $\tHeEQ$
which is spatially varying, as it is the inverse of $\Gsl(z)$. Note that the equilibration timescale $\tHeEQ\sim 10^{7}{\rm yr}$ \citep{Khrykin2016} is comparable to the transverse light crossing time, and hence our inferred lower limit on the quasar episodic lifetime from \S~\ref{Sec:Lifetime}. If quasars shine for a period comparable to our limits and the photoionization rate at a given point along the sightline is dominated by a few close foreground quasars, this suggests that \ion{He}{ii} at $z\sim3$ might not yet be in photoionization equilibrium. 

Assuming an infinite mean free path and adding up quasar contributions out to arbitrary large separations from the background sightline leads to very high $\Gsl(z)$.
Relying only on the population of discovered foreground quasars we already end up with photoionization rates in excess of the average $\Gamma^\mathrm{HeII}_\mathrm{UVB}$ predicted by UV background models \citep[e.g.][$\Gamma^\mathrm{HeII}_\mathrm{UVB}(z=3) \approx 10^{-15}\,\mathrm{s}^{-1}$]{Faucher-Giguere2009, HaardtMadau2012} and inconsistent with highly saturated \ion{He}{ii} absorption \citep{Khrykin2016}.

We therefore show in Fig~\ref{Fig:GammaModel} (green curve) the effect of a reduced mean free path. We chose $\lambda_\mathrm{mfp}=60\cMpc$ which is broadly consistent with the values from \citet{Davies2014}. However, we stress
that by creating proximity zones, the quasars heavily modify the mean free path in their vicinity \citep{Khrykin2016}.
As expected \citep[see e.g.][]{McQuinn2014}, a short mean free path not only reduces the average $\Gsl(z)$
but also enhances the amplitude of the spatial fluctuations due to individual quasars.

In a third synthesis model of the UV radiation field we include the effect of a finite quasar lifetime. We assume all foreground quasars to have a fixed age $\tTO = 30\Myr$. This is clearly a simplification since
even if all quasars would have exactly the same episodic lifetime $\tQ$, they would turn on at a random point in time and $\tTO$ would be drawn from a flat distribution between zero and $\tQ$. 
Nevertheless, limiting the quasar age decreases the photoionization rate behind bright foreground quasars since the quasars do not shine long enough for the photons to reach these regions.
The effect on the background sightline at redshifts lower than the foreground quasar redshift
is much smaller since the time retardation for these points is smaller (Figure~\ref{Fig:TimeRetardation}).
The ionization fronts of the individual foreground quasars at redshifts where $\Delta t(z) = \tTO$ are also apparent in Figure~\ref{Fig:GammaModel}.
When increasing $\tTO$, these ionization fronts shift to higher redshifts and an increasing fraction of the background sightline can be reached by the foreground quasar radiation. In reality, radiation transfer effects should substantially smooth the rapid jump of $\Gsl(z)$ at the location of the ionization front \citep[e.g.][]{Davies2017}.

An additional important factor as described in \S~\ref{Sec:Gamma_QSO} and \S~\ref{Sec:NoSpikes} is quasar obscuration.
Modeling the anisotropic emission of quasars detected as Type\,Is from Earth is straightforward -- it would simply lead to some parts of the background sightline being not illuminated \citep[e.g.][]{Furlanetto2011}.
However, if quasar emission is significantly obscured in some directions, there should also be quasars illuminating the background sightline but appear as Type\,IIs from our vantage point on Earth. These obscured AGN are certainly not in our foreground quasar sample, and would be extremely difficult to detect. 
Finally, there may be light-echoes in the \ion{He}{ii} spectra caused by quasars that have already turned off \citep[e.g.][]{Visbal2008}. Similar to obscured quasars, these objects evade detection and can not be included in the model individually, but have to be treated in a statistical sense.

These simple models illustrate that the quasar lifetime, quasar obscuration,  and $\lambda_\mathrm{mfp}$ strongly influence the \ion{He}{ii} -ionizing radiation field. Exploring how this translates into an observable \ion{He}{ii} transverse proximity effect is a challenging task, in particular when including additional complications like variable quasar orientation and extinct quasars, non-equilibrium photoionization, and a realistic IGM density structure. Our toy models, however, show the wealth of information contained in the transverse proximity effect that is in principle accessible to observations.

\section{Summary and Conclusions}
\label{Sec:Summary}

The \ion{He}{ii} transverse proximity effect provides unique insights into the relationship between \ion{He}{ii} reionization and the quasars that power it. The effect also allows one to probe the geometry and timescales of quasar emission. 
We therefore conducted a dedicated ground-based optical imaging and spectroscopic survey for quasars in the foreground of 22 background quasars with science-grade \textit{HST} FUV \ion{He}{ii} $\lya$ absorption spectra. 
Our two-tiered survey strategy, composed of a deep survey with 8\,m~class telescopes (LBT/LBC and VLT/VIMOS, $r \lesssim 24\mag$,
$\Delta\theta\lesssim10\arcmin$) and a wide survey with 4\,m~class telescopes (ESO NTT and CAHA~3.5\,m, $r \lesssim 21\mag$, $\Delta\theta\lesssim90\arcmin$)
resulted in 131 quasars, 27 of which have redshifts probed by \ion{He}{ii} absorption along the background sightline (\S~\ref{Sec:Survey}).
Adding known quasars (mainly from SDSS and BOSS), we arrive at a total of 66 usable foreground quasars. We searched for the \ion{He}{ii} transverse proximity effect as enhanced \ion{He}{ii} transmission near individual foreground quasars (\S~\ref{Sec:Highest_Gamma_Quasars} and \ref{Sec:Single_QSO_Stat}) and also statistically by stacking the data for our sample (\S~\ref{Sec:StatisticalAnalysis}).

Previous studies claimed a dramatic association of one foreground quasar with a strong \ion{He}{ii} transmission spike in the Q\,0302$-$003 sightline \citep{Heap2000, Jakobsen2003}.
Our substantially larger sample contains three new foreground quasars with even higher estimated photoionization rates at the background sightline that exceed the expected \ion{He}{ii}-ionizing background by an order of magnitude. However, none of these new sightlines exhibit a noticeable increase in \ion{He}{ii} transmission (Figure~\ref{Fig:Zooms}), suggesting that such associations are in fact rare.

Despite the large stochasticity, we find statistical evidence for the \ion{He}{ii} transverse proximity effect by stacking the \ion{He}{ii} spectra on the positions of the foreground quasars. 
The average transmission profile along the background sightlines shows increased \ion{He}{ii} transmission in the vicinity ($|R_\parallel|<15\cMpc$) of the foreground quasar positions when compared to the average transmission observed further away (Figure~\ref{Fig:FUV_Stack_20_00_00-99}).
Using a Monte Carlo method, we estimate the probability for this excess transmission to occur by chance to $0.1$\%, corresponding to a significance of $3.1\sigma$.

We show for the first time that the strength of this transverse proximity effect has the expected dependence on the photoionization rate of the foreground quasars (Figure~\ref{Fig:Gamma_Dependence}). We see a local enhancement in the \ion{He}{ii} transmission when including foreground quasars with $\Gqso > 2 \times 10^{-15}\,\mathrm{s}^{-1}$. For lower $\Gqso$, comparable to UV background estimates, no enhancement is observed.

We use the transverse light crossing time to derive a purely geometric lower limit on the quasar episodic lifetime by imposing cuts on the foreground quasar separation from the background sightline (Figure~\ref{Fig:Distance_Dependence}).
When restricting to quasars with $D_\mathrm{prop} \geq 7.7\pMpc$ the transverse proximity effect persists in our stacked spectra with a $2.6\sigma$ significance, allowing us to derive a robust lower limit on the quasar episodic lifetime of $25\Myr$.
Our analysis, based on a statistical sample and a model-independent method, puts stronger constraints on the episodic quasar lifetime than previous studies of single objects or small samples \citep[e.g][]{Jakobsen2003, Goncalves2008, Furlanetto2011, Trainor2013, Borisova2015}, and contrasts with studies finding short lifetimes \citep[e.g.][]{Kirkman2008, Schawinski2015}.
At the same time this constrains the mean free path to $\gtrsim 30\cMpc$ at $z\approx3$, consistent with current estimates \citep{McQuinn2014, Davies2014}.

The \ion{He}{ii} transverse proximity effect can reveal extensive information about the quasar opening angle, the quasar lifetime and the mean free path to \ion{He}{ii}-ionizing photons in the IGM (\S~\ref{Sec:GammaModel}). However, due to the large sightline-to-sightline variance and the weak statistical signal, there is no simple interpretation of the data beyond the lifetime constraint. Deriving more stringent constraints clearly requires extensive modeling and additional data.

It is therefore crucial to observe additional \ion{He}{ii} sightlines to reduce the sample variance (\S~\ref{Sec:Single_QSO_Stat}). 
However, due to the rarity of \ion{He}{ii}-transparent quasars and the aging of \textit{HST}/COS, a massive expansion ($> 2\times$) of the \ion{He}{ii} sample will probably only be possible with the next generation of FUV telescopes. 
Since these will only become available decades from now it is important to exploit the current \textit{HST}/COS capabilities to the limit and observe the available targets at $z\approx3$.

Furthermore one should expand the deep foreground quasar survey to \ion{He}{ii} sightlines in the northern hemisphere (see Table~\ref{Tab:Table1}), several of which show significant transmission spikes (see Figure~\ref{Fig:HeSpectra1}).  
The redshifts of the foreground quasars were inferred from UV emission lines, subject to systematic uncertainties of up to $1000\,\mathrm{km\:s^{-1}}$, corresponding to $\approx13\cMpc$. This is comparable to the width of the detected transverse proximity signal and nearly $3\times$ wider than the strong transmission spike in the Q\,0302$-$003 sightline. Improved redshifts from \ion{Mg}{ii} or [\ion{O}{iii}] ($\sigma_z =100 - 300\,\mathrm{km\:s^{-1}}$) would increase our sensitivity and would allow us to better constrain the shape of the transverse proximity profile, which could bear additional information about lifetime and anisotropic emission.

In addition, it would  be very informative, but probably infeasible, to also chart the population of Type\,II foreground AGN. An X-Ray survey comparable in sensitivity and area to our optical surveys would require several $100\,\mathrm{ks}$ exposure time per \ion{He}{ii} sightline, so covering all sightlines would exceed the allocations for the largest X-ray surveys (e.g. \textit{Chandra} \textit{COSMOS}-legacy survey \citealt{Civano2016}). Alternatively, covering the large required area ($\gtrsim 20'\times20'$) with spectroscopic IFU observations 
seems extremely expensive as well. 

In the near future, substantial progress on the physics of the \ion{He}{ii} transverse proximity effect might stem from dedicated modeling of the effect. The toy model we used in \S~\ref{Sec:GammaModel} to highlight a few effects as a showcase example is a good starting point for
more  comprehensive modeling of the \ion{He}{ii} transverse proximity effect, which would take into account the time retardation along the
sightline, non-equilibrium ionization effects, and quasar obscuration. A model tailored to our observations could for instance clarify if the absence of strong transmission spikes in the \ion{He}{ii} spectra close to bright foreground quasars is actually consistent with our current assumptions about lifetime and quasar obscuration or requires additional effects. This might also lead to tighter constraints on the quasar lifetime or the mean free path to \ion{He}{ii} ionizing photons. Evidently, the transverse proximity effect bears a wealth of information. Now that a statistical sample is available, it requires improved theoretical understanding to better interpret the signal we have observed.

\section*{Acknowledgments}

We thank Robert Simcoe for kindly supplying Magellan/Megacam imaging for the SDSS\,J1237$+$0126 field.
We would like to thank the members of the ENIGMA\footnote{\url{http://enigma.physics.ucsb.edu/}} group at the Max Planck Institute for Astronomy (MPIA) for useful discussions and support. GW has been supported by the Deutsches Zentrum f\"ur Luft- und Raumfahrt (DLR) under contracts 50\,OR\,1317 and 50\,OR\,1512.

Based on observations made with the NASA/ESA Hubble Space Telescope, obtained at the Space Telescope Science Institute, which is operated by the Association of Universities for Research in Astronomy, Inc., under NASA contract NAS 5-26555. These observations are associated with programs 11528, 11742, 12033, 12178, 12249, 13013.

The LBT is an international collaboration among institutions in the United States, Italy and Germany. LBT Corporation partners are: The University of Arizona on behalf of the Arizona Board of Regents; Istituto Nazionale di Astrofisica, Italy; LBT Beteiligungsgesellschaft, Germany, representing the Max-Planck Society, The Leibniz Institute for Astrophysics Potsdam, and Heidelberg University; The Ohio State University, and The Research Corporation, on behalf of The University of Notre Dame, University of Minnesota and University of Virginia.

This paper includes data gathered with the 6.5 meter Magellan Telescopes located at Las Campanas Observatory, Chile.

Based in part on observations at Cerro Tololo Inter-American Observatory and Kitt Peak National Observatory, National Optical Astronomy Observatory, which are operated by the Association of Universities for Research in Astronomy (AURA) under a cooperative agreement with the National Science Foundation. The authors are honored to be permitted to conduct astronomical research on Iolkam Du'ag (Kitt Peak), a mountain with particular significance to the Tohono O'odham.

Based on observations collected at the European Organisation for Astronomical Research in the Southern Hemisphere
under ESO programmes 088.A-0835(B), 090.A-0664(B), 094.A-0500(A), 094.A-0782(A).

Based on observations collected at the Centro Astronómico Hispano Alemán (CAHA) at Calar Alto, operated jointly by the Max-Planck Institut für Astronomie and the Instituto de Astrofísica de Andalucía (CSIC).

Some of the data presented herein were obtained at the W.M. Keck Observatory, which is operated as a scientific partnership among the California Institute of Technology, the University of California and the National Aeronautics and Space Administration. The Observatory was made possible by the generous financial support of the W.M. Keck Foundation. 
The authors wish to recognize and acknowledge the very significant cultural role and reverence that the summit of Mauna Kea has always had within the indigenous Hawaiian community.  We are most fortunate to have the opportunity to conduct observations from this mountain.

Funding for SDSS-III has been provided by the Alfred P. Sloan Foundation, the Participating Institutions, the National Science Foundation, and the U.S. Department of Energy Office of Science. The SDSS-III web site is \url{http://www.sdss3.org/}.
SDSS-III is managed by the Astrophysical Research Consortium for the Participating Institutions of the SDSS-III Collaboration including the University of Arizona, the Brazilian Participation Group, Brookhaven National Laboratory, Carnegie Mellon University, University of Florida, the French Participation Group, the German Participation Group, Harvard University, the Instituto de Astrofisica de Canarias, the Michigan State/Notre Dame/JINA Participation Group, Johns Hopkins University, Lawrence Berkeley National Laboratory, Max Planck Institute for Astrophysics, Max Planck Institute for Extraterrestrial Physics, New Mexico State University, New York University, Ohio State University, Pennsylvania State University, University of Portsmouth, Princeton University, the Spanish Participation Group, University of Tokyo, University of Utah, Vanderbilt University, University of Virginia, University of Washington, and Yale University.

\bibliographystyle{APJ}
\bibliography{Literature}

\LongTables
\begin{deluxetable*}{lrrrrrrrrc}
\tablecolumns{9}
\tablewidth{0pc}
\tablecaption{
Quasars discovered within our survey
}
\tablehead{
\colhead{\ion{He}{ii} QSO}   	& \colhead{$RA$ (2000)}		& \colhead{$Dec$ (2000)}		& \colhead{$z$}	& \colhead{$r$}	& \colhead{$M_{1450}$}	& \colhead{$\Delta\theta$}	& \colhead{$D_\mathrm{prop}$}		& \colhead{$\Gqso$}	& \colhead{Instrument} \\
\colhead{}			& \colhead{degree}	& \colhead{degree}	& \colhead{}	& \colhead{mag}	& \colhead{mag}		& \colhead{arcmin}		& \colhead{pMpc}	& \colhead{$\mathrm{s}^{-1}$}		& \colhead{}
}
\startdata
PC\,0058$+$0215      	 & $ 14.76370$ 	 & $  +1.87060$ 	 & $2.746$ 	 & $19.0$ 	 & $-26.1$ 	 & $48.7$ 	 & $23.1$ 	 & $6.99\e{-16}    $ 	 & NTT/EFOSC2  \\
PC\,0058$+$0215      	 & $ 15.06546$ 	 & $  +2.40075$ 	 & $2.890$ 	 & $20.7$ 	 & $-24.5$ 	 & $13.0$ 	 & $ 6.1$ 	 & $2.24\e{-15}    $ 	 & NTT/EFOSC2  \\
PC\,0058$+$0215      	 & $ 15.12711$ 	 & $  +2.50471$ 	 & $2.735$ 	 & $21.4$ 	 & $-23.7$ 	 & $ 7.1$ 	 & $ 3.4$ 	 & $3.61\e{-15}    $ 	 & VLT/VIMOS   \\
PC\,0058$+$0215      	 & $ 15.12755$ 	 & $  +2.59926$ 	 & $2.200$ 	 & $22.2$ 	 & $-22.5$ 	 & $ 8.2$ 	 & $ 4.1$ 	 & $7.84\e{-16}    $ 	 & VLT/VIMOS   \\
PC\,0058$+$0215      	 & $ 15.17677$ 	 & $  +2.28284$ 	 & $2.038$ 	 & $20.2$ 	 & $-24.3$ 	 & $15.1$ 	 & $ 7.6$ 	 & $1.21\e{-15}    $ 	 & NTT/EFOSC2  \\
PC\,0058$+$0215      	 & $ 15.28149$ 	 & $  +1.99379$ 	 & $2.620$ 	 & $19.1$ 	 & $-25.9$ 	 & $32.0$ 	 & $15.3$ 	 & $1.30\e{-15}    $ 	 & NTT/EFOSC2  \\
PC\,0058$+$0215      	 & $ 15.32373$ 	 & $  +2.54572$ 	 & $2.230$ 	 & $23.1$ 	 & $-21.6$ 	 & $ 5.0$ 	 & $ 2.5$ 	 & $9.11\e{-16}    $ 	 & VLT/VIMOS   \\
PC\,0058$+$0215      	 & $ 15.32518$ 	 & $  +2.60924$ 	 & $2.620$ 	 & $21.1$ 	 & $-24.0$ 	 & $ 7.0$ 	 & $ 3.4$ 	 & $4.56\e{-15}    $ 	 & VLT/VIMOS   \\
PC\,0058$+$0215      	 & $ 15.34928$ 	 & $  +2.50257$ 	 & $2.730$ 	 & $22.7$ 	 & $-22.5$ 	 & $ 6.5$ 	 & $ 3.1$ 	 & $1.32\e{-15}    $ 	 & VLT/VIMOS   \\
PC\,0058$+$0215      	 & $ 15.36430$ 	 & $  +2.55976$ 	 & $2.580$ 	 & $22.1$ 	 & $-22.9$ 	 & $ 7.5$ 	 & $ 3.6$ 	 & $1.48\e{-15}    $ 	 & VLT/VIMOS   \\
PC\,0058$+$0215      	 & $ 15.37384$ 	 & $  +2.74005$ 	 & $2.991$ 	 & $20.7$ 	 & $-24.7$ 	 & $15.1$ 	 & $ 7.0$ 	 & $1.93\e{-15}    $ 	 & NTT/EFOSC2  \\
PC\,0058$+$0215      	 & $ 15.44649$ 	 & $  +1.83921$ 	 & $2.751$ 	 & $20.4$ 	 & $-24.7$ 	 & $42.9$ 	 & $20.3$ 	 & $2.42\e{-16}    $ 	 & NTT/EFOSC2  \\
SDSS\,J0139$-$0847   	 & $ 24.77901$ 	 & $  -8.91407$ 	 & $2.600$ 	 & $20.0$ 	 & $-25.1$ 	 & $ 7.7$ 	 & $ 3.7$ 	 & $1.00\e{-14}    $ 	 & VLT/VIMOS   \\
SDSS\,J0139$-$0847   	 & $ 24.81955$ 	 & $  -8.73150$ 	 & $3.045$ 	 & $20.8$ 	 & $-24.6$ 	 & $ 5.2$ 	 & $ 2.4$ 	 & $1.51\e{-14}    $ 	 & VLT/VIMOS   \\
HE2QS\,J0233$-$0149  	 & $ 37.55054$ 	 & $  -1.59452$ 	 & $2.381$ 	 & $19.3$ 	 & $-25.6$ 	 & $45.7$ 	 & $22.4$ 	 & $4.34\e{-16}    $ 	 & NTT/EFOSC2  \\
HE2QS\,J0233$-$0149  	 & $ 37.85797$ 	 & $  -1.49224$ 	 & $2.498$ 	 & $19.5$ 	 & $-25.5$ 	 & $32.2$ 	 & $15.6$ 	 & $8.27\e{-16}    $ 	 & NTT/EFOSC2  \\
HE2QS\,J0233$-$0149  	 & $ 37.99089$ 	 & $  -1.58167$ 	 & $3.300$ 	 & $20.9$ 	 & $-24.6$ 	 & $22.7$ 	 & $10.2$ 	 & $8.72\e{-16}    $ 	 & NTT/EFOSC2  \\
HE2QS\,J0233$-$0149  	 & $ 38.15039$ 	 & $  -1.74063$ 	 & $2.813$ 	 & $21.5$ 	 & $-23.7$ 	 & $ 9.2$ 	 & $ 4.3$ 	 & $2.12\e{-15}    $ 	 & VLT/VIMOS   \\
HE2QS\,J0233$-$0149  	 & $ 38.16399$ 	 & $  -1.71300$ 	 & $2.230$ 	 & $21.9$ 	 & $-22.8$ 	 & $ 9.7$ 	 & $ 4.8$ 	 & $7.19\e{-16}    $ 	 & VLT/VIMOS   \\
HE2QS\,J0233$-$0149  	 & $ 38.28556$ 	 & $  -1.88356$ 	 & $2.244$ 	 & $20.8$ 	 & $-23.9$ 	 & $ 3.2$ 	 & $ 1.6$ 	 & $1.87\e{-14}    $ 	 & NTT/EFOSC2  \\
HE2QS\,J0233$-$0149  	 & $ 38.40945$ 	 & $  -1.67841$ 	 & $0.650$ 	 & $19.5$ 	 & $-22.3$ 	 & $12.2$ 	 & $ 5.1$ 	 & $4.11\e{-16}    $ 	 & NTT/EFOSC2  \\
HE2QS\,J0233$-$0149  	 & $ 38.80749$ 	 & $  -1.68010$ 	 & $2.700$ 	 & $19.6$ 	 & $-25.6$ 	 & $33.2$ 	 & $15.8$ 	 & $8.82\e{-16}    $ 	 & NTT/EFOSC2  \\
Q\,0302$-$003        	 & $ 46.08937$ 	 & $  -0.06252$ 	 & $3.152$ 	 & $24.5$ 	 & $-20.9$ 	 & $ 8.4$ 	 & $ 3.8$ 	 & $2.03\e{-16}    $ 	 & VLT/VIMOS   \\
Q\,0302$-$003        	 & $ 46.25255$ 	 & $  -0.20926$ 	 & $3.480$ 	 & $21.9$ 	 & $-23.7$ 	 & $ 5.1$ 	 & $ 2.2$ 	 & $8.03\e{-15}    $ 	 & VLT/VIMOS   \\
Q\,0302$-$003        	 & $ 46.26361$ 	 & $  -0.37372$ 	 & $3.647$ 	 & $20.4$ 	 & $-25.3$ 	 & $14.6$ 	 & $ 6.5$ 	 & $3.86\e{-15}    $ 	 & CAHA\,3.5/TWIN\\
Q\,0302$-$003        	 & $ 46.26589$ 	 & $  -0.07512$ 	 & $2.708$ 	 & $22.0$ 	 & $-23.1$ 	 & $ 5.1$ 	 & $ 2.4$ 	 & $3.98\e{-15}    $ 	 & VLT/VIMOS   \\
Q\,0302$-$003        	 & $ 46.27492$ 	 & $  -0.10460$ 	 & $3.450$ 	 & $21.3$ 	 & $-24.4$ 	 & $ 4.5$ 	 & $ 2.0$ 	 & $1.93\e{-14}    $ 	 & VLT/VIMOS   \\
Q\,0302$-$003        	 & $ 46.27830$ 	 & $  -0.01382$ 	 & $3.047$ 	 & $21.7$ 	 & $-23.7$ 	 & $ 8.5$ 	 & $ 3.9$ 	 & $2.52\e{-15}    $ 	 & VLT/VIMOS   \\
SDSS\,J0818$+$4908   	 & $124.58575$ 	 & $ +49.06787$ 	 & $2.214$ 	 & $21.4$ 	 & $-23.2$ 	 & $ 6.4$ 	 & $ 3.2$ 	 & $2.55\e{-15}    $ 	 & CAHA\,3.5/TWIN\\
SDSS\,J0818$+$4908   	 & $124.88381$ 	 & $ +49.46161$ 	 & $2.973$ 	 & $20.8$ 	 & $-24.5$ 	 & $20.6$ 	 & $ 9.5$ 	 & $9.27\e{-16}    $ 	 & CAHA\,3.5/TWIN\\
HS\,0911$+$4809      	 & $138.64102$ 	 & $ +47.91407$ 	 & $2.849$ 	 & $21.9$ 	 & $-23.4$ 	 & $ 6.4$ 	 & $ 3.0$ 	 & $3.12\e{-15}    $ 	 & CAHA\,3.5/TWIN\\
HE2QS\,J0916$+$2408  	 & $138.70401$ 	 & $ +24.15258$ 	 & $2.615$ 	 & $21.4$ 	 & $-23.7$ 	 & $21.0$ 	 & $10.1$ 	 & $3.87\e{-16}    $ 	 & NTT/EFOSC2  \\
HE2QS\,J0916$+$2408  	 & $138.72408$ 	 & $ +24.03026$ 	 & $2.903$ 	 & $20.3$ 	 & $-25.0$ 	 & $20.8$ 	 & $ 9.7$ 	 & $1.32\e{-15}    $ 	 & NTT/EFOSC2  \\
HE2QS\,J0916$+$2408  	 & $138.76413$ 	 & $ +24.30874$ 	 & $2.935$ 	 & $20.7$ 	 & $-24.6$ 	 & $20.5$ 	 & $ 9.5$ 	 & $1.02\e{-15}    $ 	 & NTT/EFOSC2  \\
HE2QS\,J0916$+$2408  	 & $138.89244$ 	 & $ +23.94818$ 	 & $2.489$ 	 & $21.0$ 	 & $-24.0$ 	 & $15.4$ 	 & $ 7.5$ 	 & $8.75\e{-16}    $ 	 & NTT/EFOSC2  \\
HE2QS\,J0916$+$2408  	 & $139.01622$ 	 & $ +23.89896$ 	 & $2.644$ 	 & $18.5$ 	 & $-26.6$ 	 & $14.7$ 	 & $ 7.1$ 	 & $1.09\e{-14}    $ 	 & CAHA\,3.5/TWIN\\
HE2QS\,J0916$+$2408  	 & $139.16456$ 	 & $ +24.19545$ 	 & $2.846$ 	 & $21.0$ 	 & $-24.3$ 	 & $ 5.6$ 	 & $ 2.6$ 	 & $9.44\e{-15}    $ 	 & NTT/EFOSC2  \\
HS\,0911$+$4809      	 & $139.27965$ 	 & $ +48.07376$ 	 & $2.630$ 	 & $20.6$ 	 & $-24.5$ 	 & $21.0$ 	 & $10.0$ 	 & $7.74\e{-16}    $ 	 & CAHA\,3.5/TWIN\\
HS\,0911$+$4809      	 & $139.48489$ 	 & $ +48.58183$ 	 & $2.793$ 	 & $18.6$ 	 & $-26.6$ 	 & $47.0$ 	 & $22.1$ 	 & $1.13\e{-15}    $ 	 & CAHA\,3.5/TWIN\\
HS\,1024$+$1849      	 & $156.21988$ 	 & $ +19.06936$ 	 & $1.600$ 	 & $19.5$ 	 & $-24.4$ 	 & $48.4$ 	 & $24.6$ 	 & $1.25\e{-16}    $ 	 & NTT/EFOSC2  \\
HS\,1024$+$1849      	 & $156.75602$ 	 & $ +18.66674$ 	 & $2.445$ 	 & $23.4$ 	 & $-21.5$ 	 & $ 9.5$ 	 & $ 4.6$ 	 & $2.45\e{-16}    $ 	 & VLT/VIMOS   \\
HS\,1024$+$1849      	 & $156.89198$ 	 & $ +19.23192$ 	 & $2.550$ 	 & $19.4$ 	 & $-25.6$ 	 & $39.5$ 	 & $19.0$ 	 & $6.05\e{-16}    $ 	 & CAHA\,3.5/TWIN\\
HS\,1024$+$1849      	 & $156.91902$ 	 & $ +18.66082$ 	 & $3.237$ 	 & $19.0$ 	 & $-26.5$ 	 & $ 5.4$ 	 & $ 2.4$ 	 & $8.70\e{-14}    $ 	 & VLT/VIMOS   \\
HS\,1024$+$1849      	 & $156.97978$ 	 & $ +18.71165$ 	 & $2.594$ 	 & $22.3$ 	 & $-22.7$ 	 & $ 9.6$ 	 & $ 4.6$ 	 & $7.30\e{-16}    $ 	 & VLT/VIMOS   \\
HS\,1024$+$1849      	 & $157.04017$ 	 & $ +18.45674$ 	 & $2.950$ 	 & $20.7$ 	 & $-24.6$ 	 & $11.0$ 	 & $ 5.1$ 	 & $3.53\e{-15}    $ 	 & NTT/EFOSC2  \\
HS\,1024$+$1849      	 & $157.05255$ 	 & $ +18.52800$ 	 & $2.200$ 	 & $21.0$ 	 & $-23.7$ 	 & $ 9.5$ 	 & $ 4.7$ 	 & $1.76\e{-15}    $ 	 & NTT/EFOSC2  \\
HS\,1024$+$1849      	 & $157.24707$ 	 & $ +18.48669$ 	 & $3.307$ 	 & $20.4$ 	 & $-25.1$ 	 & $20.9$ 	 & $ 9.3$ 	 & $1.68\e{-15}    $ 	 & NTT/EFOSC2  \\
SDSS\,J1101$+$1053   	 & $165.03416$ 	 & $ +10.89406$ 	 & $2.950$ 	 & $19.0$ 	 & $-26.3$ 	 & $26.4$ 	 & $12.3$ 	 & $2.81\e{-15}    $ 	 & NTT/EFOSC2  \\
SDSS\,J1101$+$1053   	 & $165.31383$ 	 & $ +10.93339$ 	 & $4.170$ 	 & $20.5$ 	 & $-25.3$ 	 & $10.4$ 	 & $ 4.2$ 	 & $9.12\e{-15}    $ 	 & NTT/EFOSC2  \\
SDSS\,J1101$+$1053   	 & $165.34595$ 	 & $ +10.78585$ 	 & $2.450$ 	 & $23.2$ 	 & $-21.7$ 	 & $10.0$ 	 & $ 4.8$ 	 & $2.64\e{-16}    $ 	 & VLT/VIMOS   \\
SDSS\,J1101$+$1053   	 & $165.34990$ 	 & $ +10.87617$ 	 & $3.000$ 	 & $21.4$ 	 & $-24.0$ 	 & $ 7.8$ 	 & $ 3.6$ 	 & $3.81\e{-15}    $ 	 & NTT/EFOSC2  \\
SDSS\,J1101$+$1053   	 & $165.38561$ 	 & $ +10.83785$ 	 & $2.580$ 	 & $20.2$ 	 & $-24.8$ 	 & $ 6.3$ 	 & $ 3.0$ 	 & $1.19\e{-14}    $ 	 & VLT/VIMOS   \\
SDSS\,J1101$+$1053   	 & $165.58475$ 	 & $ +10.78663$ 	 & $2.300$ 	 & $22.5$ 	 & $-22.3$ 	 & $ 8.4$ 	 & $ 4.1$ 	 & $6.08\e{-16}    $ 	 & VLT/VIMOS   \\
SDSS\,J1101$+$1053   	 & $165.60616$ 	 & $ +10.92984$ 	 & $2.490$ 	 & $23.5$ 	 & $-21.4$ 	 & $ 7.8$ 	 & $ 3.8$ 	 & $3.37\e{-16}    $ 	 & VLT/VIMOS   \\
SDSS\,J1101$+$1053   	 & $165.63311$ 	 & $ +10.76935$ 	 & $3.200$ 	 & $23.5$ 	 & $-21.9$ 	 & $11.2$ 	 & $ 5.1$ 	 & $2.96\e{-16}    $ 	 & VLT/VIMOS   \\
HS\,1157$+$3143      	 & $179.42216$ 	 & $ +31.73584$ 	 & $2.760$ 	 & $19.3$ 	 & $-25.9$ 	 & $35.5$ 	 & $17.2$ 	 & $9.67\e{-16}    $ 	 & CAHA\,3.5/TWIN\\
HS\,1157$+$3143      	 & $180.05389$ 	 & $ +31.92410$ 	 & $2.951$ 	 & $19.7$ 	 & $-25.6$ 	 & $29.0$ 	 & $13.5$ 	 & $1.20\e{-15}    $ 	 & CAHA\,3.5/TWIN\\
LBQ\,S1216$+$1656    	 & $184.62529$ 	 & $ +16.31541$ 	 & $2.857$ 	 & $20.1$ 	 & $-25.1$ 	 & $23.8$ 	 & $11.2$ 	 & $1.15\e{-15}    $ 	 & CAHA\,3.5/TWIN\\
LBQ\,S1216$+$1656    	 & $184.67681$ 	 & $ +16.91563$ 	 & $2.971$ 	 & $20.7$ 	 & $-24.6$ 	 & $17.9$ 	 & $ 8.3$ 	 & $1.30\e{-15}    $ 	 & CAHA\,3.5/TWIN\\
SDSS\,J1237$+$0126   	 & $189.30504$ 	 & $  +1.38074$ 	 & $0.624$ 	 & $19.1$ 	 & $-22.7$ 	 & $ 9.5$ 	 & $ 3.9$ 	 & $9.84\e{-16}    $ 	 & VLT/VIMOS   \\
SDSS\,J1237$+$0126   	 & $189.38441$ 	 & $  +1.45343$ 	 & $2.811$ 	 & $22.7$ 	 & $-22.6$ 	 & $ 4.3$ 	 & $ 2.0$ 	 & $3.29\e{-15}    $ 	 & VLT/VIMOS   \\
SDSS\,J1237$+$0126   	 & $189.56254$ 	 & $  +1.49484$ 	 & $0.773$ 	 & $19.3$ 	 & $-22.8$ 	 & $ 7.4$ 	 & $ 3.3$ 	 & $1.59\e{-15}    $ 	 & VLT/VIMOS   \\
SDSS\,J1237$+$0126   	 & $189.58793$ 	 & $  +1.56198$ 	 & $2.490$ 	 & $20.7$ 	 & $-24.2$ 	 & $11.1$ 	 & $ 5.4$ 	 & $2.25\e{-15}    $ 	 & VLT/VIMOS   \\
SDSS\,J1253$+$6817   	 & $192.54865$ 	 & $ +68.31659$ 	 & $2.653$ 	 & $19.8$ 	 & $-25.3$ 	 & $20.6$ 	 & $ 9.8$ 	 & $1.77\e{-15}    $ 	 & CAHA\,3.5/TWIN\\
SDSS\,J1253$+$6817   	 & $192.69830$ 	 & $ +68.30378$ 	 & $3.480$ 	 & $19.5$ 	 & $-26.2$ 	 & $17.2$ 	 & $ 7.7$ 	 & $6.83\e{-15}    $ 	 & CAHA\,3.5/TWIN\\
SDSS\,J1253$+$6817   	 & $192.75548$ 	 & $ +68.14322$ 	 & $3.653$ 	 & $20.1$ 	 & $-25.6$ 	 & $18.2$ 	 & $ 8.1$ 	 & $3.31\e{-15}    $ 	 & CAHA\,3.5/TWIN\\
SDSS\,J1253$+$6817   	 & $193.01715$ 	 & $ +68.27103$ 	 & $2.454$ 	 & $19.8$ 	 & $-25.2$ 	 & $10.2$ 	 & $ 5.1$ 	 & $5.79\e{-15}    $ 	 & CAHA\,3.5/TWIN\\
SDSS\,J1253$+$6817   	 & $193.16780$ 	 & $ +68.44593$ 	 & $2.841$ 	 & $20.7$ 	 & $-24.5$ 	 & $11.7$ 	 & $ 5.5$ 	 & $2.70\e{-15}    $ 	 & CAHA\,3.5/TWIN\\
SDSS\,J1253$+$6817   	 & $193.33580$ 	 & $ +68.27583$ 	 & $2.618$ 	 & $20.0$ 	 & $-25.1$ 	 & $ 3.1$ 	 & $ 1.5$ 	 & $6.11\e{-14}    $ 	 & CAHA\,3.5/TWIN\\
SDSS\,J1253$+$6817   	 & $193.41690$ 	 & $ +68.10003$ 	 & $2.871$ 	 & $21.2$ 	 & $-24.0$ 	 & $11.3$ 	 & $ 5.3$ 	 & $1.82\e{-15}    $ 	 & CAHA\,3.5/TWIN\\
SDSS\,J1253$+$6817   	 & $193.62625$ 	 & $ +68.22797$ 	 & $2.295$ 	 & $20.5$ 	 & $-24.3$ 	 & $ 4.9$ 	 & $ 2.5$ 	 & $1.08\e{-14}    $ 	 & CAHA\,3.5/TWIN\\
SDSS\,J1253$+$6817   	 & $193.87605$ 	 & $ +68.33807$ 	 & $3.210$ 	 & $19.4$ 	 & $-26.1$ 	 & $ 9.4$ 	 & $ 4.5$ 	 & $1.69\e{-14}    $ 	 & CAHA\,3.5/TWIN\\
SDSS\,J1253$+$6817   	 & $194.07395$ 	 & $ +68.30562$ 	 & $2.904$ 	 & $19.4$ 	 & $-25.9$ 	 & $13.4$ 	 & $ 6.2$ 	 & $7.35\e{-15}    $ 	 & CAHA\,3.5/TWIN\\
SBS\,1602$+$576      	 & $240.48315$ 	 & $ +57.59089$ 	 & $2.456$ 	 & $19.9$ 	 & $-25.0$ 	 & $16.7$ 	 & $ 8.3$ 	 & $1.96\e{-15}    $ 	 & CAHA\,3.5/TWIN\\
SBS\,1602$+$576      	 & $241.08702$ 	 & $ +57.40948$ 	 & $2.575$ 	 & $20.8$ 	 & $-24.2$ 	 & $ 7.2$ 	 & $ 3.6$ 	 & $4.95\e{-15}    $ 	 & CAHA\,3.5/TWIN\\
HE2QS\,J1630$+$0435  	 & $247.43896$ 	 & $  +4.59353$ 	 & $2.486$ 	 & $18.8$ 	 & $-26.2$ 	 & $17.7$ 	 & $ 8.0$ 	 & $5.99\e{-15}    $ 	 & CAHA\,3.5/TWIN\\
HE2QS\,J1630$+$0435  	 & $247.55734$ 	 & $  +5.15656$ 	 & $2.449$ 	 & $18.9$ 	 & $-26.0$ 	 & $35.0$ 	 & $17.5$ 	 & $1.06\e{-15}    $ 	 & CAHA\,3.5/TWIN\\
HE2QS\,J1630$+$0435  	 & $247.57873$ 	 & $  +4.64324$ 	 & $2.033$ 	 & $21.3$ 	 & $-23.2$ 	 & $ 9.7$ 	 & $ 4.4$ 	 & $1.31\e{-15}    $ 	 & CAHA\,3.5/TWIN\\
HE2QS\,J1630$+$0435  	 & $247.77057$ 	 & $  +5.02187$ 	 & $2.615$ 	 & $19.5$ 	 & $-25.6$ 	 & $25.4$ 	 & $12.5$ 	 & $1.43\e{-15}    $ 	 & CAHA\,3.5/TWIN\\
HE2QS\,J1630$+$0435  	 & $247.80360$ 	 & $  +4.54051$ 	 & $2.396$ 	 & $21.4$ 	 & $-23.5$ 	 & $ 5.4$ 	 & $ 2.5$ 	 & $5.19\e{-15}    $ 	 & CAHA\,3.5/TWIN\\
HS\,1700$+$6416      	 & $253.48743$ 	 & $ +63.56482$ 	 & $2.494$ 	 & $18.2$ 	 & $-26.8$ 	 & $60.3$ 	 & $29.5$ 	 & $7.46\e{-16}    $ 	 & CAHA\,3.5/TWIN\\
HS\,1700$+$6416      	 & $254.02245$ 	 & $ +64.12415$ 	 & $2.065$ 	 & $19.2$ 	 & $-25.3$ 	 & $32.5$ 	 & $16.4$ 	 & $6.49\e{-16}    $ 	 & CAHA\,3.5/TWIN\\
HS\,1700$+$6416      	 & $254.09336$ 	 & $ +64.59552$ 	 & $3.103$ 	 & $19.5$ 	 & $-25.9$ 	 & $38.2$ 	 & $18.2$ 	 & $8.80\e{-16}    $ 	 & CAHA\,3.5/TWIN\\
HS\,1700$+$6416      	 & $254.97282$ 	 & $ +64.02625$ 	 & $2.703$ 	 & $21.1$ 	 & $-24.1$ 	 & $12.9$ 	 & $ 6.1$ 	 & $1.43\e{-15}    $ 	 & CAHA\,3.5/TWIN\\
HS\,1700$+$6416      	 & $255.34232$ 	 & $ +64.57177$ 	 & $2.614$ 	 & $19.0$ 	 & $-26.1$ 	 & $22.3$ 	 & $11.0$ 	 & $2.94\e{-15}    $ 	 & CAHA\,3.5/TWIN\\
SDSS\,J1711$+$6052   	 & $255.37991$ 	 & $ +61.02177$ 	 & $3.352$ 	 & $18.8$ 	 & $-26.8$ 	 & $73.7$ 	 & $33.0$ 	 & $6.50\e{-16}    $ 	 & CAHA\,3.5/TWIN\\
HS\,1700$+$6416      	 & $255.57921$ 	 & $ +64.08182$ 	 & $2.628$ 	 & $20.9$ 	 & $-24.2$ 	 & $11.2$ 	 & $ 5.4$ 	 & $2.11\e{-15}    $ 	 & Keck/LRIS   \\
HS\,1700$+$6416      	 & $256.05985$ 	 & $ +64.32612$ 	 & $1.142$ 	 & $19.8$ 	 & $-23.2$ 	 & $22.3$ 	 & $11.0$ 	 & $2.00\e{-16}    $ 	 & CAHA\,3.5/TWIN\\
HS\,1700$+$6416      	 & $256.46234$ 	 & $ +65.44044$ 	 & $2.287$ 	 & $17.0$ 	 & $-27.8$ 	 & $80.4$ 	 & $39.8$ 	 & $1.06\e{-15}    $ 	 & CAHA\,3.5/TWIN\\
HS\,1700$+$6416      	 & $256.52142$ 	 & $ +64.64957$ 	 & $2.292$ 	 & $18.4$ 	 & $-26.4$ 	 & $42.4$ 	 & $20.9$ 	 & $1.05\e{-15}    $ 	 & CAHA\,3.5/TWIN\\
SDSS\,J1711$+$6052   	 & $257.36600$ 	 & $ +61.53623$ 	 & $2.402$ 	 & $18.6$ 	 & $-26.3$ 	 & $42.3$ 	 & $20.9$ 	 & $9.67\e{-16}    $ 	 & CAHA\,3.5/TWIN\\
SDSS\,J1711$+$6052   	 & $257.52998$ 	 & $ +61.07253$ 	 & $2.627$ 	 & $20.8$ 	 & $-24.3$ 	 & $15.8$ 	 & $ 7.7$ 	 & $1.17\e{-15}    $ 	 & CAHA\,3.5/TWIN\\
SDSS\,J1711$+$6052   	 & $257.55732$ 	 & $ +61.06330$ 	 & $2.611$ 	 & $20.5$ 	 & $-24.6$ 	 & $14.8$ 	 & $ 7.3$ 	 & $1.66\e{-15}    $ 	 & CAHA\,3.5/TWIN\\
SDSS\,J1711$+$6052   	 & $257.58850$ 	 & $ +61.10460$ 	 & $1.322$ 	 & $20.6$ 	 & $-22.7$ 	 & $16.2$ 	 & $ 7.3$ 	 & $2.99\e{-16}    $ 	 & CAHA\,3.5/TWIN\\
SDSS\,J1711$+$6052   	 & $257.64436$ 	 & $ +60.86447$ 	 & $2.216$ 	 & $20.9$ 	 & $-23.8$ 	 & $ 7.3$ 	 & $ 3.7$ 	 & $3.15\e{-15}    $ 	 & CAHA\,3.5/TWIN\\
SDSS\,J1711$+$6052   	 & $257.65977$ 	 & $ +61.29396$ 	 & $2.840$ 	 & $20.5$ 	 & $-24.7$ 	 & $25.9$ 	 & $12.3$ 	 & $6.53\e{-16}    $ 	 & CAHA\,3.5/TWIN\\
SDSS\,J1711$+$6052   	 & $258.06285$ 	 & $ +60.53481$ 	 & $3.171$ 	 & $20.7$ 	 & $-24.8$ 	 & $21.2$ 	 & $10.1$ 	 & $1.04\e{-15}    $ 	 & CAHA\,3.5/TWIN\\
SDSS\,J1711$+$6052   	 & $258.51527$ 	 & $ +61.51186$ 	 & $2.756$ 	 & $19.2$ 	 & $-26.0$ 	 & $42.1$ 	 & $20.4$ 	 & $7.58\e{-16}    $ 	 & CAHA\,3.5/TWIN\\
HE2QS\,J2149$-$0859  	 & $327.05627$ 	 & $  -8.96496$ 	 & $2.691$ 	 & $19.9$ 	 & $-25.2$ 	 & $18.4$ 	 & $ 9.0$ 	 & $1.98\e{-15}    $ 	 & CAHA\,3.5/TWIN\\
HE2QS\,J2149$-$0859  	 & $327.18175$ 	 & $  -9.06226$ 	 & $2.327$ 	 & $19.4$ 	 & $-25.3$ 	 & $11.9$ 	 & $ 6.0$ 	 & $4.97\e{-15}    $ 	 & CAHA\,3.5/TWIN\\
HE2QS\,J2149$-$0859  	 & $327.23032$ 	 & $  -9.02613$ 	 & $2.815$ 	 & $19.3$ 	 & $-25.9$ 	 & $ 8.4$ 	 & $ 4.1$ 	 & $1.82\e{-14}    $ 	 & CAHA\,3.5/TWIN\\
HE2QS\,J2149$-$0859  	 & $327.23682$ 	 & $  -9.05922$ 	 & $2.367$ 	 & $22.4$ 	 & $-22.4$ 	 & $ 8.9$ 	 & $ 4.3$ 	 & $6.17\e{-16}    $ 	 & VLT/VIMOS   \\
HE2QS\,J2149$-$0859  	 & $327.28763$ 	 & $  -9.03955$ 	 & $1.906$ 	 & $22.1$ 	 & $-22.2$ 	 & $ 5.7$ 	 & $ 2.9$ 	 & $1.24\e{-15}    $ 	 & VLT/VIMOS   \\
HE2QS\,J2149$-$0859  	 & $327.38216$ 	 & $  -8.90724$ 	 & $2.305$ 	 & $20.7$ 	 & $-24.1$ 	 & $ 4.7$ 	 & $ 2.3$ 	 & $9.98\e{-15}    $ 	 & VLT/VIMOS   \\
HE2QS\,J2149$-$0859  	 & $328.25071$ 	 & $  -8.83092$ 	 & $3.220$ 	 & $18.3$ 	 & $-27.2$ 	 & $53.3$ 	 & $25.0$ 	 & $1.50\e{-15}    $ 	 & CAHA\,3.5/TWIN\\
HE2QS\,J2157$+$2330  	 & $328.36882$ 	 & $ +23.48380$ 	 & $2.916$ 	 & $18.6$ 	 & $-26.6$ 	 & $58.5$ 	 & $28.1$ 	 & $7.47\e{-16}    $ 	 & CAHA\,3.5/TWIN\\
HE2QS\,J2157$+$2330  	 & $329.48269$ 	 & $ +23.16129$ 	 & $2.485$ 	 & $20.0$ 	 & $-24.9$ 	 & $21.1$ 	 & $10.5$ 	 & $1.09\e{-15}    $ 	 & CAHA\,3.5/TWIN\\
HE2QS\,J2157$+$2330  	 & $329.54253$ 	 & $ +23.82337$ 	 & $2.889$ 	 & $20.2$ 	 & $-25.1$ 	 & $19.7$ 	 & $ 9.5$ 	 & $1.54\e{-15}    $ 	 & CAHA\,3.5/TWIN\\
HE2QS\,J2157$+$2330  	 & $329.54756$ 	 & $ +23.69127$ 	 & $2.876$ 	 & $21.0$ 	 & $-24.2$ 	 & $12.6$ 	 & $ 5.9$ 	 & $1.85\e{-15}    $ 	 & VLT/VIMOS   \\
HE2QS\,J2157$+$2330  	 & $329.92955$ 	 & $ +23.45275$ 	 & $2.796$ 	 & $19.7$ 	 & $-25.5$ 	 & $27.6$ 	 & $13.2$ 	 & $1.14\e{-15}    $ 	 & CAHA\,3.5/TWIN\\
SDSS\,J2200$+$0008   	 & $330.07677$ 	 & $  +0.23124$ 	 & $2.380$ 	 & $22.6$ 	 & $-22.2$ 	 & $ 7.8$ 	 & $ 3.8$ 	 & $6.88\e{-16}    $ 	 & VLT/VIMOS   \\
SDSS\,J2200$+$0008   	 & $330.10915$ 	 & $  +0.07871$ 	 & $2.810$ 	 & $22.8$ 	 & $-22.4$ 	 & $ 5.3$ 	 & $ 2.5$ 	 & $1.82\e{-15}    $ 	 & VLT/VIMOS   \\
HE2QS\,J2157$+$2330  	 & $330.20502$ 	 & $ +23.61050$ 	 & $2.370$ 	 & $17.8$ 	 & $-27.0$ 	 & $42.9$ 	 & $21.2$ 	 & $1.86\e{-15}    $ 	 & CAHA\,3.5/TWIN\\
SDSS\,J2251$-$0857   	 & $342.75770$ 	 & $  -9.02860$ 	 & $3.590$ 	 & $21.6$ 	 & $-24.0$ 	 & $ 5.9$ 	 & $ 2.5$ 	 & $8.11\e{-15}    $ 	 & VLT/VIMOS   \\
SDSS\,J2251$-$0857   	 & $342.76396$ 	 & $  -8.87532$ 	 & $3.450$ 	 & $21.0$ 	 & $-24.7$ 	 & $ 6.0$ 	 & $ 2.7$ 	 & $1.35\e{-14}    $ 	 & VLT/VIMOS   \\
SDSS\,J2251$-$0857   	 & $342.78089$ 	 & $  -8.99061$ 	 & $2.200$ 	 & $24.0$ 	 & $-20.6$ 	 & $ 3.3$ 	 & $ 1.6$ 	 & $8.73\e{-16}    $ 	 & VLT/VIMOS   \\
SDSS\,J2251$-$0857   	 & $342.79588$ 	 & $  -8.82989$ 	 & $2.838$ 	 & $21.8$ 	 & $-23.5$ 	 & $ 7.8$ 	 & $ 3.6$ 	 & $2.38\e{-15}    $ 	 & VLT/VIMOS   \\
SDSS\,J2251$-$0857   	 & $342.96531$ 	 & $  -9.07376$ 	 & $2.352$ 	 & $24.4$ 	 & $-20.5$ 	 & $10.9$ 	 & $ 5.4$ 	 & $6.81\e{-17}    $ 	 & VLT/VIMOS   \\
SDSS\,J2346$-$0016   	 & $356.50533$ 	 & $  -0.33070$ 	 & $2.270$ 	 & $23.0$ 	 & $-21.7$ 	 & $ 7.2$ 	 & $ 3.6$ 	 & $4.97\e{-16}    $ 	 & VLT/VIMOS   \\
SDSS\,J2346$-$0016   	 & $356.58265$ 	 & $  -0.20466$ 	 & $3.294$ 	 & $23.2$ 	 & $-22.3$ 	 & $ 4.0$ 	 & $ 1.8$ 	 & $3.32\e{-15}    $ 	 & VLT/VIMOS   \\
SDSS\,J2346$-$0016   	 & $356.63806$ 	 & $  -0.17529$ 	 & $2.615$ 	 & $21.0$ 	 & $-24.0$ 	 & $ 5.8$ 	 & $ 2.8$ 	 & $6.95\e{-15}    $ 	 & NTT/EFOSC2  \\
HE\,2347$-$4342      	 & $357.23866$ 	 & $ -43.44527$ 	 & $2.241$ 	 & $20.7$ 	 & $-24.0$ 	 & $17.6$ 	 & $ 8.7$ 	 & $7.09\e{-16}    $ 	 & NTT/EFOSC2  \\
HE\,2347$-$4342      	 & $357.27585$ 	 & $ -43.65106$ 	 & $2.246$ 	 & $19.3$ 	 & $-25.4$ 	 & $20.6$ 	 & $10.2$ 	 & $1.78\e{-15}    $ 	 & NTT/EFOSC2  \\
HE\,2347$-$4342      	 & $357.28091$ 	 & $ -43.45866$ 	 & $2.389$ 	 & $20.3$ 	 & $-24.5$ 	 & $15.8$ 	 & $ 7.7$ 	 & $1.38\e{-15}    $ 	 & NTT/EFOSC2  \\
HE\,2347$-$4342      	 & $357.32826$ 	 & $ -43.15734$ 	 & $1.560$ 	 & $20.5$ 	 & $-23.3$ 	 & $21.5$ 	 & $10.9$ 	 & $2.25\e{-16}    $ 	 & NTT/EFOSC2  \\
HE\,2347$-$4342      	 & $357.38097$ 	 & $ -43.64325$ 	 & $1.682$ 	 & $19.8$ 	 & $-24.2$ 	 & $17.0$ 	 & $ 8.6$ 	 & $8.51\e{-16}    $ 	 & NTT/EFOSC2  \\
HE\,2347$-$4342      	 & $357.60328$ 	 & $ -43.57593$ 	 & $2.270$ 	 & $22.6$ 	 & $-22.1$ 	 & $ 8.7$ 	 & $ 4.3$ 	 & $4.94\e{-16}    $ 	 & VLT/VIMOS   \\
HE\,2347$-$4342      	 & $357.71931$ 	 & $ -43.51556$ 	 & $1.790$ 	 & $21.2$ 	 & $-23.0$ 	 & $ 6.0$ 	 & $ 3.0$ 	 & $2.31\e{-15}    $ 	 & VLT/VIMOS   \\
HE\,2347$-$4342      	 & $357.77952$ 	 & $ -43.37940$ 	 & $0.704$ 	 & $19.2$ 	 & $-22.8$ 	 & $ 6.8$ 	 & $ 2.9$ 	 & $1.99\e{-15}    $ 	 & NTT/EFOSC2  \\
HE\,2347$-$4342      	 & $357.82798$ 	 & $ -43.52736$ 	 & $1.730$ 	 & $23.0$ 	 & $-21.1$ 	 & $ 9.9$ 	 & $ 5.0$ 	 & $1.42\e{-16}    $ 	 & VLT/VIMOS 
\enddata
\tablenotetext{}{For the sightlines of SDSS\,J0139$-$0847, LBQS\,1216$+$1656, SDSS\,J2200$+$0008 and J\,2251$-$0857 no \ion{He}{ii} spectra are available. Despite that, we still list discovered forground quasars.}
\label{Tab:Objects}
\end{deluxetable*}

\end{document}